\documentclass[10pt,a4paper]{article} 
\input{mystyle.sty}

\begin{document}
	\title{Shirac: A linear algebra for event-based system modeling}
	\author{Iwan Feras Fattohi, feras.fattohi@uni-ulm.de, \\Christian Prehofer, c.prehofer@eu.denso.com, \\ Frank Slomka, frank.slomka@uni-ulm.de}
	\date{\today} % no date
	\maketitle
	\newpage
	\tableofcontents
	\newpage
	\section{Introduction}
Digital signal theory is an extension of the analysis of continuous signals.
This extension is provided by discretization and sampling \cite{bracewell2000fourier}.
The sampling of signals can be mathematically described by a series of Dirac impulses \cite{bracewell2000fourier} and is well known.
Properties of the Dirac impulse, such as sampling, are derived in distribution theory \cite{grubb2008distributions} \cite{strichartz2003guide}.
The theory generalizes differential calculus to functions that are not differentiable in the classical sense such as the Heaviside step function.
Therefore, distribution theory allows one to adopt analog analysis concepts to digital signals.

The performance of computing systems is often modeled by automata as presented in the domain of model checking \cite{clarke1997model} or in queuing theory \cite{shortle2018fundamentals} of computer networks. 
Other approaches such as network \cite{le2001network} and real-time calculus \cite{thiele2000real} are working on stepwise, monotone and linear functions. Both models are abstract.
In our opinion, a clear mathematical concept to compute arbitrary event occurrences in real-time analysis is missing.

In this report, we extend the concept of Dirac combs, a series of Dirac impulses as known from signal theory, to performance analysis of computers. 
The goal is to connect methods from electrical engineering or physics to different models of computation such as graphs, and network as well as real-time calculus.

Our goal is to build the mathematical foundation on an extended algebra and calculus of impulses that we call the Shirac (\underline{Shi}fted Di\underline{rac}) algebra. To this end, we develop a linear space of any combination of digital signals, messages or performance requests.
The idea is that it should be possible to compute on any kind of combination of execution traces before they are being analyzed.
This means complicated message patterns as known from networking or real-time system analysis can be mathematically constructed by combinations of simple elements.

Such an approach enables a general performance analysis of computing systems.
The main advantage of this work is that we only need operations already known by every engineer from their mathematics courses known as linear algebra and calculus.

We first give a short introduction to distributions and the concept to compute with them, in Section \ref{section:distributiontheory}. 
Based on this foundation, we prove the linear space of traces or impulses, in Section \ref{section:impulsespace}, and provide a list of symbols in Table \ref{table:listofsymbols}. 
Lastly, we present a calculus of the linear space in Section \ref{section:operations}.
The goal of the report is to build a base of this theory which can be later used by computer science applications.

\section{Distribution theory}\label{section:distributiontheory}
\begin{figure}[t!]
	\centering
	\includegraphics[scale=0.43]{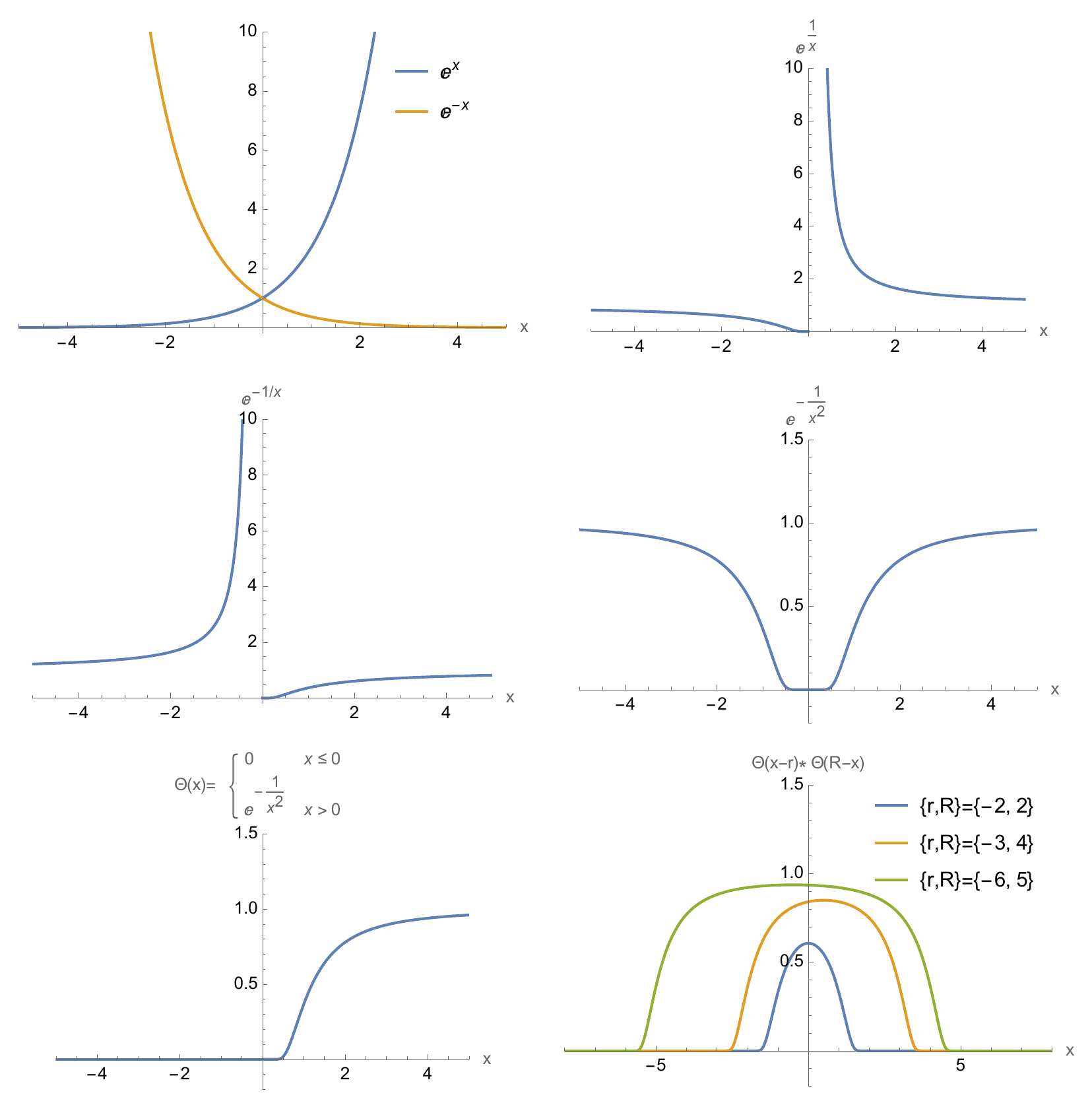}
	\caption{The test function $\testfunction_{r,R}(x)  = \theta(x - r) \cdot \theta(R- x)$ is constructed from exponential functions. The construction starts with the function $\operatorname{exp}(-1/x^2)$. From $\lim_{x \searrow 0} \operatorname{exp}(-1/x^2) = 0$ and  $\lim_{x \searrow 0} \frac{d}{dx} \operatorname{exp}(-1/x^2) = 0$, it follows that $\theta(x)$ is differentiable (in fact it is infinitely differentiable \cite{grubb2008distributions}). Thus, we get a function $\theta(x)$ that is zero for $x \leq 0$, non-zero for $x > 0$ and infinitely often differentiable. By linear combination of $\theta$ with itself, we can construct a function $\testfunction_{r,R}(x)  = \theta(x - r) \cdot \theta(R- x)$  that is differentiable everywhere. On top of that, we can choose by parameters $r$ and $R$ where $\testfunction_{r,R}(x)$ shall be non-zero and zero.}
	\label{fig:testfunctionconstruction}
\end{figure}
As we use results of distribution theory in our proofs we give a short-hand introduction to distributions. A complete view to the theory can be found in \cite{grubb2008distributions} and  \cite{strichartz2003guide}.
\begin{definition}[Notation for function sets \cite{grubb2008distributions}]
	Let $\Omega \subseteq \realnumbers^{n}$ be an open subset of $\realnumbersvector{n}$. 
	The spaces of continuous, continuously differentiable, $n$-times continuously differentiable and infinitely continuous differentiable functions on $\Omega$ are respectively denoted by $C(\Omega)$, $C^{1}(\Omega)$, $C^{n}(\Omega)$ and $C^{\infty}(\Omega)$.
\end{definition}
\begin{definition}[Support of a function \cite{grubb2008distributions}]
	The \textbf{support} of a function $f\colon \Omega \to \realnumbers$ where $\Omega \subseteq \realnumbers$ is the complement of the set of inputs $x \in \Omega$ to $f$ that are mapped to $0$. Formally,
	\begin{equation} 
		\operatorname{supp}(f) = \{x \in \Omega\,| \, f(x) \neq 0 \}
	\end{equation}
	Function $f$ has a \textbf{compact support} if $\operatorname{supp}(f)$ is closed and bounded. $C^{\infty}_{0}(\Omega)$ denotes the set of infinitely continuous differentiable functions on $\Omega$.
\end{definition}

\begin{definition}[Test function \cite{grubb2008distributions}]\label{def:testfunction}
	A \textbf{test function} is an $\infty$-times continuously differentiable function with compact support in $\Omega$ denoted by $\testfunction \in C^{\infty}_{0}(\Omega)$. $\mathcal{D}(\Omega) = C^{\infty}_{0}(\Omega)$ is called the \textbf{space of test functions}.
\end{definition}
A test function can be constructed by a combination of exponential functions.
A well-known test function is \cite{strichartz2003guide}
\begin{alignat}{1}
	\theta(x) &= 
	\left\{
	\begin{array}{ll}
		e^{-1/x^{2}} &~~ x > 0 \\~
		0 & ~~x \leq 0
	\end{array}
	\right.
	\\
	\testfunction_{r,R}(x)  & = \theta(x - r) \cdot \theta(R- x) ~~~~r,R \in \realnumbers
\end{alignat}
whose construction based on exponential functions is depicted in Figure \ref{fig:testfunctionconstruction}.
%\begin{figure}
%	\centering
%	\includegraphics[scale=0.9]{img/testfunction.pdf}
%	\caption{Test function 	$\testfunction_{-3,3}(x)$ }
%	\label{fig:testfunction}
%\end{figure}
We can construct further test functions based on $\testfunction_{r,R}(x)$ by linear combinations of $\testfunction_{r,R}(x)$ with itself since the space of test functions is known to be a vector space \cite{grubb2008distributions} on $\realnumbers$.

Let $f \colon \realnumbers \to \realnumbers$ be a function with a jump discontinuity at $x_0 \in \realnumbers$. 
A goal in distribution theory is to find a derivative of $f$ at $x_0$ by using a test function. The concept of a function is generalized and is called a distribution.
\begin{definition}[Distribution \cite{grubb2008distributions}]
	A \textbf{distribution} $f \colon C^{\infty}_{0}(\Omega) \to \Omega$ is a continuous linear functional. The vector space of distributions on $\Omega$ is denoted $\mathcal{D'}(\Omega)$. When $f \in \mathcal{D'}(\Omega)$, we denote the value of $f$ on $\testfunction \in C^{\infty}_{0}(\Omega)$ by $f(\testfunction)$ or $\langle f, \testfunction \rangle$.
\end{definition}
A subset of the distributions are the locally integrable functions which means that certain functions can be extended to distributions.
\begin{theorem}[Locally integrable functions \cite{strichartz2003guide}]\label{theorem:locallyintegrable}
	If $f\colon \Omega \to \realnumbers$ is a \textbf{locally integrable} function, i.e. $\int_{K} |f(x)|dx < \infty$ holds for compact sets $K \subset \Omega$, then $f \in \mathcal{D'}(\Omega)$. 
\end{theorem}
Let $f$ be locally integrable. To evaluate $f$ as a distribution, we have to apply $f$ to a test function $\testfunction$. $\langle f, \testfunction \rangle$ denotes that $f$ is applied to $\testfunction$. $\langle f, \testfunction \rangle$ is usually computed using partial integration. For example, distribution theory presents an approach to solve $\langle f', \testfunction \rangle$ where $f'$ denotes the derivative of $f$.  
More precisely, let us consider a function $f$ that is locally integrable and that has jump discontinuities. 
In contrast to classical calculus, distribution theory presents a method to compute a derivative for $f$.
To this end, assume there exists a sequence of differentiable and locally integrable functons $f_n$ approaching $f$ for $n \to \infty$. Then, the derivative $f'$ of $f$ can be determined by applying a test function $\testfunction$ to $f'$ as follows \cite{grubb2008distributions} \cite{strichartz2003guide}:
\begin{alignat}{1}
	\langle f', \testfunction \rangle &
	=
	\int_{-\infty}^{\infty} f'(x) \cdot \testfunction(x)\, dx 
	\label{eq:distributionalderivative1}
	\\
	&
	\lim\limits_{n \to \infty} \int_{-\infty}^{\infty} f_{n}'(x) \cdot \testfunction(x)\, dx
	\\
	&
	=
	\lim\limits_{n \to \infty}
	[f_{n}(x) \, \testfunction(x)]_{-\infty}^{\infty} - \int_{-\infty}^{\infty} f_{n}(x) \, \testfunction'(x) \, dx
	\\
	&
	=
	\lim\limits_{n \to \infty}
	[f_{n}(x) \, \testfunction(x)]_{-\infty}^{\infty} - \int_{-\infty}^{\infty} f_{n}(x) \, \testfunction'(x) \, dx
	\\
	&
	=
	\lim\limits_{n \to \infty}
	[\underbrace{f_{n}(\infty)}_{< \, \infty}  \underbrace{\testfunction(\infty)}_{= \, 0} - \underbrace{f_{n}(-\infty)}_{ < \, \infty} \underbrace{\testfunction(-\infty)}_{ = \, 0}] - \int_{-\infty}^{\infty} f_{n}(x) \, \testfunction'(x) \, dx
	\\
	&
	=
	\lim\limits_{n \to \infty}
	- \int_{-\infty}^{\infty} f_{n}(x) \, \testfunction'(x) \, dx
	\\
	&
	=
	- \int_{-\infty}^{\infty} f(x) \, \testfunction'(x) \, dx
	\\
	&
	=
	- \langle f, \testfunction' \rangle
	\label{eq:distributionalderivative2}
\end{alignat}
Equations \eqref{eq:distributionalderivative1} to \eqref{eq:distributionalderivative2} show that the derivative $f'$ of a discontinuous function $f$ is computed by applying $f$ to the derivative $\testfunction'$ of the test function $\testfunction$. Informally, the problem of differentiating $f$ is moved to the differentiation of $\testfunction$. Since $\testfunction$ is infinitely differentiable ($\testfunction$ is e.g. a combination of exponential functions), $\testfunction'$ exists. Thus, $\testfunction'$ can be used to define the derivative of $f$ acording to Equations \eqref{eq:distributionalderivative1} to \eqref{eq:distributionalderivative2} which is called the distributional derivative of $f$ \cite{grubb2008distributions} \cite{strichartz2003guide}. This is a basic approach in distribution theory to describe derivatives of discontinuous functions.

A fundamental discontinuous function is the Heaviside function as it describes a single jump. It is named after Oliver Heaviside who used this function to describe an electric circuit when it is switched on \cite{josephs1950heaviside}. We derive in the following the derivative of the Heaviside function which is known to be the Dirac impulse  as a special case of the equation $\langle f', \testfunction \rangle = - \langle f, \testfunction' \rangle$.

\subsection{Heaviside function and Dirac delta}\label{section:heavisidefunctionanddiracdelta}
The goal of this report is to compute on impulses. 
To this end, a mathematical concept of an impulse is needed.
Such a concept is the Dirac impulse first introduced by Paul Dirac \cite{dirac1981principles}.
There exist a lot of text books introducing the Dirac impulse to signal theory, communication theory and circuit analysis.
However, to support the linear algebraic approach presented in this work, we revise the introduction of the Dirac impulse in the way it was originally given by Paul Dirac in \cite{dirac1981principles}. 
He already applied partial integration as in distribution theory to compute the derivative of a Heaviside function which we also apply in our proofs. 
Therefore, we start with the Heaviside function.
\begin{definition}[Heaviside function \cite{strichartz2003guide}]\label{def:heavisidefunction}
	The \textbf{Heaviside function} $\heaviside$ is defined by
	\begin{alignat}{1}
		\heaviside(x) \coloneqq 
		\left\{ 
		\begin{array}{ll}
			1,& x > 0 \\
			0,& x < 0 \\
			\heaviside(0), & x = 0
		\end{array}
		\right.
	\end{alignat}
	where $\heaviside(0) \in [0,1]$.
\end{definition}
\begin{figure}
	\centering
	\includegraphics[scale=0.7]{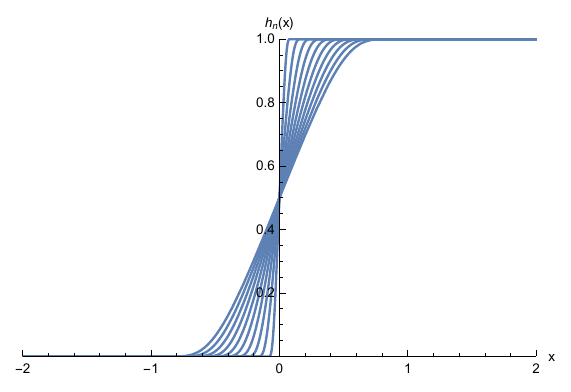}
	\caption{Family of functions $h_{k}(x)$ as defined by Equation \eqref{eq:heavisideapprox} for $k = 0.1, 0.2, \dots, 1$}
	\label{fig:heavisideapprox}
\end{figure}
An explanation for defining $\heaviside(0) \in [0,1]$ is provided by the derivative $ \frac{d}{dx} (\heaviside(x)) |_{x=0}$  of the Heaviside function at $x = 0$.
This derivative can be computed by using partial integration and by approximating the Heaviside function for $k \to 0$ with a family of continuously differentiable functions
and the properties
\begin{alignat}{1}\label{eq:heavisideapproxproperties}
	h_{k}(x) &= 1,~~~ x > k \\
	h_{k}(x) &= 0, ~~~ x < -k
\end{alignat}
and $h_{k}(x) \in C^{\infty}$ for all $k \in \naturalnumbers$. An exemplary approximation is the family of functions
\begin{equation}\label{eq:heavisideapprox}
	h_{k}(x) = \left\{\begin{array}{lll}
		\mathlarger{\frac{1}{1 + \operatorname{e}^{\frac{4kx}{x^{2}-k^{2}}}}}   &,|x| &< k 
		\vspace{3pt}
		\\
		0  &,x &< - k 
		\\
		1  &,x &> k
	\end{array}
	\right.
\end{equation}
that is depicted in Figure \ref{fig:heavisideapprox}.
Based on the Heaviside approximation $h_{k}$, we compute the derivative $ \frac{d}{dx} (\heaviside(x)) $ as follows

%This derivative can be computed using partial integration if it is locally integrable in intervals not including 0.
%This is a key idea in distribution theory.
%Since $	\int_{a}^{0} \frac{d}{dx} (\heaviside(x)) \, dx < \infty$ and $	\int_{0}^{b} \frac{d}{dx} (\heaviside(x)) \, dx < \infty$ for $a \in \realnumbers^{-}$ and $b \in \realnumbers^{+}$, it follows that $ \frac{d}{dx} (\heaviside(x))$ is locally integrable in $[a,0]$ and $[0,b]$, so that we can compute its value by partial integration. Let $\testfunction \in \mathcal{D}$ be a test function. Then,

\begin{alignat}{1}
	\langle \heaviside', \testfunction \rangle 
	&
	=
	\lim_{k \to 0} \, \langle h_{k}', \testfunction \rangle
	\\
	&
	=
	\lim_{k \to 0} 
	\int_{-\infty}^{\infty} h_{k}'(x) \, \testfunction(x) \, dx
	\\
	&
	=
	\lim_{k \to 0} 
	[h_{k}(x) \, \testfunction(x)]_{-\infty}^{\infty} - \int_{-\infty}^{\infty} h_{k}(x) \,  \testfunction'(x) \, dx
	\\
	&
	=
	\lim_{k \to 0} 
	(1\cdot 0 - 0 \cdot 0)- \int_{-\infty}^{-k} \underbrace{h_{k}(x)}_{= \, 0}  \testfunction'(x) \, dx - \int_{-k}^{k} h_{k}(x) \, \testfunction'(x) \, dx 
	\nonumber
	\\
	& ~~~~~~~~~~~~~~~~~~~~~~~~~~~~~~\,
	-   \int_{k}^{\infty}\underbrace{ h_{k}(x)}_{= \, 1}  \testfunction'(x) \, dx
	\label{eq:heavisideintegralzero}
	\\
	&
	=
	 -  \int_{0}^{\infty} \testfunction'(x) \, dx		\label{eq:heavisideintegralzer2}
	\\
	&
	=
	-[\testfunction(x)]_{0}^{\infty}
	\\
	&
	=
	\testfunction(0)
	\\
	&
	\eqqcolon
	\langle \diracdelta, \testfunction \rangle
\end{alignat}
which is defined as the Dirac impulse. Note that Equation \eqref{eq:heavisideintegralzer2} is a special case of $\langle f', \testfunction \rangle = - \langle f, \testfunction' \rangle$ from Equations \eqref{eq:distributionalderivative1} to \eqref{eq:distributionalderivative2} when $f = \heaviside$.
Furthermore, note that we could define $\heaviside(0) = 0$ or $\heaviside(0) = 1$. But neither of these definitions impact the integral in Equation \eqref{eq:heavisideintegralzero}. 
Both definitions lead to the result $\langle \heaviside', \testfunction \rangle = 	\testfunction(0)$. This is a well-known property of distributions \cite{strichartz2003guide}: There may exist multiple functions that are mapped to the same distribution. 
In case of the Heaviside function, both definitions $\heaviside(0) = 0$ and $\heaviside(0) = 1$ yield the same distribution. 
As a result of this partial integration, Paul Dirac derived in \cite{dirac1981principles} the Dirac impulse as the derivative of the Heaviside function.
\begin{definition}[Dirac delta function]\label{def:diracdelta}
	The \textbf{Dirac delta function} $\delta$ is defined
	\begin{alignat}{1}
		\diracdelta(\timepoint) = \frac{d\heaviside(\timepoint)}{d\timepoint} \,.
	\end{alignat}
\end{definition}
Note that Dirac delta distribution, impulse function, Dirac delta, Dirac impulse and Dirac delta impulse are synonyms of the Dirac delta function.
\subsection{Calculus of distributions}\label{section:calculusofdistribution}
Besides of generalizing differential calculus, distribution theory presents concepts to calculate with derivatives of discontinuous functions.
In other words, operations on distributions can be defined.
%The well-known sampling property of the Dirac impulse $\int \diracdelta(t-a)\phi(t)dt = \phi(a)$ is based on such an operation.
%If we change the shift of $\diracdelta(t-a)$, which means we operate on it, then $\phi$ is sampled at a different shift.
%In other words, an operation on a distribution causes an operation on the test function.
The theory presents the extension of operations such as addition and convolution from functions to distributions.
For this reason, they are also called generalized functions \cite{grubb2008distributions}.
In this section, we introduce operations on distributions which later
 allow us to construct a vector space.
 
An operation on a distribution can be defined based on the concept of \textbf{adjoint maps} \cite{grubb2008distributions}, also called \textbf{adjoint identities} \cite{strichartz2003guide}. 
Informally, the idea is to find the result of an operation on a distribution by finding a mathematically equal result of applying a different (adjoint) operation to a test function. 
This equality is called the adjoint map. 
An example of an adjoint map is $\langle f', \testfunction \rangle = \langle f, \testfunction' \rangle$ from Equation \eqref{eq:distributionalderivative1} to \eqref{eq:distributionalderivative2} where differentiation as an operation on a distribution $f$ is defined by differentation of the test function $\testfunction$. 
We present its general definition followed up by a textbook example of an adjoint map known as the sampling property of the Dirac impulse.

An adjoint map is described by two continuous linear transformations $T, T^{\prime} \colon \mathcal{D}(\Omega) \to \mathcal{D}(\Omega)$ on the vector space of test functions.
$T$ and $T^{\prime}$ being linear means for 
$\testfunction_1, \testfunction_2 \in \mathcal{D}(\Omega), \lambda \in \realnumbers$ that
\begin{alignat}{1}
	T(\testfunction_1 + \testfunction_2) &= T(\testfunction_1) + T(\testfunction_2) \\
	T(\lambda \cdot \testfunction_1) &= \lambda \cdot T(\testfunction)
\end{alignat}
hold
and accordingly for $T^{\prime}$ too.
An adjoint map between $T$ and $T^{\prime}$ means that
\begin{equation}\label{eq:adjoinmaptestfunctions}
	\int_{-\infty}^{\infty} T (\testfunction(x)) \, \theta(x) \, dx = \int_{-\infty}^{\infty}  \testfunction(x) \, T^{\prime} (\theta(x)) \, dx
\end{equation}
holds for any $\testfunction, \theta \in \spaceoftestfunctions$ \cite{strichartz2003guide}.
Equation \ref{eq:adjoinmaptestfunctions} states that the integral of $T (\testfunction(x)) \cdot \theta(x)$ is equal to the integral of $\testfunction(x) \cdot T^{\prime} (\theta(x))$.
The operation $T$ on a test function $\testfunction$ can be reformulated to another (adjoint) operation $T^{\prime}$ that is applied to the test function $\theta$.
In this way, the result of the operator $T$ is defined.
$T (\testfunction(x))$ is defined to have an impact on a test function $\theta$. 
If $\testfunction$ were a distribution, then an operation on $\testfunction$ would be defined by finding an operator $T'$ that manipulates the test function $\theta$.
As a result, an operation on a distribution would be defined by finding its impact on a test function.
%Let us assume that Equation \ref{eq:adjoinmaptestfunctions} holds. 
%Furthermore, assume that we do not know how to solve $\int_{-\infty}^{\infty} T (\testfunction(x)) \, \theta(x) \, dx$.
%If we know the solution of $\int_{-\infty}^{\infty}  \testfunction(x) \, T^{\prime} (\theta(x)) \, dx$, then we also know the solution of $\int_{-\infty}^{\infty} T (\testfunction(x)) \, \theta(x) \, dx$ by Equation \ref{eq:adjoinmaptestfunctions}.
%The original problem $\int_{-\infty}^{\infty} T (\testfunction(x)) \, \theta(x) \, dx$ is solved by solving another (adjoint) problem $\int_{-\infty}^{\infty}  \testfunction(x) \, T^{\prime} (\theta(x)) \, dx$.

In fact,  adjoint maps can be extended from test functions to distributions based on the following theorem stating that any distribution can be approximated by a sequence of test functions.
\begin{theorem}[Distribution from test functions \cite{strichartz2003guide}]\label{theorem:distributionfromtestfunctions}
	Given any distribution $f \in \mathcal{D'}(\Omega)$, there exists a sequence $(\testfunction_n)$ of test functions such that $f = \lim_{n \to \infty} (\testfunction_n)$ as distributions.
\end{theorem}
In this way, new operations on distributions can be defined by finding an adjoint map on test functions.
More formally, the idea is to define an operation $T$ on a distribution $f$ as a result of a limit process in which $f$ is approximated by a sequence of test functions $\testfunction_n$:
\begin{equation}\label{eq:operatorintegral}
	\langle T f, \testfunction \rangle = \int_{-\infty}^{\infty} T(f(x)) \, \testfunction(x) \, dx =  \lim_{n \to \infty}  \int_{-\infty}^{\infty} T(\testfunction_{n}(x)) \, \testfunction(x) \, dx
	=
	\lim_{n \to \infty} \langle T \testfunction_{n}, \testfunction\rangle \, .
\end{equation}
Note that a test function is a distribution since it is locally integrable (Theorem \ref{theorem:locallyintegrable}) which implies $\mathcal{D} \subseteq \mathcal{D'}$ and therefore we can write 	$\langle T f, \testfunction \rangle  = 	\lim_{n \to \infty} \langle T \testfunction_{n}, \testfunction\rangle$.
We observe that the problem $T(f)$ is reduced to the problem $T(\testfunction_{n})$ based on Theorem \ref{theorem:distributionfromtestfunctions}. 
The reason is that we know how to solve $T(\testfunction_{n})$ using the well-known algebra of functions.
Therefore, we can derive $T(f)$ by solving $T(\testfunction_{n})$ since $\testfunction_{n}$ approaches $f$ in the limit process.
This means an operation on a distribution can be defined by an operation on a function. 
In other words, Theorem \ref{theorem:distributionfromtestfunctions} enables the extension of operations on functions to distributions.

Let us now see how we can determine the limit in Equation \ref{eq:operatorintegral}.
Assume that there exists an operator $T^{\prime}$ that constitutes an adjoint map with $T$.
Then, by Equation \ref{eq:operatorintegral}, it follows that
\begin{alignat}{1}\label{eq:operatorintegral2}
	\langle T f, \testfunction \rangle = \int_{-\infty}^{\infty} T(f(x)) \, \testfunction(x) \, dx 
	&
	=  \lim_{n \to \infty}  \int_{-\infty}^{\infty} T(\testfunction_{n}(x)) \, \testfunction(x) \, dx 
	\\
	&
	= 
	\lim_{n \to \infty}  \int_{-\infty}^{\infty} \testfunction_{n}(x) \, T^{\prime}(\testfunction(x)) \, dx
	\\
	&
	=
	\lim_{n \to \infty}	\langle  \testfunction_{n}, T^{\prime}\testfunction \rangle 
	\\
	&
	=
	\langle  f, T^{\prime}\testfunction \rangle \, . \label{eq:operatorintegral3}
\end{alignat}
The adjoint map $	\langle T f, \testfunction \rangle = \langle  f, T^{\prime}\testfunction \rangle$ means that appyling $T$ to $f$ will have an effect (described by $T^{\prime})$ on the test function $\testfunction$.
$T^{\prime}$ manipulates the test function $\testfunction$ which is the input to $f$.
By assumption, $T$ and $T^{\prime}$ form an adjoint map for any test function $\testfunction$, so that any outcome $\langle T f, \testfunction \rangle$ can be described by $\langle  f, T^{\prime}\testfunction \rangle$ independent of the choice of $\testfunction$.
For this reason, $T$ is defined as an operation on the distribution $f$ by the adjoint map $	\langle T f, \testfunction \rangle = \langle  f, T^{\prime}\testfunction \rangle$.

The approach from Equation \eqref{eq:operatorintegral2} to \eqref{eq:operatorintegral3} reduces the problem of deriving $T(f)$ to finding the (adjoint) operator $T^{\prime}$ which can be usually computed using partial integration.
We demonstrate this approach and derive the adjoint map for the well-known translation of the Dirac impulse $\diracdelta_{a}(\testfunction) = \testfunction(a) $ where $\diracdelta_{a} = \diracdelta(x-a)$ or in distributional notation $\langle T_{-a} \, \diracdelta, \testfunction \rangle = \langle \diracdelta, T_{a} \, \testfunction \rangle $ where 
the translations $T_a$ and $T_{-a}$ are defined as the linear transformations
\begin{alignat}{1}
	T_a \colon \mathcal{D} \to \mathcal{D}, \testfunction(x) \mapsto \testfunction(x+a) 
	\\
	T_{-a} \colon \mathcal{D} \to \mathcal{D}, \testfunction(x) \mapsto \testfunction(x-a) 
\end{alignat}
with $a \in \realnumbers$.
We derive the adjoint map $\langle T_{-a} \, \diracdelta, \testfunction \rangle = \langle \diracdelta, T_{a} \, \testfunction \rangle $
%As a part of this problem, we have to find out how to extend the operation $T_a$, that is defined on test functions, to distributions.
%The idea is that we can describe a distribution as the limit of a sequence of test functions, so that an operation on a distribution can be defined by such a limit process.
%Since a locally integrable function is a distribution (Theorem \ref{theorem:locallyintegrable}), it follows that test functions are distributions, as they are locally integrable. This means $\mathcal{D} \subseteq \mathcal{D'}$.
\begin{figure}
	\centering
	\includegraphics[scale=0.7]{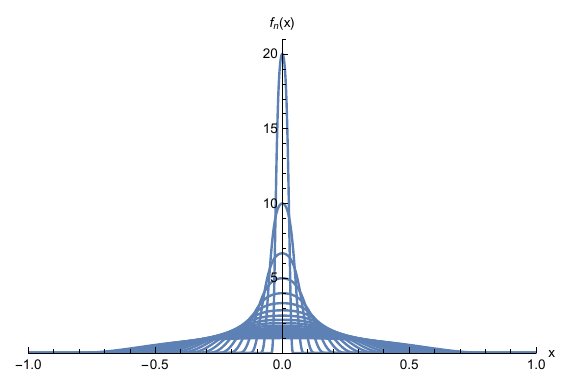}
	\caption{Family of functions $f_{n}(x)$ as defined by Equation \eqref{eq:diracapprox} for $n = 0.05, 0.1, \dots, 1$}
	\label{fig:diracapprox}
\end{figure}
%\begin{theorem}[Distribution from test functions \cite{strichartz2003guide}]\label{theorem:distributionfromtestfunctions}
%	Given any distribution $f \in \mathcal{D'}(\Omega)$, there exists a sequence $(\testfunction_n)$ of test functions such that $f = \lim_{n \to \infty} (\testfunction_n)$ as distributions.
%\end{theorem}
%This means an operation on test functions can be extended to distributions. If $f = \lim_{n \to \infty} (\testfunction_n)$ is a distribution approximated by a sequence of test functions $ (\testfunction_n)$, the adjoint identity can be written as \cite{strichartz2003guide}
%\begin{alignat}{1}
%	\langle T f, \theta \rangle 
%	&
%	=
%	\lim_{n \to \infty} \langle T \testfunction_n, \theta \rangle = \lim_{n \to \infty} \int_{-\infty}^{\infty} T(\testfunction_n(x))  \,  \theta(x) \, dx \\
%	&
%	=
%	\lim_{n \to \infty}\int_{-\infty}^{\infty} \testfunction_n(x)  \, T^{\prime}(\theta(x)) \, dx = \lim_{n \to \infty} \langle \testfunction_n, T^{\prime}(\theta) \rangle = \langle f, T^{\prime}(\theta) \rangle
%\end{alignat}
%where $\theta \in \mathcal{D}$.
using the following family of test functions to approximate the Dirac impulse:
\begin{alignat}{2}\label{eq:diracapprox}
	f_{k}(x) & = \frac{1}{k} f\left(\frac{x}{k}\right) \\
	f(x) &= \left\{\begin{array}{lll} 
		\frac{4(x^{2}+1)\cdot \operatorname{e}^{\frac{4x}{x^2-1}}} {\left((x^{2}-1)\left(1 + \operatorname{e}^{\frac{4x}{x^{2}-1}}\right)\right)^{2} }   &,|x| &< 1 
		\\
		0  &, |x| & \geq 1 
	\end{array}
	\right.
\end{alignat}
for $k \in \naturalnumbers$ that has the properties 
\begin{alignat}{1}\label{eq:diracapprox2}
	\int_{-1/k}^{1/k}f_k(x) \, dx &= 1 \\
	f_k(x) &= 0 ~~~ x \geq |1/k| \label{eq:diracapprox3}
\end{alignat}
and is depicted in Figure \ref{fig:diracapprox}.
Based on the sequence $f_{k}(x)$ and its properties shown in Equation \ref{eq:diracapprox2} and \ref{eq:diracapprox3}, we derive the translation of the Dirac impulse.
\begin{lemma}[Translation of Dirac impulse \cite{strichartz2003guide}]\label{lemma:translationdirac}
	Let $f_{k}(x)$ be a sequence of test functions approximating the Dirac impulse with the properties $	\int_{-1/k}^{1/k}f_k(x) \, dx = 1 $ and $	f_k(x) = 0, x \geq |1/k|$. Then, $\langle T_{-a}  \, \diracdelta, \testfunction \rangle = \langle \diracdelta, T_{a} \, \testfunction \rangle$ is the adjoint identity of translating the Dirac impulse.
\end{lemma}
\begin{proof}
\begin{alignat}{1}
	\langle T_{-a} \, \diracdelta, \testfunction \rangle
	&
	=
	\lim\limits_{k \to \infty} \int_{-\infty}^{\infty} f_{k}(x-a)  \, \testfunction(x) \, dx
	\\
	&
	=
	\lim\limits_{k \to \infty} 
	[F_{k}(x-a)  \, \testfunction(x)]_{-\infty}^{\infty} - \int_{-\infty}^{\infty} F_{k}(x-a) \,  \testfunction'(x) \, dx \label{eq:diractranslation}
	\\
	&
	=
	\lim\limits_{k \to \infty} 
	1 \cdot 0 - 0 \cdot 0 - \int_{-\infty}^{-1/k+a} \underbrace{F_{k}(x-a)}_{= \, 0, \, \mathrm{by} \, \eqref{eq:diracapprox2}} \testfunction'(x) \, dx - \int_{-1/k+a}^{1/k+a} F_{k}(x+a) \,  \testfunction'(x) \, dx \nonumber
	\\
	& ~~~~~~~~~~~~~~~~~~~~~~~~~~~~~~
	- \int_{1/k+a}^{\infty} \underbrace{F_{k}(x-a)}_{= \, 1, \, \mathrm{by} \, \eqref{eq:diracapprox2}} \testfunction'(x) \, dx
	\\
	&
	=
	\lim\limits_{k \to \infty} 
	- \int_{-1/k+a}^{1/k+a} F_{k}(x-a) \,  \testfunction'(x) \, dx - \int_{1/k+a}^{\infty} \testfunction'(x) \, dx
	\\
	&
	=
	- \int_{a}^{\infty} \testfunction'(x) \, dx
	\\
	&
	=
	-[\testfunction(x)]_{a}^{\infty}
	\\
	&
	=
	\testfunction(a)
	\\
	&
	=
	\langle \diracdelta, T_{a} \, \testfunction \rangle
\end{alignat}
\end{proof}
Hence, we get the adjoint map $\langle T_{-a} \, \diracdelta, \testfunction \rangle = \langle \diracdelta, T_{a} \, \testfunction \rangle$ which defines the translation of the Dirac impulse as the negated translation of the test function.
This translation is known as the sampling property $\diracdelta_{a}( \testfunction) = \testfunction(-a)$ of the Dirac impulse.
Informally, an adjoint map defines an operation on a distribution by applying the operation on the test function instead of the distribution. In this way, it is usually easier to find a solution since test functions have useful properties such as differentiability and compact supports.
%wenn translation fertig ist, dann kann man mit impuls gruppe anfangen, weil man da die translation schon braucht. die heaviside maske kann man dann später herleiten bei caluclus of impulse algebra.
%hier auch noch die methode mit der approximation der distribution durch testfunktion einführen.
We apply adjoint maps in the following to derive further operations on Dirac impulses to construct a vector space.

\section{Impulse algebra}\label{section:impulsespace}
\begin{figure}[t!] \centering
\begin{tikzpicture}
	\pgfmathsetmacro{\sone}{0.5}
	\pgfmathsetmacro{\stwo}{1.3}
	\pgfmathsetmacro{\sthree}{3}
	\pgfmathsetmacro{\sfour}{4}
	\pgfmathsetmacro{\sfive}{5}
	\pgfmathsetmacro{\cone}{0.1}
	\pgfmathsetmacro{\ctwo}{0.25}
	\pgfmathsetmacro{\cthree}{0.32}
	\pgfmathsetmacro{\cfour}{0.45}
	\begin{axis}[%
		,xlabel=$t$
		,ylabel=$R(t)$
		,y=4.5cm
		,axis x line = bottom,axis y line = left
		,x label style={at={(axis description cs:0.99,0.17)},anchor=north}
		,y label style={at={(axis description cs:0.1,.9)},rotate=270,anchor=south}
		,xtick={0,\sone,\stwo,\sthree,\sfour}
		,xticklabels={,$s_1$,$s_2$,$s_3$,$s_4$}
		,ytick={0,\cone,\ctwo,\cthree,\cfour}
		,yticklabels={,$c_1$,$c_1 + c_2$,$c_1 + c_2 + c_3$,$c_1 + c_2 + c_3 + c_4$}
		,yticklabel style={right,xshift = 3.5}
		,xmax=5
		,ymax=0.6 % or enlarge y limits=upper
		]
		\addplot+[jump mark left, no marks, thick,black] coordinates {(0,0) (\sone,\cone) (\stwo,\ctwo) (\sthree,\cthree) (\sfour,\cfour) (\sfive,\cfour) };
	\end{axis}
	\node (yzero) at (-0.2, 0) {0};
	\node (xzero) at (0, -0.25) {0};
	\begin{axis}[%
		,yshift=-2cm
		,xlabel=$t$
		,ylabel=$\frac{d}{dt}R(t)$
		,x label style={at={(axis description cs:0.99,0.33)},anchor=north}
		,y label style={at={(axis description cs:0.1,.8)},rotate=270,anchor=south}
		,y=4.5cm
		,axis x line = bottom,axis y line = left
		,xmin=0, xmax=5, ymin=0, ymax=0.3
		,xtick={0,\sone,\stwo,\sthree,\sfour}
		,xticklabels={0,$s_1$,$s_2$,$s_3$,$s_4$}
		,ymajorticks=false
		]
		\addplot+[ycomb,mark=triangle, thick,black] plot coordinates {(\sone,\cone) (\stwo,\ctwo - \cone) (\sthree,\cthree - \ctwo) (\sfour,\cfour - \cthree)}
		node[above=-0.12cm,pos=0.05,black] {$c_{1}$}
		node[above=0.cm,pos=0.27,black] {$c_{2}$}
		node[above=-0.11cm,pos=0.76,black] {$c_{3}$}
		node[above=-0.12cm,pos=1.05,black] {$c_{4}$};
	\end{axis}
\end{tikzpicture}
\caption{Cumulative workload $R(t)$ by the jobs $d/dt R(t)$ \cite{fattohigeneral}.}
\label{fig:requestfunction}
\end{figure}
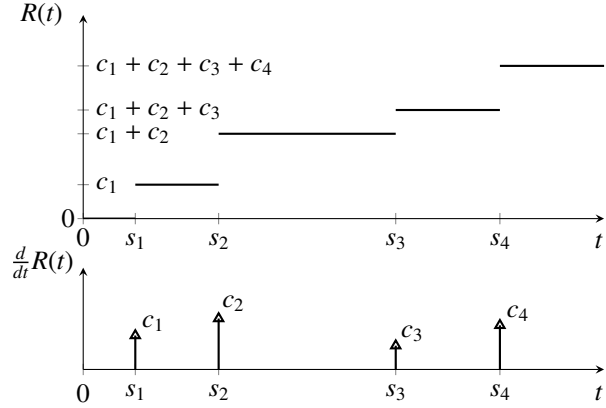
The Dirac impulse is well-known in digital signal theory for its sampling property.
A continuous signal of a physical quantity is the input to a digital system using an analog-to-digital converter.
It samples the input with a given period, so that the output is a digital signal \cite{buttazzo2011hard}. 
The digital signal is the input to a processing unit of an embedded real-time system. 
Any change in the digital signal represented by a Dirac impulse may cause a task of the embedded system to be requested for execution. 

In real-time analysis, it is abstractly said that events request the execution of tasks.
More precisely, an event generates a job of a task at the time point of its occurrence, called the request time.
A job has a certain workload to be processed that is modeled by execution time.
Furthermore, a job must be completed before a specified deadline representing e.g. the stability requirement of a controller \cite{buttazzo2011hard}.
%Therefore, a job is characterized by a request time, an execution time and a deadline.

A goal in real-time analysis is to check whether deadlines of jobs sharing a processor with respect to a given scheduling policy  are violated \cite{buttazzo2011hard}. Task models are the basis for analysis. 
%Different models exist dependent on the event pattern or data and control flow of the task \cite{baruah1999generalized}, \cite{baruah2003dynamic}, \cite{stigge2011digraph}, \cite{albers2006hierarchical}.
The models assume that a sequence of jobs, or \textbf{job train}, is generated. 
However, a model of job trains is often not described  \cite{albers2006hierarchical}, \cite{richter2004compositional},\cite{stigge2011digraph}, \cite{thiele2000real} although jobs are the objects to be scheduled.

For example, the request curve $R(t)$ of the real-time calculus \cite{thiele2000real} shown in Figure \ref{fig:requestfunction} describes the cumulative workload of a job train over time. 
A single job is not defined by $R(t)$ but the sum of the workloads of the job train is described.
In contrast to this, the derivative of $R(t)$ shown in Figure \ref{fig:requestfunction} is a series of Dirac impulses describing the jump discontinuities of the request curve at time points $s_n$ with $n \in \{1,2,3,4\}$. 
A change in the request curve describes the workload $c_n$ that is requested at the time point of the jump. 
If the workload $c_n$ is described by a job, then a Dirac impulse represents a job that is requested at $s_n$ with workload $c_n$ equal to the amplitude of the impulse.
This means a job train is represented by a series of Dirac impulses, or, an \textbf{impulse train} $$\sum_{n = 1}^{N} c_n \, \diracdelta(t - s_n)$$ 

%
%the digital signal can be the source of a sequence of events or impulses (that we call an \textbf{impulse train}) in an embedded real-time system that causes the instantiation and execution of tasks. An instance of a task is a job that is described by its request time (the time point of an event), execution time and deadline. For example, a periodic digital signal may induce a periodic impulse train which in turn causes a periodic generation of jobs. We call a sequence of jobs a job train. Periodic task models are well-studied in the field of real-time analysis \cite{buttazzo2011hard}, \cite{liu1973scheduling}.
%However, if we want to analyze a computing system for an arbitrarily given job train, then we need a model of job trains where request times are arbitrary or constitute multiple nested periods and where each job may have a different execution time.

Our approach in modeling any job train is to construct a vector space of Dirac impulses, so that any job train can be described by a linear combination of impulses, i.e. an impulse train.
It is well-known that the space of distributions is a vector space which means that Dirac impulses, which are distributions, can be linearly combined to form new distributions \cite{grubb2008distributions}.
However, to the best of our knowledge, there is no literature on the vector space of Dirac impulses which we could apply to present a performance analysis of computers. 
Therefore, we present the impulse algebra which is a linear algebra based on the Dirac impulse.
Later, we apply the algebra to introduce new operations on the Dirac impulse based on adjoint maps for analysis purposes.
As the algebra requires many symbols we provide a list of them in Table \ref{table:listofsymbols}.
%The basic element of the presented algebra is the Dirac impulse impulse which was introduced by Paul Dirac \cite{dirac1981principles} to describe the derivative of the Heaviside function. 
%It is also applied to describe a single event in a digital signal, so that the whole signal can be described by a linear combination of Dirac impulses.
%To construct any linear combination, we need to define a vector space of Dirac impulses. To this end, we derive the vector space operations based on distribution theory.
%The list of symbols of the algebra is in Table \ref{table:listofsymbols}.
\subsection{Impulse group}\label{section:impulsegroups}
The vector space of Dirac impulses, called the Dirac space or impulse space, is constructed based on an abelian group of Dirac impulses that is endowed by scalar multiplication of real numbers. Hence, the Dirac space is a vector space on the field of the real numbers that is denoted by the triple $(\realnumbers, +_{\realnumbers}, \cdot_{\realnumbers})$ where $\realnumbers$ is the set of real numbers and $+_{\realnumbers}$ and $\cdot_{\realnumbers}$ are addition and multiplication of real numbers.
To construct this abelian group, we defined the shifted impulse.
\begin{definition}[Shifted impulse]\label{def:shiftedimpulse}
	Let $\impulsevariable \in \realnumbers$ be a variable, $\impulseshift \in \realnumbers$ and $\impulseindex \in \naturalnumberswithzeroandinfinity$.
	Then, the function
	\begin{equation}
		\shiftedimpulse{\impulseindex}{\impulsevariable} = \diracdelta(\impulsevariable - \impulseindex \impulseshift)
	\end{equation}
describes a right-shifted Dirac impulse and is called a \textbf{\shiftedimpulsename} and $ \impulseindex \impulseshift$ is called the \textbf{\shiftname} of the \shiftedimpulsename{} $\shiftedimpulse{\impulseindex}{\impulsevariable}$. 
In short-form, we write
\begin{equation}
	\shiftedimpulsesf{\impulseindex}{\impulsevariable} \coloneqq \shiftedimpulse{\impulseindex}{\impulsevariable}
\end{equation}
\end{definition}
As we present in the following a linear space of \shiftedimpulsename{s}, we introduce additional short-form notation for the summation of Dirac impulses in Appendix \ref{appendix:notation}.  
The \shiftedimpulsename{} can be an element of an algebraic structure, called the \impulsegroupname.
\begin{definition}[\impulsegroupnamecapital]\label{def:impulsegroup}
	The pair $\impulsegroup{\impulsevariable}  \coloneqq (\shiftedimpulseset{\impulsevariable}, \impulsegroupoperation)$ is called \textbf{\impulsegroupname} (IG) where
	$\shiftedimpulseset{\impulsevariable} \coloneqq \{ \shiftedimpulse{\impulseindex}{\impulsevariable}, - \shiftedimpulse{\impulseindex}{\impulsevariable} \}$ is the \textbf{\shiftedimpulsesetname} and 
	\begin{alignat}{1}
		\impulsegroupoperation \colon \shiftedimpulseset{\impulsevariable} \times \shiftedimpulseset{\impulsevariable} \to \shiftedimpulseset{\impulsevariable}, (u_{n}^{ \, x}, v_{m}^{ \, x}) \mapsto u_{n}^{ \, x} +_{\realnumbers} v_{m}^{ \, x}
	\end{alignat}
	is called \textbf{impulse addition}.
\end{definition}
In the following, we show that the \impulsegroupname{} satisfies the axioms of an abelian group.
\begin{lemma}[Impulse group]\label{lemma:impulsegroup}
	The impulse group $\impulsegroup{\impulsevariable}$ is an abelian group.
\end{lemma}
\begin{proof}
	Let $\Omega \subseteq \realnumbers$ be an open set and let $K \subseteq \Omega$ be a compact subset of $\Omega$. Let  $\testfunction(\impulsevariable), \varphi(\impulsevariable) \in \spaceoftestfunctions(\Omega)$ be test functions such that $\operatorname{supp} (\testfunction) \in K$ and $\operatorname{supp} (\varphi) \in K$. 
	Let $n,m,o \in \naturalnumbers, v,w,z \in \realnumbers$. 
	Let $(f_{k})_{k=1}^{\infty}$ be the sequence of test functions approximating the Dirac impulse for $k \to \infty$ with the properties $f_{k}(x) = 0, x > |1/k|$ for all $k$ and $\int_{-1/k}^{1/k} f_{k}(x) = 1$ \cite{strichartz2003guide}.
	
	By the sampling property of the Dirac impulse (Lemma \ref{lemma:translationdirac}), the Dirac impulse maps a test function on a function value at the shift of the impulse: $\langle \diracdelta_{\impulseindex \impulseshift}, \testfunction \rangle = \testfunction(\impulseindex \impulseshift) \in \realnumbers$, where $\diracdelta_{\impulseindex \impulseshift} = \diracdelta(\impulsevariable - \impulseindex \impulseshift)$. We use this property to show that an addition of impulses can be realized by an addition of real numbers.
	
	Closure of $\impulsegroupoperation$:
	\begin{alignat}{1}
		v_{n}^{ \, x} \impulsegroupoperation w_{m}^{ \, x}
		&
		=
		\langle \diracdelta_{nv} +_{\realnumbers} \diracdelta_{mw}, \testfunction \rangle
		\label{eq:impulsegroupclosure1}
		\\
		&
		=
		\lim_{k \to \infty}
		\int_{-\infty}^{\infty} (f_{k}(\impulsevariable - nv) +_{\realnumbers} f_{k}(\impulsevariable -mw)) \cdot \testfunction(\impulsevariable) \, d\impulsevariable
		\\
		&
		=
		\lim_{k \to \infty}
		\int_{-\infty}^{\infty} f_{k}(\impulsevariable - nv) \cdot \testfunction(\impulsevariable) \, d\impulsevariable 
		+_{\realnumbers} 
		\int_{-\infty}^{\infty} f_{k}(\impulsevariable - mw) \cdot \testfunction(\impulsevariable) \, d\impulsevariable
		\\
		&
		=
		\testfunction(nv) + \testfunction(mw)
		\\
		&
		=
		\lim_{k \to \infty}
		\int_{-\infty}^{\infty} f_{k}(\impulsevariable) \cdot(\underbrace{\testfunction(x + nv) +_{\realnumbers} \testfunction(\impulsevariable + mw)}_{\in \spaceoftestfunctions}) \, d\impulsevariable
		\\
		&
		=
		\langle \diracdelta, \testfunction_{nv} +_{\realnumbers} \testfunction_{mw} \rangle
		\label{eq:impulsegroupclosure2}
	\end{alignat}
	where $+_{\realnumbers}$ is the addition of real numbers.
	Equations \ref{eq:impulsegroupclosure1} to \ref{eq:impulsegroupclosure2} state that there exists a pair of test functions such that applying the Dirac impulse on their sum is equal to applying the sum of two Dirac impulses to a single test function, so that $\shiftedimpulseset{\impulsevariable}$ is closed under $\impulsegroupoperation$.

	Associativity:
	\begin{alignat}{1}
		(v_{n}^{ \, x} \impulsegroupoperation w_{m}^{ \, x}) \impulsegroupoperation z_{o}^{ \, x} 
		&
		=
		\langle (\diracdelta_{nv} +_{\realnumbers} \diracdelta_{mw}) +_{\realnumbers} \diracdelta_{oz} , \testfunction \rangle
		\\
		&
		=
		\lim_{k \to \infty}
		\int_{-\infty}^{\infty} ((f_{k}(\impulsevariable - nv) +_{\realnumbers} f_{k}(\impulsevariable -mw))+_{\realnumbers} f_{k}(\impulsevariable - oz)) \cdot \testfunction(\impulsevariable) \, d\impulsevariable
		\\
		&
		\oset[1.5ex]{\mathclap{\mathrm{asso. \, in \,}\realnumbers}}{=}
		\lim_{k \to \infty}
		\int_{-\infty}^{\infty} (f_{k}(\impulsevariable - nv) +_{\realnumbers} (f_{k}(\impulsevariable -mw)+_{\realnumbers} f_{k}(\impulsevariable - oz))) \cdot \testfunction(\impulsevariable) \, d\impulsevariable
		\\
		&
		=
		\langle \diracdelta_{nv} +_{\realnumbers} (\diracdelta_{mw} +_{\realnumbers} \diracdelta_{oz}) , \testfunction \rangle
		\\
		&
		=
		v_{n}^{ \, x} \impulsegroupoperation (w_{m}^{ \, x} \impulsegroupoperation z_{o}^{ \, x} )
	\end{alignat}

	Commutativity:
	\begin{alignat}{1}
		v_{n}^{ \, x} \impulsegroupoperation w_{m}^{ \, x}
		&
		=
		\langle \diracdelta_{nv} +_{\realnumbers} \diracdelta_{mw} , \testfunction \rangle
		\\
		&
		=
		\lim_{k \to \infty}
		\int_{-\infty}^{\infty} (f_{k}(\impulsevariable - nv) +_{\realnumbers} f_{k}(\impulsevariable -mw)) \cdot \testfunction(\impulsevariable) \, d\impulsevariable
		\\
		&
		\oset[1.5ex]{\mathclap{\mathrm{comm. \, in \,}\realnumbers}}{=}
		\lim_{k \to \infty}
		\int_{-\infty}^{\infty} (f_{k}(\impulsevariable - mw) +_{\realnumbers} f_{k}(\impulsevariable -nv)) \cdot \testfunction(\impulsevariable) \, d\impulsevariable
		\\
		&
		=
		\langle \diracdelta_{mw} +_{\realnumbers} \diracdelta_{nv}  , \testfunction \rangle
		\\
		&
		=
		w_{m}^{ \, x} \impulsegroupoperation v_{n}^{ \, x} 
	\end{alignat}

	Neutral element: %Let $mw = \max\{ \operatorname{Bd}_{\realnumbers} \Omega\}$ be the maximum of the boundary of the open set $\Omega$. By definition of the test function, it follows that $\testfunction(mw) = 0$. Thus,
	Let $w_{K} = \max\{ K \}$ be the maximum of the compact set $K$. Let $w_{K}^{\impulsevariable} \coloneqq \diracdelta(\impulsevariable - w_{K})$ denote the Dirac impulse at the shift $w_{K}$. By definition of $\testfunction$ it follows that $\testfunction(w_{K}) = 0$. Thus,
	\begin{alignat}{1}
		v_{n}^{ \, x} \impulsegroupoperation w_{K}^{ \, x}
		&
		=
		\langle \diracdelta_{nv} +_{\realnumbers} \diracdelta_{w_{K}} , \testfunction \rangle
		\\
		&
		=
		\lim_{k \to \infty}
		\int_{-\infty}^{\infty} (f_{k}(\impulsevariable - nv) +_{\realnumbers} f_{k}(\impulsevariable -w_{K})) \cdot \testfunction(\impulsevariable) \, d\impulsevariable
		\\
		&
		=
		\lim_{k \to \infty}
		\int_{-\infty}^{\infty} f_{k}(\impulsevariable - nv) \cdot \testfunction(\impulsevariable) \, d\impulsevariable 
		+_{\realnumbers} 
		\int_{-\infty}^{\infty} f_{k}(\impulsevariable - w_{K}) \cdot \testfunction(\impulsevariable) \, d\impulsevariable
		\\
		&
		=
		\langle \diracdelta_{nv}, \testfunction \rangle +_{\realnumbers} \langle \diracdelta_{w_{K}}, \testfunction \rangle
		\\
		&
		=
		v_{n}^{ \, x}  +_{\realnumbers} \testfunction(w_{K})
		\\
		&
		=
		v_{n}^{ \, x}  +_{\realnumbers} 0
		\\
		&
		\oset[1.5ex]{\mathclap{\mathrm{neutral. \, in \,}\realnumbers}}{=}
		v_{n}^{ \, x}
	\end{alignat}
	Thus, $w_{K}^{ \, x}$ is the neutral element of $\shiftedimpulseset{\impulsevariable}$.
	
	Inverse element: 	Let $w_{K}^{ \, x}$ with $w_{K} = \max\{ K \}$ be the neutral element. Then,
	\begin{alignat}{1}
		v_{n}^{ \, x} \impulsegroupoperation -v_{n}^{ \, x}
		&
		=
		\langle \diracdelta_{nv} +_{\realnumbers} - \diracdelta_{nv} , \testfunction \rangle
		\\
		&
		=
		\lim_{k \to \infty}
		\int_{-\infty}^{\infty} (f_{k}(\impulsevariable - nv) +_{\realnumbers} - f_{k}(\impulsevariable -nv)) \cdot \testfunction(\impulsevariable) \, d\impulsevariable
		\\
		&
		=
		0
		\\
		&
		=
		w_{K}^{ \, x}
	\end{alignat}
	As $	v_{n}^{ \, x} \impulsegroupoperation -v_{n}^{ \, x}$ results in the neutral element $w_{K}^{ \, x}$ , it follows that $	- v_{n}^{ \, x}$ is the inverse element of $	v_{n}^{ \, x} $.
\end{proof}
We set the neutral element to the Dirac impulse that samples a test function at the maximum of the compact set $K$ since all test functions in $\spaceoftestfunctions(\Omega)$ have their support in $K$. By Definition \ref{def:testfunction}, $\testfunction$ is zero outside its support, so that the Shirac occurring at $mw = \max\{ K \}$ samples $\testfunction(mw) = 0$ which is neutral in $\realnumbers$. 
The impulse group describes any addition of Dirac impulses. In other words, we can construct Dirac combs using the \impulsegroupoperationname.
\begin{example}[Impulse group]\label{example:impulsegroup}
	Let us consider the \shiftedimpulsenameplural{} $\shiftedimpulse{1}{\impulsevariable},\shiftedimpulse{2}{\impulsevariable},\shiftedimpulse{3}{\impulsevariable} \in \shiftedimpulseset{\impulsevariable}$. By closure of \impulsegroupoperationname, we can construct a finite series of  \shiftedimpulsenameplural{}
	\begin{equation}
		\shiftedimpulseset{\impulsevariable} \ni \shiftedimpulse{0}{\impulsevariable} \impulsegroupoperation \shiftedimpulse{1}{\impulsevariable} \impulsegroupoperation \shiftedimpulse{2}{\impulsevariable}  = \sum_{\impulseindex = 0}^{2} \diracdelta(\impulsevariable - \impulseindex s) = \sum_{\impulseindex = 0}^{2} \shiftedimpulse{\impulseindex}{\impulsevariable}
	\end{equation}
	that is an element of $\shiftedimpulseset{\impulsevariable}$. This means we can apply \impulsegroupoperationname{} on $\sum_{\impulseindex = 0}^{2} \shiftedimpulse{\impulseindex}{\impulsevariable}$ and another group element, e.g. $\sum_{\impulseindex = 0}^{1} q_{\impulseindex}(\impulsevariable)$ to construct the series of  \shiftedimpulsenameplural{}
	\begin{alignat}{1}
		\sum_{\impulseindex = 0}^{2} \shiftedimpulse{\impulseindex}{\impulsevariable} \impulsegroupoperation \sum_{\impulseindex = 0}^{1} q_{\impulseindex}(\impulsevariable)
		&
		=
		\shiftedimpulse{0}{\impulsevariable} \impulsegroupoperation \shiftedimpulse{1}{\impulsevariable} \impulsegroupoperation  \shiftedimpulse{2}{\impulsevariable}
		\impulsegroupoperation
		q_{0}(\impulsevariable)\impulsegroupoperation q_{1}(\impulsevariable) \in \shiftedimpulseset{\impulsevariable}
	\end{alignat}
	that is element of the impulse group by closure of \impulsegroupoperationname. Summarizing, it is possible to construct an arbitrary  \shiftedimpulsename{} series.
\end{example}
\subsection{Impulse spectral space}
In this section, we present a vector space of impulses on the field of the real numbers, called the \impulsespectralspacename, by combining the abelian group of impulses with the scalar multiplication of real numbers.
To this end, we show that the \impulsespectralspacename{} satisfies the axioms of a vector space \cite{hefferon2016linear}.
\begin{definition}[Impulse spectral space]\label{def:impulsespectralspace}
	The triple $\impulsespectralspace{\impulsevariable} = (\shiftedimpulseset{\impulsevariable}, \impulsegroupoperation, \impulsespectralspacemult)$ is called \textbf{\impulsespectralspacename}{} where $\shiftedimpulseset{\impulsevariable}$ and $\impulsegroupoperation$ are the impulse set and impulse addition of Definition \ref{def:impulsegroup} and
	\begin{alignat}{1}
		\impulsespectralspacemult \colon \realnumbers \times \shiftedimpulseset{\impulsevariable} \to \shiftedimpulseset{\impulsevariable}, (\lambda, v_{n}^{ \, x}) \mapsto \lambda \cdot_{\realnumbers} v_{n}^{ \, x}
	\end{alignat}
	is called \textbf{\impulsespectralspacemultname} and $\cdot_{\realnumbers}$ is multiplication of real numbers.
\end{definition}
\begin{theorem}[Impulse spectral vector space]\label{theorem:impulsespectralspace}
	The \impulsespectralspacename{} $\impulsespectralspace{\impulsevariable} = (\shiftedimpulseset{\impulsevariable}, \impulsegroupoperation, \impulsespectralspacemult)$ is a vector space on the field of  the real numbers where $\shiftedimpulseset{\impulsevariable}$ is the set of vectors, and $\impulsegroupoperation$ and $\impulsespectralspacemult$ are respectively vector addition and scalar multiplication.
\end{theorem}
\begin{proof}
	By Lemma \ref{lemma:impulsegroup}, we know that the \shiftedimpulsesetname{} $\shiftedimpulseset{\impulsevariable}$ and \impulsegroupoperationname{} $\impulsegroupoperation$ are an abelian group.
	Hence, we prove the vector space axioms of scalar multiplication.
	Let $\lambda,\alpha,v,w \in \realnumbers, n,m \in \naturalnumbers$.
	Let $(f_{k})_{k=1}^{\infty}$ be a sequence of functions approximating the Dirac distribution for $k \to \infty$ with the properties $f_{k}(x) = 0, x > |1/k|$ for all $k$ and $\int_{-1/k}^{1/k} f_{k}(x) = 1$.
	
	Closure of scalar multiplication:
	\begin{alignat}{1}
		\lambda \impulsespectralspacemult v_{n}^{ \, \impulsevariable} 
		&
		= 
		\langle \lambda \cdot_{\realnumbers} \diracdelta_{nv}, \testfunction \rangle
		\\
		&
		=
		\lim_{k \to \infty} \int_{-\infty}^{\infty} \lambda \cdot_{\realnumbers} f_{k}(\impulsevariable - nv) \cdot_{\realnumbers} \testfunction(\impulsevariable) \, d\impulsevariable
		\\
		&
		=
		\lim_{k \to \infty} \int_{-\infty}^{\infty} f_{k}(\impulsevariable - nv) \cdot_{\realnumbers} \underbrace{\lambda \cdot_{\realnumbers} \testfunction(\impulsevariable)}_{\in \spaceoftestfunctions} \, d\impulsevariable
		\\
		&
		=
		\langle \diracdelta_{nv}, \lambda \cdot_{\realnumbers} \testfunction \rangle
	\end{alignat}
	where $\cdot_{\realnumbers}$ is the multiplication of real numbers.
	
	Associativity of scalar multiplication:
	\begin{alignat}{1}
		\alpha \impulsespectralspacemult (\lambda \impulsespectralspacemult v_{n}^{ \, \impulsevariable} ) \label{eq:associativityscalarmultiplication1}
		&
		=
		\langle \alpha \cdot_{\realnumbers}(\lambda \cdot_{\realnumbers} \diracdelta_{nv}), \testfunction \rangle
		\\
		&
		=
		\lim_{k \to \infty} \int_{-\infty}^{\infty} \alpha \cdot_{\realnumbers} (\lambda \cdot_{\realnumbers} f_{k}(\impulsevariable - nv)) \cdot_{\realnumbers} \testfunction(\impulsevariable) \, d\impulsevariable
		\\
		&
		\oset[1.5ex]{\mathclap{\mathrm{asso. \, in \,}\realnumbers}}{=}
		\lim_{k \to \infty} \int_{-\infty}^{\infty} (\alpha \cdot_{\realnumbers} \lambda) \cdot_{\realnumbers} f_{k}(\impulsevariable - nv) \cdot_{\realnumbers} \testfunction(\impulsevariable) \, d\impulsevariable
		\\
		&
		=
		\langle (\alpha \cdot_{\realnumbers} \lambda) \cdot_{\realnumbers} \diracdelta_{nv}, \testfunction \rangle
		\\
		&
		=
		(\alpha \cdot_{\realnumbers} \lambda ) \impulsespectralspacemult v_{n}^{ \, \impulsevariable} 
	\end{alignat}

	Distributivity of scalar multiplication:
	\begin{alignat}{1}
		\lambda \impulsespectralspacemult (v_{n}^{ \, x} \impulsegroupoperation w_{m}^{ \, x}) 		\label{eq:distributivityscalarmultiplication1}
		&
		=
		\langle \lambda \cdot_{\realnumbers} (\diracdelta_{nv} +_{\realnumbers} \diracdelta_{mw}), \testfunction \rangle
		\\
		&
		=
		\lim_{k \to \infty} \int_{-\infty}^{\infty} \lambda \cdot_{\realnumbers} (f_{k}(\impulsevariable - nv) +_{\realnumbers} f_{k}(\impulsevariable - mw)) \cdot_{\realnumbers} \testfunction(\impulsevariable) \, d\impulsevariable
		\\
		&
		\oset[1.5ex]{\mathclap{\mathrm{distr. \, in \,}\realnumbers}}{=}
		\lim_{k \to \infty} \int_{-\infty}^{\infty}  (\lambda \cdot_{\realnumbers} f_{k}(\impulsevariable - nv) +_{\realnumbers} \lambda \cdot_{\realnumbers} f_{k}(\impulsevariable - mw)) \cdot_{\realnumbers} \testfunction(\impulsevariable) \, d\impulsevariable
		\\
		&
		=
		\langle \lambda \cdot_{\realnumbers}  \diracdelta_{nv} +_{\realnumbers} \lambda \cdot_{\realnumbers}  \diracdelta_{mw}, \testfunction \rangle
		\\
		&
		=
		\lambda \impulsespectralspacemult  v_{n}^{ \, x} \impulsegroupoperation  \lambda \impulsespectralspacemult  w_{m}^{ \, x}
	\end{alignat}

	Distributivity of scalar addition:
	\begin{alignat}{1}
		(\lambda +_{\realnumbers} \alpha) \impulsespectralspacemult v_{n}^{ \, x}
		&
		=
		\langle (\lambda +_{\realnumbers} \alpha) \cdot_{\realnumbers} \diracdelta_{nv} , \testfunction \rangle 
		\label{eq:scalaraddition1}
		\\
		&
		=
		\lim_{k \to \infty} \int_{-\infty}^{\infty} (\lambda +_\realnumbers \alpha) \cdot_{\realnumbers} f_{k}(\impulsevariable - nv) \cdot_{\realnumbers} \testfunction(\impulsevariable) \, d\impulsevariable
		\\
		&
		\oset[1.5ex]{\mathclap{\mathrm{distr. \, in \,}\realnumbers}}{=}
		\lim_{k \to \infty} \int_{-\infty}^{\infty} \lambda \cdot_{\realnumbers} f_{k}(\impulsevariable - nv) +_{\realnumbers} \alpha \cdot_{\realnumbers} f_{k}(\impulsevariable - nv)  \cdot_{\realnumbers} \testfunction(\impulsevariable) \, d\impulsevariable
		\\
		&
		=
		\langle \lambda \cdot_{\realnumbers} \diracdelta_{nv} +_{\realnumbers} \alpha \cdot_{\realnumbers} \diracdelta_{nv} , \testfunction \rangle 
		\\
		&
		=
		\lambda \impulsespectralspacemult v_{n}^{ \, x} +_{\realnumbers} \alpha \impulsespectralspacemult v_{n}^{ \, x}
		\label{eq:scalaraddition2}
	\end{alignat}
	
	Neutral element of scalar multiplication:
	\begin{alignat}{1}
		1 \impulsespectralspacemult v_{n}^{\, \impulsevariable} 
		&
		= 
		\langle 1 \cdot_{\realnumbers} \diracdelta_{nv}, \testfunction \rangle
		\\
		&
		=
		\lim_{k \to \infty} \int_{-\infty}^{\infty} 1 \cdot_{\realnumbers} f_{k}(\impulsevariable - nv) \cdot_{\realnumbers} \testfunction(\impulsevariable) \, d\impulsevariable
		\\
		&
		\oset[1.5ex]{\mathclap{\mathrm{neutral \, in \,}\realnumbers}}{=}
		\lim_{k \to \infty} \int_{-\infty}^{\infty} f_{k}(\impulsevariable - nv) \cdot_{\realnumbers} \testfunction(\impulsevariable) \, d\impulsevariable
		\\
		&
		=
		\langle \diracdelta_{nv},  \testfunction \rangle
		\\
		&
		=
		v_{n}^{ \, \impulsevariable} 
	\end{alignat}
\end{proof}
The combination of the impulse group with the real numbers as a vector space enables the scaling of Dirac impulses by real numbers to e.g. describe the weight of an impulse which can represent the execution time of a job.
\begin{example}[\impulsespectralspacenamecapital]
	Let $\sum_{\impulseindex = 1}^{3} \shiftedimpulse{\impulseindex}{\impulsevariable}$ be the finite sequence of \shiftedimpulsenameplural{} of Example \ref{example:impulsegroup}. 
	If we want to model that all  \shiftedimpulsenameplural{} are amplified by a factor of $5$, then we can multiply a scalar to the \shiftedimpulsenameplural{} using the distributivity of scalar multiplication:
	\begin{equation}
		5 \impulsespectralspacemult \sum_{\impulseindex = 0}^{2} \shiftedimpulse{\impulseindex}{\impulsevariable} \oset[1.5ex]{\eqref{eq:distributivityscalarmultiplication1}}{=} \sum_{\impulseindex = 0}^{2} 5 \impulsespectralspacemult  \shiftedimpulse{\impulseindex}{\impulsevariable}
	\end{equation}
	which is a vector of the \impulsespectralspacename{}, so that we can further manipulate this vector by using \impulsegroupoperationname{} and \impulsespectralspacemultname{} to describe any combination of different impulses.
\end{example}
%The \impulsespectralspacename{} allows us to model an arbitrary series of Dirac impulses where each of them can have any amplitude, i.e. we can construct any linear combination of impulses.
%We present a vector subspace of the \impulsespectralspacename{} where impulse addition is restricted two the addition of impulses of equal shifts. The result of such an impulse addition is a new impulse of the same shift and with added amplitudes by the distributivity of scalar addition, see \eqref{eq:scalaraddition1} - \eqref{eq:scalaraddition2}.
\subsection{Monovariate impulse spectral space}%todo ändere shirac impulse zu shirac
As a special case, the \impulsegroupoperationname{} of two \shiftedimpulsenameplural{} with equal shifts is a new \shiftedimpulsename{} of the same shift and the sum of their amplitudes.
\begin{example}[Addition of equal \shiftedimpulsenameplural{}]
	Let $2 \, \impulsespectralspacemult \, \diracdelta(\impulsevariable - \impulseindex \phase) \in \impulsespectralspace{\impulsevariable}$ and $3 \, \impulsespectralspacemult \, \diracdelta(\impulsevariable - \impulseindex \phase) \in \impulsespectralspace{\impulsevariable}$. Then, the \impulsegroupoperationname
	\begin{equation}
		2 \impulsespectralspacemult \diracdelta(\impulsevariable - \impulseindex \phase) \impulsegroupoperation 3 \impulsespectralspacemult \diracdelta(\impulsevariable - \impulseindex \phase) \oset[1.5ex]{\eqref{eq:scalaraddition1}}{=} (2 +_{\realnumbers} 3) \impulsespectralspacemult  \diracdelta(\impulsevariable - \impulseindex \phase) = 5 \impulsespectralspacemult \diracdelta(\impulsevariable - \impulseindex \phase) 
	\end{equation}
	adds up the amplitudes of the impulses and does not change their shift.
\end{example}
In fact, the set of  \shiftedimpulsenameplural{} $\fixedshiftedimpulseset{\impulsevariable}{
\impulseshift} \coloneqq \{ \shiftedimpulse{\impulseindex}{\impulsevariable}, - \shiftedimpulse{\impulseindex}{\impulsevariable} \}$ that have shifts being multiples of a fixed value $\impulseshift \in \realnumbers$ forms together with the \impulsegroupoperationname{} and \impulsespectralspacemultname{} a vector subspace \cite{hefferon2016linear} of the \impulsespectralspacename{}.
\begin{theorem}[\monovariateimpulsespectralspacenamecapital]\label{corollary:monovariateimpulsespectralspace}
	Let $\impulseshift \in \realnumbers$. Then, the triple $\monovariateimpulsespectralspace{\impulsevariable}{\impulseshift} = (\fixedshiftedimpulseset{\impulsevariable}{
		\impulseshift}, \impulsegroupoperation, \impulsespectralspacemult)$ describes a vector subspace of the \impulsespectralspacename{} and is called the \textbf{\monovariateimpulsespectralspacename}.
\end{theorem}
\begin{proof}
	To show that $\monovariateimpulsespectralspace{\impulsevariable}{\impulseshift} $ is a vector subspace of $ \impulsespectralspace{\impulsevariable}$, we show the closure of \impulsegroupoperationname{} and \impulsespectralspacemultname{} with respect to $\fixedshiftedimpulseset{\impulsevariable}{
		\impulseshift}$ \cite{hefferon2016linear}.
	Let $n,m,N,M \in \naturalnumbers, \impulseshift \in \realnumbers$ and consider the two vectors
	\begin{equation}
		\sum_{i = n}^{n +N} a_{i} \, \diracdelta(x - is) \in \fixedshiftedimpulseset{\impulsevariable}{
			\impulseshift}
	\end{equation}
	and
	\begin{equation}
		\sum_{i = m}^{m +M} b_{i} \, \diracdelta(x - is) \in \fixedshiftedimpulseset{\impulsevariable}{
			\impulseshift}
	\end{equation}
	Let $\{z, z +1, \ldots, z + Z\} =  \{ n, n + 1, \ldots , n + N\} \cap \{ m, m + 1, \ldots, m + M \}$.
	
	Case 1:
	Assume that $n \leq m \leq n + N \leq m + M$. Then, the \impulsegroupoperationname{} of the two vectors is
	\begin{alignat}{1}
		\sum_{i = n}^{m - 1} a_{i} \, \diracdelta(x - is) \impulsegroupoperation \sum_{i = z}^{z + Z} (a_{i} +_{\realnumbers} b_{i}) \cdot \diracdelta(x - is) \impulsegroupoperation \sum_{i = n + N + 1}^{m + M} a_{i} \, \diracdelta(x - is) \in \fixedshiftedimpulseset{\impulsevariable}{
			\impulseshift}
	\end{alignat}

	Case 2: 
	Assume that $n \leq m \leq m + M \leq n + N$. Then, the \impulsegroupoperationname{} of the two vectors is
	\begin{alignat}{1}
		\sum_{i = n}^{m - 1} a_{i} \, \diracdelta(x - is) \impulsegroupoperation \sum_{i = z}^{z + Z} (a_{i} +_{\realnumbers} b_{i}) \cdot \diracdelta(x - is) \impulsegroupoperation \sum_{i = m + M + 1}^{n + N} a_{i} \, \diracdelta(x - is) \in \fixedshiftedimpulseset{\impulsevariable}{
			\impulseshift} 
	\end{alignat}

	Case 3: 
	Assume that $n \leq n + N < m \leq m + M$. Then, the \impulsegroupoperationname{} of the two vectors is
	\begin{alignat}{1}
		\sum_{i = n}^{n + N} a_{i} \, \diracdelta(x - is) \impulsegroupoperation \sum_{i = n + N + 1}^{m - 1} 0 \cdot \diracdelta(x - is) \impulsegroupoperation \sum_{i = m }^{m + M} a_{i} \, \diracdelta(x - is) \in \fixedshiftedimpulseset{\impulsevariable}{
			\impulseshift} 
	\end{alignat}
	Therefore, $\fixedshiftedimpulseset{\impulsevariable}{
	\impulseshift}$ is closed under \impulsegroupoperationname{}.
	To show the closure of \impulsespectralspacemultname{}, consider that for $\lambda \in \realnumbers$, we have
	\begin{alignat}{1}
		\lambda \impulsespectralspacemult \sum_{i = n}^{n +N} a_{i} \, \diracdelta(x - is) 
		\oset[1.5ex]{\eqref{eq:distributivityscalarmultiplication1}}{=}
		\sum_{i = n}^{n +N} \lambda \cdot a_{i} \, \diracdelta(x - is) 
		\in \fixedshiftedimpulseset{\impulsevariable}{\impulseshift}
	\end{alignat}
\end{proof}
The vector space property of the  \shiftedimpulsenameplural{} does not only serve the purpose of knowing that any impulse train can be described by a linear combination of \shiftedimpulsenameplural{}.
The closure of the vector space operations also imply a general specification of impulse trains.
More precisely, we observe that the shift (request time) and the amplitude (execution time) are the parameters required to specify any vector in the \impulsespectralspacename{}.
The reason is that a general vector consists of a series of Dirac impulses described by these two parameters.
One can also define the degree $N$ as the third parameter to describe the number of impulses in case of a periodic impulse train.
Therefore, we derive in the following a method (a matrix operation) constructing any sequence of \shiftedimpulsenameplural{} given a specification of the shifts, amplitudes and degrees of the impulses.
In this way, we can describe any impulse train in the \impulsespectralspacename{} if its specification is known.

	\subsection{Multivariate impulse spectral space}\label{section:impulseoperator}
In digital systems, we often find nested periodic impulse trains that are e.g. generated by a controller or a by computer program with loops \cite{albers2006hierarchical}. 
For example, consider a message that is periodically sent over a network with period $p_{1} \in \positivrealnumbers$ and each message is divided into $\degreevectorcomponent{2} \in \naturalnumbers$ packets that are sent in a burst with period $p_{2}$. This means the periodic message generates a burst of packets where the burst is itself a periodic impulse train, i.e. we have a nested periodic impulse train. Formally, we describe a nested impulse train by the convolution of \shiftedimpulsenameplural{}.
The result of the convolution of two Diracs impulses is well-known, however, we show their convolution in Lemma \ref{lemma:convolution} to see that this result is also based on the concept of adjoint maps. 
\begin{lemma}[Convolution of Dirac impulses]\label{lemma:convolution}
	Let $a,b \in \realnumbers$. Then $\langle T_{-a} \diracdelta \ast T_{-b} \diracdelta, \phi \rangle = \langle \diracdelta, T_{a+b} \phi \rangle$ which means $\diracdelta(t-a) \ast \diracdelta(t-b) = \diracdelta(t-a-b)$.
\end{lemma}
\begin{proof}
	Let $(f_{k})_{k \in \naturalnumbers}$ be a sequence of test functions approximating the Dirac impulse, i.e. the $f_{k}(\impulsevariable)$ have the properties $f_{k}(\impulsevariable) = 0, |\impulsevariable| \geq 1/k$ and $\int_{-1/k}^{1/k} f_{k}(\impulsevariable) \, d \impulsevariable = 1$ for all $k \in \naturalnumbers$. 
	Then,
	\begin{alignat}{1}
		\langle T_{-a} \diracdelta \ast T_{-b} \diracdelta, \phi \rangle
		&
		=
		\lim_{k \to \infty} \langle T_{-a} f_{k} \ast T_{-b} \diracdelta, \phi \rangle
		\\	
		&
		= \lim\limits_{n \to \infty} \int_{-\infty}^{\infty} (T_{-a} f_{k} \ast T_{-b} \diracdelta)(x) \cdot \phi(x) \, dx
		\\
		&
		=
		\lim\limits_{n \to \infty} \int_{-\infty}^{\infty} \int_{-\infty}^{\infty}  \delta(y - a) \cdot f_{k}(\impulsevariable - y - b) \, dy \cdot \phi(x) \, dx
		\\
		&
		\oset[1.5ex]{\mathclap{\mathrm{Le.} \ref{lemma:translationdirac}}}{=}
		\lim\limits_{n \to \infty} \int_{-\infty}^{\infty}  f_{k}(\impulsevariable - a - b) \cdot \phi(x) \, dx
		\\
		&
		\oset[1.5ex]{\mathclap{\mathrm{Le.} \ref{lemma:translationdirac}}}{=}
		\phi(a + b)
		\\
		&
		=
		\langle \diracdelta, T_{a+b} \, \phi \rangle
	\end{alignat}
\end{proof}
The idea is to describe one of the Dirac impulses as the limit of test functions which is convolved with the other Dirac impulse. 
Applying Lemma \ref{lemma:convolution}, we can describe any $n-$times nested impulse train.
\begin{example}\label{example:nestedtrain}
	Let $\degreevectorcomponent{1},\degreevectorcomponent{2} \in \naturalnumbers$, then a two-times nested impulse train is described by 
	\begin{alignat}{1} 
		& 
		\sum_{\impulseindex_{1} = 0}^{\degreevectorcomponent{1}-1} \amplitudevector{\impulseindex_{1}}\diracdelta(\impulsevariable - \impulseindex_{1} \phasevectorcomponent{1}) 
		\ast 
		\sum_{\impulseindex_{2} = 0}^{\degreevectorcomponent{2}-1} \amplitudevector{\impulseindex_{2}}\diracdelta(\impulsevariable - \impulseindex_{2} \phasevectorcomponent{2})  
		\\
		=
		&
		\sum_{\impulseindex_{1} = 0}^{\degreevectorcomponent{1}-1}  \sum_{\impulseindex_{2} = 0}^{\degreevectorcomponent{2}-1}  \amplitudevector{\impulseindex_{1} }\amplitudevector{\impulseindex_{2} } \diracdelta(\impulsevariable - \impulseindex_{1} \phasevectorcomponent{1}- \impulseindex_{2} \phasevectorcomponent{2} ) 
		\\
		\eqqcolon
		& \,
		\impulsespectraldensityeinsteinmulti{\amplitude}{\impulseindex}{\impulseshift}{\impulsevariable}
	\end{alignat}
\end{example}
\begin{example}[Superposition]
	If two impulse trains occur independently from each other and share the same network, then a receiver in the network observes a superposition of their sent impulse trains.  
	%The superposition or addition of the two signals at every point is called the interference in physics.
	In the impulse algebra, superposition is represented by \impulsegroupoperationname{}:
	\begin{alignat}{1}
		&
		\underbrace{
		\sum_{\impulseindex_{1} = 0}^{\degreevectorcomponent{1} - 1}  \sum_{\impulseindex_{2} = 0}^{\degreevectorcomponent{2} - 1}  \amplitudevector{\impulseindex_{1} }\amplitudevector{\impulseindex_{2} } \diracdelta(\impulsevariable - \impulseindex_{1} \phasevectorcomponent{1}- \impulseindex_{2} \phasevectorcomponent{2} ) 
		}_{\eqqcolon \, {\impulsespectraldensityeinsteinmulti{\amplitude}{\impulseindex}{\impulseshift}{\impulsevariable}}_{1}}
		\nonumber
		\impulsegroupoperation
		\underbrace{
		\sum_{\impulseindex_{3} = 0}^{\degreevectorcomponent{3} - 1}  \sum_{\impulseindex_{4} = 0}^{\degreevectorcomponent{4} - 1}  \amplitudevector{\impulseindex_{3} }\amplitudevector{\impulseindex_{4} } \diracdelta(\impulsevariable - \impulseindex_{3} \phasevectorcomponent{3}- \impulseindex_{4} \phasevectorcomponent{4} ) 
		}_{\eqqcolon \, {\impulsespectraldensityeinsteinmulti{\amplitude}{\impulseindex}{\impulseshift}{\impulsevariable}}_{2}}
		\eqqcolon
		\sum_{\impulseinterferenceindexone = 1}^{2}
		{\impulsespectraldensityeinsteinmulti{\amplitude}{\impulseindex}{\impulseshift}{\impulsevariable}}_{\impulseinterferenceindexone}
	\end{alignat}
\end{example} 
%Generalizing, we can describe an arbitrary superposition of impulses according to the closure of the \impulsespectralspacename{}. 
%We describe impulses that shift other impulses using convolution. Superposed impulses are described by \impulsegroupoperationname{}. The amplitude of an impulse can be scaled by using \impulsespectralspacemultname{}.
%This means we have three basic operations to construct vectors in the impulse algebra: convolution, \impulsegroupoperationname{} and \impulsespectralspacemultname.
%Based on them, we present in the following an algorithm to construct any vector using known matrix operations.

Up to this point, we can conclude that any superposition of periodic impulse trains can be described by a simple addition in a vector space.
Scalar multiplication on the other hand modulates the amplitude of the impulses, while convolution allows any shift in time.
As a result, it is possible to construct any discrete trace of impulses representing any impulse train by well-known algebraic operations.

\subsection{Linear construction of complex multivariate spectral trains}
As mentioned in the last section, let us investigate how any kind of impulse train can be built by a seqence of linear vector operations.
Starting from a simple modulated impulse spectral train, a complex vector of multiple superpositions of impulse trains is constructed.

A periodic sequence of \shiftedimpulsenameplural{} having the shift $\impulseshift \in \realnumbers$, the degree $\impulsedegree \in \naturalnumbers$ (the number of impulses), and the amplitudes $\amplitudevector{\impulseindex}$ for all $\impulseindex \in \{ 1, 2, \ldots, \impulsedegree \}$ is described by the
 \textbf{\impulsespectraltrainname} (\impulsespectraltrainshort)
\begin{alignat}{1}
	\impulsespectraltrain{\impulseshift}{\impulsedegree}{\impulsevariable} 
	&
	\coloneqq
	\sum_{\impulseindex = 0}^{\impulsedegree - 1} 
	\amplitudevector{\impulseindex} \diracdelta(\impulsevariable - \impulseindex \impulseshift)
	\eqqcolon
	\impulsespectraldensityeinsteinmono{\amplitude}{\impulseindex}{\impulseshift}{\impulsevariable}
	\end{alignat}
If $\naturalnumbers \ni \multiperiodicimpulsedimension$ \impulsespectraltrainnameplural{} are nested (see Example \ref{example:nestedtrain} for the case $\multiperiodicimpulsedimension =2 $), we describe their series of impulses by the \textbf{\impulsespectraldensityname} (\impulsespectraldensitynameshort)
\begin{alignat}{1}
	\impulsespectraldensity{\impulseshift}{\impulsedegree}{\impulsevariable}
	&
	\coloneqq
	\impulsespectraldensityeinsteinmonoindexed{\amplitude}{\impulseindex}{\impulseshift}{\impulsevariable}{1} \ast \impulsespectraldensityeinsteinmonoindexed{\amplitude}{\impulseindex}{\impulseshift}{\impulsevariable}{2} \ast \ldots \ast \impulsespectraldensityeinsteinmonoindexed{\amplitude}{\impulseindex}{\impulseshift}{\impulsevariable}{\multiperiodicimpulsedimension}
	\\
	&
	\coloneqq
	\sum_{\impulseindex_{1} = 0}^{\degreevectorcomponent{1} - 1} 
	\amplitudevector{\impulseindex_{1}} \diracdelta(\impulsevariable - \impulseindex_{1}  \phasevectorcomponent{1}) 
	\ast
	\sum_{\impulseindex_{2} = ´0}^{\degreevectorcomponent{2} - 1} 
	\amplitudevector{\impulseindex_{2}} \diracdelta(\impulsevariable - \impulseindex_{2} \phasevectorcomponent{2})  \ast
	 \ldots 
	 \nonumber
	 \\
	 &~~~~~~~~~~~~~~~~~~~~~~~~~~~~~~~~~~~~~~~~~~~~
	 \ast 
	 \sum_{\impulseindex_{\multiperiodicimpulsedimension} = 0}^{\degreevectorcomponent{\multiperiodicimpulsedimension}-1} 
	\amplitudevector{\impulseindex_{\multiperiodicimpulsedimension}} \diracdelta(\impulsevariable - \impulseindex_{\multiperiodicimpulsedimension} \phasevectorcomponent{\multiperiodicimpulsedimension})
	\\
	&
	=
	\sum_{\impulseindex_{1} = 0}^{\degreevectorcomponent{1} - 1} \dots \sum_{\impulseindex_{\multiperiodicimpulsedimension} = 0}^{\degreevectorcomponent{\multiperiodicimpulsedimension}-1}
	\amplitudevector{\impulseindex_{1}} \cdot \ldots \cdot \amplitudevector{\impulseindex_{\multiperiodicimpulsedimension}}  \diracdelta(\impulsevariable - \impulseindex_{1} \phasevectorcomponent{1}  - \ldots - \impulseindex_{\multiperiodicimpulsedimension} \phasevectorcomponent{\multiperiodicimpulsedimension} ) 
	\\
	&
	\coloneqq
	\impulsespectraldensityeinsteinmulti{\amplitude}{\impulseindex}{\impulseshift}{\impulsevariable}
\end{alignat}

The superposition of $\numberofcomponents \in \naturalnumbers$ \impulsespectraldensitynameshort s is described by the \impulsegroupoperationname{}

\begin{alignat}{1}
		\impulseinterference{\impulseshift}{\impulsedegree}{\impulsevariable}
		\coloneqq
	&
	\sum_{\impulseindex_{1,1} = 0}^{\degreevectorcomponent{1,1} - 1} \dots \sum_{\impulseindex_{1,\multiperiodicimpulsedimension} = 0}^{\degreevectorcomponent{1,\multiperiodicimpulsedimension} - 1}
	\amplitudevector{\impulseindex_{1,1}} \cdot \ldots \cdot \amplitudevector{\impulseindex_{1,\multiperiodicimpulsedimension}}  \diracdelta(\impulsevariable - \impulseindex_{1,1} \phasevectorcomponent{1,1}  - \ldots - \impulseindex_{1,\multiperiodicimpulsedimension} \phasevectorcomponent{1,\multiperiodicimpulsedimension} ) 
	\nonumber
	\\
	\impulsegroupoperation
	&
	\nonumber
	\\
	\vdots ~
	\nonumber
	&
	\\
	\impulsegroupoperation
	&
	\sum_{\impulseindex_{\numberofcomponents,1} = 0}^{\degreevectorcomponent{\numberofcomponents,1} - 1} \dots \sum_{\impulseindex_{\numberofcomponents,\multiperiodicimpulsedimension} = 0}^{\degreevectorcomponent{\numberofcomponents,\multiperiodicimpulsedimension} - 1}
	\amplitudevector{\impulseindex_{\numberofcomponents,1}} \cdot \ldots \cdot \amplitudevector{\impulseindex_{\numberofcomponents,\multiperiodicimpulsedimension}}  \diracdelta(\impulsevariable - \impulseindex_{\numberofcomponents,1} \phasevectorcomponent{\numberofcomponents,1}  - \ldots - \impulseindex_{\numberofcomponents,\multiperiodicimpulsedimension} \phasevectorcomponent{\numberofcomponents,\multiperiodicimpulsedimension} ) 
	\\
	=
	&
	\sum_{\impulseinterferenceindexone = 1}^{\numberofcomponents }
	\sum_{\impulseindex_{\impulseinterferenceindexone,1} = 0}^{\degreevectorcomponent{\impulseinterferenceindexone,1} - 1} \dots \sum_{\impulseindex_{\impulseinterferenceindexone,\multiperiodicimpulsedimension} = 0}^{\degreevectorcomponent{\impulseinterferenceindexone,\multiperiodicimpulsedimension} - 1}
	\amplitudevector{\impulseindex_{\impulseinterferenceindexone,1}} \cdot \ldots \cdot \amplitudevector{\impulseindex_{\impulseinterferenceindexone,\multiperiodicimpulsedimension}}  \diracdelta(\impulsevariable - \impulseindex_{\impulseinterferenceindexone,1} \phasevectorcomponent{\impulseinterferenceindexone,1}  - \ldots - \impulseindex_{\impulseinterferenceindexone,\multiperiodicimpulsedimension} \phasevectorcomponent{\impulseinterferenceindexone,\multiperiodicimpulsedimension} ) 
	\\
	\eqqcolon
	&
\sum_{\impulseinterferenceindexone = 1}^{\numberofcomponents}
{	\impulsespectraldensityeinsteinmulti{\amplitude}{\impulseindex}{\impulseshift}{\impulsevariable}}_{\impulseinterferenceindexone}
\end{alignat}
that is called \textbf{\impulseinterferencename} (\impulseinterferencenameshort).
The \impulseinterferencenameshort{} can be written out as a summation of convolved \impulsespectraltrainname s
\begin{alignat}{1}
	\impulseinterference{\impulseshift}{\impulsedegree}{\impulsevariable} 
	&
	=
	\sum_{\impulseinterferenceindexone = 1}^{\numberofcomponents}
	\sum_{\impulseindex_{\impulseinterferenceindexone,1} = 0}^{\degreevectorcomponent{\impulseinterferenceindexone,1} - 1} \dots \sum_{\impulseindex_{\impulseinterferenceindexone,\multiperiodicimpulsedimension} = 0}^{\degreevectorcomponent{\impulseinterferenceindexone,\multiperiodicimpulsedimension} -1}
	\amplitudevector{\impulseindex_{\impulseinterferenceindexone,1}} \cdot \ldots \cdot \amplitudevector{\impulseindex_{\impulseinterferenceindexone,\multiperiodicimpulsedimension}}  \diracdelta(\impulsevariable - \impulseindex_{\impulseinterferenceindexone,1} \phasevectorcomponent{\impulseinterferenceindexone,1}  - \ldots - \impulseindex_{\impulseinterferenceindexone,\multiperiodicimpulsedimension} \phasevectorcomponent{\impulseinterferenceindexone,\multiperiodicimpulsedimension} ) 
	\\
	&
	=	
	\sum_{\impulseinterferenceindexone = 1}^{\numberofcomponents}
	\impulsespectraldensityeinsteinmonoindexed{\amplitude}{\impulseindex}{\impulseshift}{\impulsevariable}{\impulseinterferenceindexone,1} \ast \impulsespectraldensityeinsteinmonoindexed{\amplitude}{\impulseindex}{\impulseshift}{\impulsevariable}{\impulseinterferenceindexone,2} \ast \ldots \ast \impulsespectraldensityeinsteinmonoindexed{\amplitude}{\impulseindex}{\impulseshift}{\impulsevariable}{\impulseinterferenceindexone,\multiperiodicimpulsedimension}
	\nonumber
	\\
	&
	=
	\impulsespectraldensityeinsteinmonoindexed{\amplitude}{\impulseindex}{\impulseshift}{\impulsevariable}{1,1} \ast \impulsespectraldensityeinsteinmonoindexed{\amplitude}{\impulseindex}{\impulseshift}{\impulsevariable}{1,2} \ast \ldots \ast \impulsespectraldensityeinsteinmonoindexed{\amplitude}{\impulseindex}{\impulseshift}{\impulsevariable}{1,\multiperiodicimpulsedimension}
	\nonumber
	\\
	&
	\impulsegroupoperation
	\impulsespectraldensityeinsteinmonoindexed{\amplitude}{\impulseindex}{\impulseshift}{\impulsevariable}{2,1} \ast \impulsespectraldensityeinsteinmonoindexed{\amplitude}{\impulseindex}{\impulseshift}{\impulsevariable}{2,2} \ast \ldots \ast \impulsespectraldensityeinsteinmonoindexed{\amplitude}{\impulseindex}{\impulseshift}{\impulsevariable}{2,\multiperiodicimpulsedimension}
	\nonumber
	\\
	& ~
	\impulsegroupoperation
	\nonumber
	\\
	& ~ \,
	\vdots
	\nonumber
	\\
	&
	\impulsegroupoperation
	\impulsespectraldensityeinsteinmonoindexed{\amplitude}{\impulseindex}{\impulseshift}{\impulsevariable}{\numberofcomponents,1} \ast \impulsespectraldensityeinsteinmonoindexed{\amplitude}{\impulseindex}{\impulseshift}{\impulsevariable}{\numberofcomponents,2} \ast \ldots \ast \impulsespectraldensityeinsteinmonoindexed{\amplitude}{\impulseindex}{\impulseshift}{\impulsevariable}{\numberofcomponents,\multiperiodicimpulsedimension}
	\label{eq:interferenceflattened}
\end{alignat}
Equation \ref{eq:interferenceflattened} can be constructed by matrix manipulations as follows.
Let
\begin{alignat}{1}
	\amplitudematrix
	=
	\begin{pmatrix}
		\amplitudematrixentry{\impulseindex}{1}{1} & \amplitudematrixentry{\impulseindex}{1}{2}  & \dots & \amplitudematrixentry{\impulseindex}{1}{\impulseinterferencedegreetwo} \\
		\amplitudematrixentry{\impulseindex}{2}{1}  & \amplitudematrixentry{\impulseindex}{2}{2}  & \dots & \amplitudematrixentry{\impulseindex}{2}{\impulseinterferencedegreetwo}  \\
		\vdots & \vdots & \ddots & \vdots \\
		\amplitudematrixentry{\impulseindex}{\impulseinterferencedegreeone}{1}  & \amplitudematrixentry{\impulseindex}{\impulseinterferencedegreeone}{2}  & \dots & \amplitudematrixentry{\impulseindex}{\impulseinterferencedegreeone}{\impulseinterferencedegreetwo} \\
	\end{pmatrix}
	~~~
	\shiftmatrix{\degreevector}{\impulsevariable}
	=
 	\begin{pmatrix}
 	 \shiftmatrixentry{\impulseindex}{\impulsevariable}{1}{1} &  \shiftmatrixentry{\impulseindex}{\impulsevariable}{1}{2}& \dots & \shiftmatrixentry{\impulseindex}{\impulsevariable}{1}{\impulseinterferencedegreetwo} \\
 	 \shiftmatrixentry{\impulseindex}{\impulsevariable}{2}{1} &  \shiftmatrixentry{\impulseindex}{\impulsevariable}{2}{2} & \dots &  \shiftmatrixentry{\impulseindex}{\impulsevariable}{2}{\impulseinterferencedegreetwo} \\
 	\vdots & \vdots & \ddots & \vdots \\
 	 \shiftmatrixentry{\impulseindex}{\impulsevariable}{\impulseinterferencedegreeone}{1} &  \shiftmatrixentry{\impulseindex}{\impulsevariable}{\impulseinterferencedegreeone}{2} & \dots &  \shiftmatrixentry{\impulseindex}{\impulsevariable}{\impulseinterferencedegreeone}{\impulseinterferencedegreetwo} \\
 \end{pmatrix}_{\boldsymbol{N}}
\end{alignat}
specify the \textbf{\amplitudematrixname} and the \textbf{\shiftmatrixname} of the \impulseinterferencenameshort{} where $(\amplitudevector{\impulseindex})_{\impulseinterferenceindexone,\impulseinterferenceindextwo} = ( \amplitudevector{1}, \amplitudevector{2}, \dots, \amplitudevector{\numberofrepetitionsvectorcomponent{\impulseinterferenceindexone,\impulseinterferenceindextwo}})$ is the \textbf{\amplitudevectorname},  $\shiftmatrixentry{\impulseindex}{\impulsevariable} {\impulseinterferenceindexone}{\impulseinterferenceindextwo} = ( \diracdelta(\impulsevariable), \diracdelta(\impulsevariable - \phasevectorcomponent{\impulseinterferenceindexone,\impulseinterferenceindextwo}), \dots, \diracdelta(\impulsevariable - (\degreevectorcomponent{\impulseinterferenceindexone,\impulseinterferenceindextwo}-1) \phasevectorcomponent{\impulseinterferenceindexone,\impulseinterferenceindextwo}))$ is the \textbf{\shiftvectorname} and $\degreematrix \in \naturalnumbers^{\impulseinterferencedegreeone \times \impulseinterferencedegreetwo}$ is the \textbf{\degreematrixname} where $\degreematrixentry{\impulseinterferenceindexone}{\impulseinterferenceindextwo}$ is the \degreename{} of the \shiftvectorname{} $\shiftmatrixentry{\impulseindex}{\impulsevariable}{\impulseinterferenceindexone}{\impulseinterferenceindextwo}$.  
Then, we define the  \textbf{\dotproductname} $\matrixdotproduct$ of an \amplitudevectorname{} and an \shiftvectorname{}
\begin{alignat}{1}
	(\amplitudevector{\impulseindex})_{\impulseinterferenceindexone,\impulseinterferenceindextwo}
	\matrixdotproduct
	(\phasevectorcomponent{\impulseindex})_{\impulseinterferenceindexone,\impulseinterferenceindextwo} 
	&
	=
	\begin{pmatrix}
		\amplitudevector{1} \\  \amplitudevector{2} \\ \vdots \\ \amplitudevector{\numberofrepetitionsvectorcomponent{\impulseinterferenceindexone,\impulseinterferenceindextwo}}
	\end{pmatrix}
\matrixdotproduct
	\begin{pmatrix}
	\diracdelta(\impulsevariable) \\ \diracdelta(\impulsevariable - \phasevectorcomponent{\impulseinterferenceindexone,\impulseinterferenceindextwo}) \\ \vdots \\ \diracdelta(\impulsevariable - (\degreevectorcomponent{\impulseinterferenceindexone,\impulseinterferenceindextwo}-1) \phasevectorcomponent{\impulseinterferenceindexone,\impulseinterferenceindextwo})
\end{pmatrix}
\\
&
	\coloneqq
	\begin{pmatrix}
		\amplitudevector{1} \impulsespectralspacemult \diracdelta(\impulsevariable) \\  \amplitudevector{2} \impulsespectralspacemult \diracdelta(\impulsevariable - \phasevectorcomponent{\impulseinterferenceindexone,\impulseinterferenceindextwo})  \\ \vdots \\ \amplitudevector{\numberofrepetitionsvectorcomponent{\impulseinterferenceindexone,\impulseinterferenceindextwo}} \impulsespectralspacemult  \diracdelta(\impulsevariable - (\degreevectorcomponent{\impulseinterferenceindexone,\impulseinterferenceindextwo}-1) \phasevectorcomponent{\impulseinterferenceindexone,\impulseinterferenceindextwo})
	\end{pmatrix}
	\cdot 
	\begin{pmatrix}
		1 \\ 1 \\ \vdots \\ 1_{\degreevectorcomponent{\impulseinterferenceindexone,\impulseinterferenceindextwo}}
	\end{pmatrix}
	\\
	&
	=
	\amplitudevector{1} \impulsespectralspacemult \diracdelta(\impulsevariable) + \amplitudevector{2} \impulsespectralspacemult \diracdelta(\impulsevariable - \phasevectorcomponent{\impulseinterferenceindexone,\impulseinterferenceindextwo}) + \dots + \amplitudevector{\numberofrepetitionsvectorcomponent{\impulseinterferenceindexone,\impulseinterferenceindextwo}} \impulsespectralspacemult  \diracdelta(\impulsevariable - (\degreevectorcomponent{\impulseinterferenceindexone,\impulseinterferenceindextwo}-1) \phasevectorcomponent{\impulseinterferenceindexone,\impulseinterferenceindextwo})
	\\
	&
	=
	\impulsespectraldensityeinsteinmonoindexed{\amplitude}{\impulseindex}{\impulseshift}{\impulsevariable}{\impulseinterferenceindexone,\impulseinterferenceindextwo}
\end{alignat}
and extend it to the \textbf{\matrixdotproductname} of the \amplitudematrixname{} and the \shiftmatrixname{}
\begin{alignat}{1}
	\amplitudematrix \matrixdotproduct \shiftmatrix{\degreevector}{\impulsevariable}  
	&
	= 
		\begin{pmatrix}
		(\amplitudevector{\impulseindex})_{1,1} & (\amplitudevector{\impulseindex})_{1,2} & \dots & (\amplitudevector{\impulseindex})_{1,\impulseinterferencedegreetwo} \\
		(\amplitudevector{\impulseindex})_{2,1} & (\amplitudevector{\impulseindex})_{2,2} & \dots & (\amplitudevector{\impulseindex})_{2,\impulseinterferencedegreetwo} \\
		\vdots & \vdots & \ddots & \vdots \\
		(\amplitudevector{\impulseindex})_{\impulseinterferencedegreeone,1} & (\amplitudevector{\impulseindex})_{\impulseinterferencedegreeone,2} & \dots & (\amplitudevector{\impulseindex})_{\impulseinterferencedegreeone,\impulseinterferencedegreetwo} \\
	\end{pmatrix}
	\matrixdotproduct
	\begin{pmatrix}
		(\phasevectorcomponent{\impulseindex})_{1,1} & (\phasevectorcomponent{\impulseindex})_{1,2} & \dots &(\phasevectorcomponent{\impulseindex})_{1,\impulseinterferencedegreetwo} \\
		(\phasevectorcomponent{\impulseindex})_{2,1} & (\phasevectorcomponent{\impulseindex})_{2,2} & \dots & (\phasevectorcomponent{\impulseindex})_{2,\impulseinterferencedegreetwo} \\
		\vdots & \vdots & \ddots & \vdots \\
		(\phasevectorcomponent{\impulseindex})_{\impulseinterferencedegreeone,1} & (\phasevectorcomponent{\impulseindex})_{\impulseinterferencedegreeone,2} & \dots & (\phasevectorcomponent{\impulseindex})_{\impulseinterferencedegreeone,\impulseinterferencedegreetwo} \\
	\end{pmatrix}
	\\
	&
	=
	\begin{pmatrix}
		(\amplitudevector{\impulseindex})_{1,1}
		\matrixdotproduct
		(\phasevectorcomponent{\impulseindex})_{1,1}   & 	(\amplitudevector{\impulseindex})_{1,2}
		\matrixdotproduct
		(\phasevectorcomponent{\impulseindex})_{1,2}   &  \dots & 	(\amplitudevector{\impulseindex})_{1,\impulseinterferencedegreetwo}
		\matrixdotproduct
		(\phasevectorcomponent{\impulseindex})_{1,\impulseinterferencedegreetwo} 
		\\
		(\amplitudevector{\impulseindex})_{2,1}
		\matrixdotproduct
		(\phasevectorcomponent{\impulseindex})_{2,1}  & 	(\amplitudevector{\impulseindex})_{2,2}
		\matrixdotproduct
		(\phasevectorcomponent{\impulseindex})_{2,2}   &  \dots & 	(\amplitudevector{\impulseindex})_{2,\impulseinterferencedegreetwo}
		\matrixdotproduct
		(\phasevectorcomponent{\impulseindex})_{2,\impulseinterferencedegreetwo} 
		\\
		\vdots & \vdots & \ddots & \vdots 
		\\
		(\amplitudevector{\impulseindex})_{\impulseinterferencedegreeone,1}
		\matrixdotproduct
		(\phasevectorcomponent{\impulseindex})_{\impulseinterferencedegreeone,1} & 	(\amplitudevector{\impulseindex})_{\impulseinterferencedegreeone,2}
		\matrixdotproduct
		(\phasevectorcomponent{\impulseindex})_{\impulseinterferencedegreeone,2}  &  \dots & 	(\amplitudevector{\impulseindex})_{\impulseinterferencedegreeone,\impulseinterferencedegreetwo}
		\matrixdotproduct
		(\phasevectorcomponent{\impulseindex})_{\impulseinterferencedegreeone,\impulseinterferencedegreetwo} 
	\end{pmatrix}
	\\
	&
	=
	\begin{pmatrix}
		(\impulsespectraldensityeinsteinmono{\amplitude}{\impulseindex}{\impulseshift}{\impulsevariable})_{1,1}  & (\impulsespectraldensityeinsteinmono{\amplitude}{\impulseindex}{\impulseshift}{\impulsevariable})_{1,2}  &  \dots & (\impulsespectraldensityeinsteinmono{\amplitude}{\impulseindex}{\impulseshift}{\impulsevariable})_{1,\impulseinterferencedegreetwo}  
		\\
		(\impulsespectraldensityeinsteinmono{\amplitude}{\impulseindex}{\impulseshift}{\impulsevariable})_{2,1}  & (\impulsespectraldensityeinsteinmono{\amplitude}{\impulseindex}{\impulseshift}{\impulsevariable})_{2,2}  &  \dots & (\impulsespectraldensityeinsteinmono{\amplitude}{\impulseindex}{\impulseshift}{\impulsevariable})_{2,\impulseinterferencedegreetwo}
		\\
		\vdots & \vdots & \ddots & \vdots 
		\\
		(\impulsespectraldensityeinsteinmono{\amplitude}{\impulseindex}{\impulseshift}{\impulsevariable})_{\impulseinterferencedegreeone,1}  & (\impulsespectraldensityeinsteinmono{\amplitude}{\impulseindex}{\impulseshift}{\impulsevariable})_{\impulseinterferencedegreeone,2}  &  \dots & (\impulsespectraldensityeinsteinmono{\amplitude}{\impulseindex}{\impulseshift}{\impulsevariable})_{\impulseinterferencedegreeone ,\impulseinterferencedegreetwo}
	\end{pmatrix} 
\end{alignat}
that is called \textbf{\impulsespectralmatrixname}.
It contains all \impulsespectraltrainshort s $(\impulsespectraldensityeinsteinmono{\amplitude}{\impulseindex}{\impulseshift}{\impulsevariable})_{\impulseinterferenceindexone,\impulseinterferenceindextwo} $ of the \impulseinterferencenameshort{} for $\impulseinterferenceindexone \in \{1,2,\dots, \impulseinterferencedegreeone\}$ and $\impulseinterferenceindextwo \in \{1,2,\dots, \impulseinterferencedegreetwo\}$. More precisely, the $\impulseinterferenceindexone$-th row describes an \impulsespectraldensityname{} $\impulsespectraldensityeinsteinmultiindexed{\amplitude}{\impulseindex}{\impulseshift}{\impulsevariable}{\impulseinterferenceindexone}$, so that the convolution of the \impulsespectraltrainshort s of a row result in one \impulsespectraldensitynameshort{} of the \impulseinterferencenameshort{}. Then, the addition of the \impulsespectraldensitynameshort s, i.e. the addition of the rows results in the \impulseinterferencenameshort. We realize this algorithm by two matrix operations $\innerconvolution{}{}$ and $	\impulsespectralinterferenceeinstein{}{}$ for the convolution of the columns and the addition of the rows.

The \textbf{\innerconvolutionname} $\innerconvolution{}{}$ convolves the columns of $\amplitudematrix \matrixdotproduct \shiftmatrix{\degreevector}{\impulsevariable} $.
Let $\unitvector{\impulseinterferenceindextwo} \in \{0,1\}^{\impulseinterferencedegreetwo}$ be a unit vector where the entry of index $\impulseinterferenceindextwo$ is 1 and all others are 0.
 Then,
\begin{alignat}{1}
	\innerconvolution{\amplitudematrix}{\shiftmatrix{\degreevector}{\impulsevariable} }
	&
	\coloneqq
	\begin{pmatrix}
		(((\amplitudematrix \matrixdotproduct \shiftmatrix{\degreevector}{\impulsevariable}  )\cdot \unitvector{1}) \cdot \unitvector{1}) \ast (((\amplitudematrix \matrixdotproduct \shiftmatrix{\degreevector}{\impulsevariable} ) \cdot \unitvector{1}) \cdot \unitvector{2} )\ast \dots \ast (((\amplitudematrix \matrixdotproduct \shiftmatrix{\degreevector}{\impulsevariable}  )\cdot \unitvector{1}) \cdot \unitvector{\impulseinterferencedegreetwo}) \\
		(((\amplitudematrix \matrixdotproduct \shiftmatrix{\degreevector}{\impulsevariable}  )\cdot \unitvector{2}) \cdot \unitvector{1}) \ast (((\amplitudematrix \matrixdotproduct \shiftmatrix{\degreevector}{\impulsevariable}  )\cdot \unitvector{2}) \cdot \unitvector{2}) \ast \dots \ast (((\amplitudematrix \matrixdotproduct \shiftmatrix{\degreevector}{\impulsevariable}  ) \cdot \unitvector{2}) \cdot \unitvector{\impulseinterferencedegreetwo}) 
		\\
		\vdots
		\\
		(((\amplitudematrix \matrixdotproduct \shiftmatrix{\degreevector}{\impulsevariable}  )\cdot \unitvector{\impulseinterferencedegreeone}) \cdot \unitvector{1}) \ast (((\amplitudematrix \matrixdotproduct \shiftmatrix{\degreevector}{\impulsevariable} ) \cdot \unitvector{\impulseinterferencedegreeone}) \cdot \unitvector{2}) \ast \dots \ast (((\amplitudematrix \matrixdotproduct \shiftmatrix{\degreevector}{\impulsevariable} ) \cdot \unitvector{\impulseinterferencedegreeone}) \cdot \unitvector{\impulseinterferencedegreetwo}) \\
	\end{pmatrix}
	\\
	&
	=
	\begin{pmatrix}
		(\impulsespectraldensityeinsteinmono{\amplitude}{\impulseindex}{\impulseshift}{\impulsevariable})_{1,1}  \ast (\impulsespectraldensityeinsteinmono{\amplitude}{\impulseindex}{\impulseshift}{\impulsevariable})_{1,2}  \ast  \ldots \ast (\impulsespectraldensityeinsteinmono{\amplitude}{\impulseindex}{\impulseshift}{\impulsevariable})_{1,\impulseinterferencedegreetwo}  
		\\
		(\impulsespectraldensityeinsteinmono{\amplitude}{\impulseindex}{\impulseshift}{\impulsevariable})_{2,1}  \ast (\impulsespectraldensityeinsteinmono{\amplitude}{\impulseindex}{\impulseshift}{\impulsevariable})_{2,2}  \ast  \ldots \ast (\impulsespectraldensityeinsteinmono{\amplitude}{\impulseindex}{\impulseshift}{\impulsevariable})_{2,\impulseinterferencedegreetwo}
		\\
		\vdots 
		\\
		(\impulsespectraldensityeinsteinmono{\amplitude}{\impulseindex}{\impulseshift}{\impulsevariable})_{\impulseinterferencedegreeone,1}  \ast (\impulsespectraldensityeinsteinmono{\amplitude}{\impulseindex}{\impulseshift}{\impulsevariable})_{\impulseinterferencedegreeone,2}  \ast  \ldots \ast (\impulsespectraldensityeinsteinmono{\amplitude}{\impulseindex}{\impulseshift}{\impulsevariable})_{\impulseinterferencedegreeone ,\impulseinterferencedegreetwo}
	\end{pmatrix} 
\end{alignat}
which is called \textbf{\impulseconvolutionvectorname} or \textbf{\impulseconvolutionvectornamesf}.
	Furthermore, the \textbf{\innersumname} sums up the rows of the convolved columns and the result is the \impulseinterferencenameshort:

\begin{alignat}{1}
	\innerconvolution{\amplitudematrix}{\shiftmatrix{\degreevector}{\impulsevariable} } \cdot \mathbbm{1}
	&
	\coloneqq
	\begin{pmatrix}
		(\impulsespectraldensityeinsteinmono{\amplitude}{\impulseindex}{\impulseshift}{\impulsevariable})_{1,1}  \ast (\impulsespectraldensityeinsteinmono{\amplitude}{\impulseindex}{\impulseshift}{\impulsevariable})_{1,2}  \ast  \ldots \ast (\impulsespectraldensityeinsteinmono{\amplitude}{\impulseindex}{\impulseshift}{\impulsevariable})_{1,\impulseinterferencedegreetwo}  
		\\
		(\impulsespectraldensityeinsteinmono{\amplitude}{\impulseindex}{\impulseshift}{\impulsevariable})_{2,1}  \ast (\impulsespectraldensityeinsteinmono{\amplitude}{\impulseindex}{\impulseshift}{\impulsevariable})_{2,2}  \ast  \ldots \ast (\impulsespectraldensityeinsteinmono{\amplitude}{\impulseindex}{\impulseshift}{\impulsevariable})_{2,\impulseinterferencedegreetwo}
		\\
		\vdots 
		\\
		(\impulsespectraldensityeinsteinmono{\amplitude}{\impulseindex}{\impulseshift}{\impulsevariable})_{\impulseinterferencedegreeone,1}  \ast (\impulsespectraldensityeinsteinmono{\amplitude}{\impulseindex}{\impulseshift}{\impulsevariable})_{\impulseinterferencedegreeone,2}  \ast  \ldots \ast (\impulsespectraldensityeinsteinmono{\amplitude}{\impulseindex}{\impulseshift}{\impulsevariable})_{\impulseinterferencedegreeone ,\impulseinterferencedegreetwo}
	\end{pmatrix} 
	\cdot 
	\begin{pmatrix}
		1 \\ 1 \\ \vdots \\ 1_{\impulseinterferencedegreeone}
	\end{pmatrix}
	\\
	&
	= 
	\sum_{\impulseinterferenceindexone = 0}^{\impulseinterferencedegreeone}
	{	\impulsespectraldensityeinsteinmulti{\amplitude}{\impulseindex}{\impulseshift}{\impulsevariable}}_{\impulseinterferenceindexone}
	\\
	&
	=
	\impulsespectralinterferenceeinstein{ \hspace{1pt} \amplitudematrix \hspace{-1.5pt} \matrixdotproduct \hspace{-1.5pt} \shiftmatrix{\impulsedegree}{\impulsevariable} \hspace{1pt}}
	\\
	&
	=
	\impulseinterference{\impulseshift}{\impulsedegree}{\impulsevariable} 
\end{alignat}

\subsection{Example of operator}
Let 
\begin{alignat}{2}
	\amplitudematrixentry{\impulseindex}{1}{1} &= \begin{pmatrix}
		2 & 4 & 5
	\end{pmatrix}^{\intercal}
	~~~
	\shiftmatrixentry{\impulseindex}{\impulsevariable}{1}{1} = \begin{pmatrix}
		\diracdelta(\impulsevariable) & 	\diracdelta(\impulsevariable - 5) & 		\diracdelta(\impulsevariable - 10)
	\end{pmatrix}^{\intercal}
	\\
	\amplitudematrixentry{\impulseindex}{1}{2} &= \begin{pmatrix}
		3 & 1 & 8
	\end{pmatrix}^{\intercal}
	~~~
	\shiftmatrixentry{\impulseindex}{\impulsevariable}{1}{2} = \begin{pmatrix}
		\diracdelta(\impulsevariable) & 	\diracdelta(\impulsevariable - 3) & 		\diracdelta(\impulsevariable - 6)
	\end{pmatrix}^{\intercal}
	\\
	\amplitudematrixentry{\impulseindex}{2}{1} &= \begin{pmatrix}
		6 & 2 & 3
	\end{pmatrix}^{\intercal}
	~~~
	\shiftmatrixentry{\impulseindex}{\impulsevariable}{2}{1} = \begin{pmatrix}
		\diracdelta(\impulsevariable) & 	\diracdelta(\impulsevariable - 7) & 		\diracdelta(\impulsevariable - 14)
	\end{pmatrix}^{\intercal}
	\\
	\amplitudematrixentry{\impulseindex}{2}{2} &= \begin{pmatrix}
		9 & 4 & 2
	\end{pmatrix}^{\intercal}
	~~~
	\shiftmatrixentry{\impulseindex}{\impulsevariable}{2}{2} = \begin{pmatrix}
		\diracdelta(\impulsevariable) & 	\diracdelta(\impulsevariable - 4) & 		\diracdelta(\impulsevariable - 8)
	\end{pmatrix}^{\intercal}
\end{alignat}
be \amplitudevectorname s and \shiftvectorname s that are elements of the \amplitudematrixname{} 
\begin{alignat}{1}
	\amplitudematrix 
	= 
	\begin{pmatrix}
		\amplitudematrixentry{\impulseindex}{1}{1} & \amplitudematrixentry{\impulseindex}{1}{2} \\
		\amplitudematrixentry{\impulseindex}{2}{1} & \amplitudematrixentry{\impulseindex}{2}{2}   
	\end{pmatrix}
	=
	\begin{pmatrix}
		\begin{pmatrix}
			2 \\ 4 \\ 5
		\end{pmatrix}
		 & 
		 \begin{pmatrix}
		 	3 \\ 1 \\ 8
		 \end{pmatrix}
	  \\
		  \begin{pmatrix}
		  	6 \\ 2 \\ 3
		  \end{pmatrix}
	  &
	   \begin{pmatrix}
	  	9 \\ 4 \\ 2
	  \end{pmatrix}
	\end{pmatrix}
\end{alignat}
and the \shiftmatrixname
\begin{alignat}{1}
	\shiftmatrix{\impulseindex}{\impulsevariable} 
	= 
	\begin{pmatrix}
		\shiftmatrixentry{\impulseindex}{\impulsevariable}{1}{1} & 	\shiftmatrixentry{\impulseindex}{\impulsevariable}{1}{2}  \\
		\shiftmatrixentry{\impulseindex}{\impulsevariable}{2}{1} & 	\shiftmatrixentry{\impulseindex}{\impulsevariable}{2}{2} 
	\end{pmatrix}
	=
	\begin{pmatrix}
		\begin{pmatrix}
			\diracdelta(\impulsevariable) \\	\diracdelta(\impulsevariable - 5) \\ \diracdelta(\impulsevariable - 10)
		\end{pmatrix}
		&
		\begin{pmatrix}
			\diracdelta(\impulsevariable) \\	\diracdelta(\impulsevariable - 3) \\ \diracdelta(\impulsevariable - 6)
		\end{pmatrix}
		\\
		\begin{pmatrix}
			\diracdelta(\impulsevariable) \\	\diracdelta(\impulsevariable - 7) \\ \diracdelta(\impulsevariable - 14)
		\end{pmatrix}
		&
		\begin{pmatrix}
			\diracdelta(\impulsevariable) \\	\diracdelta(\impulsevariable - 4) \\	\diracdelta(\impulsevariable - 8)
		\end{pmatrix}
	\end{pmatrix}
\end{alignat}
We compute the \dotproductname{} of the \amplitudevectorname s and the \shiftvectorname s to compute the \matrixdotproductname{} of the \amplitudematrixname{} and the \shiftmatrixname:
\begin{alignat}{1}
	\amplitudematrixentry{\impulseindex}{1}{1} \matrixdotproduct \shiftmatrixentry{\impulseindex}{\impulsevariable}{1}{1} 
	&
	=   
	\begin{pmatrix}
		2 \\ 4 \\ 5
	\end{pmatrix}
	\matrixdotproduct 
	\begin{pmatrix}
		\diracdelta(\impulsevariable)  \\	\diracdelta(\impulsevariable - 5) \\	\diracdelta(\impulsevariable - 10)
	\end{pmatrix}
	=
	\begin{pmatrix}
		2 \impulsespectralspacemult \diracdelta(\impulsevariable)  \\	4 \impulsespectralspacemult \diracdelta(\impulsevariable - 5) \\	5 \impulsespectralspacemult\diracdelta(\impulsevariable - 10)
	\end{pmatrix}
	\cdot 
	\begin{pmatrix}
		1 \\ 1 \\ 1
	\end{pmatrix}
	\\
	&
	=
	2 \, \diracdelta(\impulsevariable) + 4 \, \diracdelta(\impulsevariable - 5) + 5  \, \diracdelta(\impulsevariable - 10)
	\\
	&
	=
	(\impulsespectraldensityeinsteinmono{\amplitude}{\impulseindex}{\impulseshift}{\impulsevariable})_{1,1} 
	\\
	\amplitudematrixentry{\impulseindex}{1}{2} \matrixdotproduct \shiftmatrixentry{\impulseindex}{\impulsevariable}{1}{2} 
	&
	=   
	\begin{pmatrix}
		3 \\ 1 \\ 8
	\end{pmatrix}
	\matrixdotproduct 
	\begin{pmatrix}
		\diracdelta(\impulsevariable)  \\	\diracdelta(\impulsevariable - 3) \\	\diracdelta(\impulsevariable - 6)
	\end{pmatrix}
	=
	\begin{pmatrix}
		3 \impulsespectralspacemult \diracdelta(\impulsevariable)  \\	1 \impulsespectralspacemult \diracdelta(\impulsevariable - 3) \\	8 \impulsespectralspacemult\diracdelta(\impulsevariable - 6)
	\end{pmatrix}
	\cdot 
	\begin{pmatrix}
		1 \\ 1 \\ 1
	\end{pmatrix}
	\\
	&
	=
	3 \,  \diracdelta(\impulsevariable) + \diracdelta(\impulsevariable - 3) + 8  \, \diracdelta(\impulsevariable - 6)
	\\
	&
	=
	(\impulsespectraldensityeinsteinmono{\amplitude}{\impulseindex}{\impulseshift}{\impulsevariable})_{1,2} 
	\\
	\amplitudematrixentry{\impulseindex}{2}{1} \matrixdotproduct \shiftmatrixentry{\impulseindex}{\impulsevariable}{2}{1} 
	&
	=   
	\begin{pmatrix}
		6 \\ 2 \\ 3
	\end{pmatrix}
	\matrixdotproduct 
	\begin{pmatrix}
		\diracdelta(\impulsevariable)  \\	\diracdelta(\impulsevariable - 7) \\	\diracdelta(\impulsevariable - 14)
	\end{pmatrix}
	=
	\begin{pmatrix}
		6 \impulsespectralspacemult \diracdelta(\impulsevariable)  \\	2 \impulsespectralspacemult \diracdelta(\impulsevariable - 7) \\	3 \impulsespectralspacemult\diracdelta(\impulsevariable - 14)
	\end{pmatrix}
	\cdot 
	\begin{pmatrix}
		1 \\ 1 \\ 1
	\end{pmatrix}
	\\
	&
	=
	6  \, \diracdelta(\impulsevariable) + 2 \,   \diracdelta(\impulsevariable - 7) + 3  \, \diracdelta(\impulsevariable - 14)
	\\
	&
	=
	(\impulsespectraldensityeinsteinmono{\amplitude}{\impulseindex}{\impulseshift}{\impulsevariable})_{2,1} 
	\\
	\amplitudematrixentry{\impulseindex}{2}{2} \matrixdotproduct \shiftmatrixentry{\impulseindex}{\impulsevariable}{2}{2} 
	&
	=   
	\begin{pmatrix}
		9 \\ 4 \\ 2
	\end{pmatrix}
	\matrixdotproduct 
	\begin{pmatrix}
		\diracdelta(\impulsevariable)  \\	\diracdelta(\impulsevariable - 4) \\	\diracdelta(\impulsevariable - 8)
	\end{pmatrix}
	=
	\begin{pmatrix}
		9 \impulsespectralspacemult \diracdelta(\impulsevariable)  \\	4 \impulsespectralspacemult \diracdelta(\impulsevariable - 4) \\	2 \impulsespectralspacemult\diracdelta(\impulsevariable - 8)
	\end{pmatrix}
	\cdot 
	\begin{pmatrix}
		1 \\ 1 \\ 1
	\end{pmatrix}
	\\
	&
	=
	9  \, \diracdelta(\impulsevariable) + 4   \, \diracdelta(\impulsevariable - 4) + 2 \,  \diracdelta(\impulsevariable - 8)
	\\
	&
	=
	(\impulsespectraldensityeinsteinmono{\amplitude}{\impulseindex}{\impulseshift}{\impulsevariable})_{2,2} 
\end{alignat}
\begin{alignat}{1}
	\amplitudematrix \matrixdotproduct \shiftmatrix{\impulseindex}{\impulsevariable} 
	&
	=
	\begin{pmatrix}
			\amplitudematrixentry{\impulseindex}{1}{1} \matrixdotproduct \shiftmatrixentry{\impulseindex}{\impulsevariable}{1}{1}
			&
				\amplitudematrixentry{\impulseindex}{1}{2} \matrixdotproduct \shiftmatrixentry{\impulseindex}{\impulsevariable}{1}{2}  
		\\
			\amplitudematrixentry{\impulseindex}{2}{1} \matrixdotproduct \shiftmatrixentry{\impulseindex}{\impulsevariable}{2}{1} 
		&
			\amplitudematrixentry{\impulseindex}{2}{2} \matrixdotproduct \shiftmatrixentry{\impulseindex}{\impulsevariable}{2}{2} 
	\end{pmatrix}
\\
	&
	=
	\begin{pmatrix}
		\begin{pmatrix}
			2 \\ 4 \\ 5
		\end{pmatrix} 
		\matrixdotproduct
		\begin{pmatrix}
			\diracdelta(\impulsevariable) \\	\diracdelta(\impulsevariable - 5) \\ \diracdelta(\impulsevariable - 10)
		\end{pmatrix}
		& 
		\begin{pmatrix}
			3 \\ 1 \\ 8
		\end{pmatrix}
		\matrixdotproduct
		\begin{pmatrix}
			\diracdelta(\impulsevariable) \\	\diracdelta(\impulsevariable - 3) \\ \diracdelta(\impulsevariable - 6)
		\end{pmatrix}
		\\
		\begin{pmatrix}
			6 \\ 2 \\ 3
		\end{pmatrix}
		\matrixdotproduct
		\begin{pmatrix}
			\diracdelta(\impulsevariable) \\	\diracdelta(\impulsevariable - 7) \\ \diracdelta(\impulsevariable - 14)
		\end{pmatrix}
		&
		\begin{pmatrix}
			9 \\ 4 \\ 2
		\end{pmatrix}
		\matrixdotproduct
		\begin{pmatrix}
			\diracdelta(\impulsevariable) \\	\diracdelta(\impulsevariable - 4) \\	\diracdelta(\impulsevariable - 8)
		\end{pmatrix}
	\end{pmatrix}
\\
&
=
	\begin{pmatrix}
		\begin{pmatrix}
			2 \impulsespectralspacemult \diracdelta(\impulsevariable)  \\	4 \impulsespectralspacemult \diracdelta(\impulsevariable - 5) \\	5 \impulsespectralspacemult\diracdelta(\impulsevariable - 10)
		\end{pmatrix}
		\cdot 
		\begin{pmatrix}
			1 \\ 1 \\ 1
		\end{pmatrix}
		&
		\begin{pmatrix}
			3 \impulsespectralspacemult \diracdelta(\impulsevariable)  \\	1 \impulsespectralspacemult \diracdelta(\impulsevariable - 3) \\	8 \impulsespectralspacemult\diracdelta(\impulsevariable - 6)
		\end{pmatrix}
		\cdot 
		\begin{pmatrix}
			1 \\ 1 \\ 1
		\end{pmatrix}
		\\
		\begin{pmatrix}
			6 \impulsespectralspacemult \diracdelta(\impulsevariable)  \\	2 \impulsespectralspacemult \diracdelta(\impulsevariable - 7) \\	3 \impulsespectralspacemult\diracdelta(\impulsevariable - 14)
		\end{pmatrix}
		\cdot 
		\begin{pmatrix}
			1 \\ 1 \\ 1
		\end{pmatrix}
		&
		\begin{pmatrix}
			9 \impulsespectralspacemult \diracdelta(\impulsevariable)  \\	4 \impulsespectralspacemult \diracdelta(\impulsevariable - 4) \\	2 \impulsespectralspacemult\diracdelta(\impulsevariable - 8)
		\end{pmatrix}
		\cdot 
		\begin{pmatrix}
			1 \\ 1 \\ 1
		\end{pmatrix}
	\end{pmatrix}
\\
	&
	= 
	\begin{pmatrix}
			2 \, \diracdelta(\impulsevariable) + 4   \, \diracdelta(\impulsevariable - 5) + 5  \, \diracdelta(\impulsevariable - 10)
		&
			3  \, \diracdelta(\impulsevariable) + \diracdelta(\impulsevariable - 3) + 8 \,  \diracdelta(\impulsevariable - 6)
		\\
			6 \,  \diracdelta(\impulsevariable) + 2  \,  \diracdelta(\impulsevariable - 7) + 3 \,  \diracdelta(\impulsevariable - 14) 
			&
			9  \, \diracdelta(\impulsevariable) + 4  \,  \diracdelta(\impulsevariable - 4) + 2  \, \diracdelta(\impulsevariable - 8)
	\end{pmatrix}
\\
&
=
\begin{pmatrix}
		(\impulsespectraldensityeinsteinmono{\amplitude}{\impulseindex}{\impulseshift}{\impulsevariable})_{1,1} 
		&
		(\impulsespectraldensityeinsteinmono{\amplitude}{\impulseindex}{\impulseshift}{\impulsevariable})_{1,2} 
		\\
		(\impulsespectraldensityeinsteinmono{\amplitude}{\impulseindex}{\impulseshift}{\impulsevariable})_{2,1} 
		&
		(\impulsespectraldensityeinsteinmono{\amplitude}{\impulseindex}{\impulseshift}{\impulsevariable})_{2,2} 
\end{pmatrix}
\end{alignat}
Next, we convolve the columns of $	\amplitudematrix \matrixdotproduct \shiftmatrix{\impulseindex}{\impulsevariable} $ as follows

\begin{alignat}{1}
	\innerconvolution{\amplitudematrix}{\shiftmatrix{\degreevector}{\impulsevariable} }
&
=
\begin{pmatrix}
	(((\amplitudematrix \matrixdotproduct \shiftmatrix{\degreevector}{\impulsevariable}  )\cdot \unitvector{1}) \cdot \unitvector{1}) \ast (((\amplitudematrix \matrixdotproduct \shiftmatrix{\degreevector}{\impulsevariable} ) \cdot \unitvector{2}) \cdot \unitvector{1} )\\
	(((\amplitudematrix \matrixdotproduct \shiftmatrix{\degreevector}{\impulsevariable}  )\cdot \unitvector{1}) \cdot \unitvector{2}) \ast (((\amplitudematrix \matrixdotproduct \shiftmatrix{\degreevector}{\impulsevariable}  )\cdot \unitvector{2}) \cdot \unitvector{2})
\end{pmatrix}
\\
&
=
\begin{pmatrix}
	(\impulsespectraldensityeinsteinmono{\amplitude}{\impulseindex}{\impulseshift}{\impulsevariable})_{1,1}  \ast (\impulsespectraldensityeinsteinmono{\amplitude}{\impulseindex}{\impulseshift}{\impulsevariable})_{1,2}  
	\\
	(\impulsespectraldensityeinsteinmono{\amplitude}{\impulseindex}{\impulseshift}{\impulsevariable})_{2,1}  \ast (\impulsespectraldensityeinsteinmono{\amplitude}{\impulseindex}{\impulseshift}{\impulsevariable})_{2,2}  
\end{pmatrix} 
\\
&
=
\begin{pmatrix}
	(2 \diracdelta(\impulsevariable) + 4  \diracdelta(\impulsevariable - 5) + 5 \diracdelta(\impulsevariable - 10))
	\ast
	(3 \diracdelta(\impulsevariable) + \diracdelta(\impulsevariable - 3) + 8 \diracdelta(\impulsevariable - 6))
	\\
	(6 \diracdelta(\impulsevariable) + 2  \diracdelta(\impulsevariable - 7) + 3 \diracdelta(\impulsevariable - 14) )
	\ast
	(9 \diracdelta(\impulsevariable) + 4  \diracdelta(\impulsevariable - 4) + 2 \diracdelta(\impulsevariable - 8))
\end{pmatrix}
\\
&
=
\begin{pmatrix}
	\begin{smallmatrix}
		& 6 \diracdelta(\impulsevariable) &+ 12 \diracdelta(\impulsevariable - 5) &+ 15 \diracdelta(\impulsevariable - 10) \\
		+
		& 2 \diracdelta(\impulsevariable - 3) &+  4 \diracdelta(\impulsevariable - 8) &+  5 \diracdelta(\impulsevariable - 13) \\
		+
		& 16 \diracdelta(\impulsevariable - 6) &+  32 \diracdelta(\impulsevariable - 11) &+  40 \diracdelta(\impulsevariable - 16)
	\end{smallmatrix}
	\\
	\\
	\begin{smallmatrix}
		& 54 \diracdelta(\impulsevariable) & + 18 \diracdelta(\impulsevariable - 7) &+ 27 \diracdelta(\impulsevariable - 14) \\
		+&
		24 \diracdelta(\impulsevariable - 4) &+ 8 \diracdelta(\impulsevariable - 11) &+ 12 \diracdelta(\impulsevariable - 18) \\
		+&
		12 \diracdelta(\impulsevariable - 8) &+ 4 \diracdelta(\impulsevariable - 15) &+ 6 \diracdelta(\impulsevariable - 22)
	\end{smallmatrix}
\end{pmatrix}
\\
&
=
\begin{pmatrix}
	\sum_{\impulseindex_{1,1} = 1}^{3} \sum_{\impulseindex_{1,2} = 1}^{3} \amplitude^{\impulseindex_{1,1}} \amplitude^{\impulseindex_{1,2}} \diracdelta(\impulsevariable - (\impulseindex_{1,1}  - 1) \phasevectorcomponent{1,1} - (\impulseindex_{1,2}  - 1) \phasevectorcomponent{1,2})
	\\
	\sum_{\impulseindex_{2,1} = 1}^{3} \sum_{\impulseindex_{2,2} = 1}^{3} \amplitude^{\impulseindex_{2,1}} \amplitude^{\impulseindex_{2,2}} \diracdelta(\impulsevariable - (\impulseindex_{2,1}  - 1) \phasevectorcomponent{2,1} - (\impulseindex_{2,2}  - 1) \phasevectorcomponent{2,2})
\end{pmatrix}
\end{alignat}
Lastly, we multiply the \impulseconvolutionvectornamesf{} with the one vector to get the \impulseinterferencename{}:
\begin{alignat}{1}
	\impulsespectralinterferenceeinstein{\amplitudematrix \matrixdotproduct \shiftmatrix{\impulsedegree}{\impulsevariable}}
	&
	=
	\innerconvolution{\amplitudematrix}{\shiftmatrix{\degreevector}{\impulsevariable} } \cdot \mathbbm{1}
	\\
	&
	=
	\begin{pmatrix}
		(\impulsespectraldensityeinsteinmono{\amplitude}{\impulseindex}{\impulseshift}{\impulsevariable})_{1,1}  \ast (\impulsespectraldensityeinsteinmono{\amplitude}{\impulseindex}{\impulseshift}{\impulsevariable})_{1,2} 
		\\
		(\impulsespectraldensityeinsteinmono{\amplitude}{\impulseindex}{\impulseshift}{\impulsevariable})_{2,1}  \ast (\impulsespectraldensityeinsteinmono{\amplitude}{\impulseindex}{\impulseshift}{\impulsevariable})_{2,2}  
	\end{pmatrix} 
	\cdot 
	\begin{pmatrix}
		1 \\ 1
	\end{pmatrix}
\\
&
=
 6   \, \diracdelta(\impulsevariable) + 12   \, \diracdelta(\impulsevariable - 5) + 15  \,  \diracdelta(\impulsevariable - 10) 
  \nonumber
 \\
 &
+
 2  \,  \diracdelta(\impulsevariable - 3) +  4  \,  \diracdelta(\impulsevariable - 8) +  5  \,  \diracdelta(\impulsevariable - 13) 
 \nonumber
 \\
 &
+
 16   \, \diracdelta(\impulsevariable - 6) +  32  \,  \diracdelta(\impulsevariable - 11) +  40   \,  \diracdelta(\impulsevariable - 16)
  \nonumber
 \\
 &
 +
  54   \, \diracdelta(\impulsevariable) + 18   \, \diracdelta(\impulsevariable - 7) + 27  \,  \diracdelta(\impulsevariable - 14) 
  \nonumber
 \\
 &
 +
 24  \,  \diracdelta(\impulsevariable - 4) + 8   \, \diracdelta(\impulsevariable - 11) + 12   \, \diracdelta(\impulsevariable - 18) 
  \nonumber
 \\
 &
 +
 12   \, \diracdelta(\impulsevariable - 8) + 4  \,  \diracdelta(\impulsevariable - 15) + 6   \, \diracdelta(\impulsevariable - 22)
	\\
	&
	= 
		\sum_{\impulseindex_{1,1} = 1}^{3} \sum_{\impulseindex_{1,2} = 1}^{3} \amplitude^{\impulseindex_{1,1}} \amplitude^{\impulseindex_{1,2}} \diracdelta(\impulsevariable - (\impulseindex_{1,1}  - 1) \phasevectorcomponent{1,1} - (\impulseindex_{1,2}  - 1) \phasevectorcomponent{1,2})
		\nonumber
	\\
	&
	+
		\sum_{\impulseindex_{2,1} = 1}^{3} \sum_{\impulseindex_{2,2} = 1}^{3} \amplitude^{\impulseindex_{2,1}} \amplitude^{\impulseindex_{2,2}} \diracdelta(\impulsevariable - (\impulseindex_{2,1}  - 1) \phasevectorcomponent{2,1} - (\impulseindex_{2,2}  - 1) \phasevectorcomponent{2,2})
	\\
	&
	=
	\sum_{\impulseinterferenceindexone = 1}^{2}
	\sum_{\impulseindex_{\impulseinterferenceindexone,1} = 1}^{3} \sum_{\impulseindex_{\impulseinterferenceindexone,2} = 1}^{3} \amplitude^{\impulseindex_{\impulseinterferenceindexone,1}} \amplitude^{\impulseindex_{\impulseinterferenceindexone,2}} \diracdelta(\impulsevariable - (\impulseindex_{\impulseinterferenceindexone,1}  - 1) \phasevectorcomponent{\impulseinterferenceindexone,1} - (\impulseindex_{\impulseinterferenceindexone,2}  - 1) \phasevectorcomponent{\impulseinterferenceindexone,2})
	\\
	&
	=
	\impulseinterference{\impulseshift}{\impulsedegree}{\impulsevariable} 
\end{alignat}
We can also directly compute the \impulseinterferencename{} as follows:

\begin{alignat}{1}
	\impulsespectralinterferenceeinstein{\amplitudematrix \matrixdotproduct \shiftmatrix{\impulsedegree}{\impulsevariable}}
	&
	=
	\innerconvolution{\amplitudematrix}{\shiftmatrix{\degreevector}{\impulsevariable} } \cdot \mathbbm{1}
	\\
	&
	=
	\begin{pmatrix}
		(((\amplitudematrix \matrixdotproduct \shiftmatrix{\degreevector}{\impulsevariable}  )\cdot \unitvector{1}) \cdot \unitvector{1}) \ast (((\amplitudematrix \matrixdotproduct \shiftmatrix{\degreevector}{\impulsevariable} ) \cdot \unitvector{1}) \cdot \unitvector{2} )\\
		(((\amplitudematrix \matrixdotproduct \shiftmatrix{\degreevector}{\impulsevariable}  )\cdot \unitvector{2}) \cdot \unitvector{1}) \ast (((\amplitudematrix \matrixdotproduct \shiftmatrix{\degreevector}{\impulsevariable}  )\cdot \unitvector{2}) \cdot \unitvector{2})
	\end{pmatrix}
	\cdot 
	\begin{pmatrix}
		1 \\ 1
	\end{pmatrix}
	\\
	&
	=
	(((\amplitudematrix \matrixdotproduct \shiftmatrix{\degreevector}{\impulsevariable}  )\cdot \unitvector{1}) \cdot \unitvector{1}) \ast (((\amplitudematrix \matrixdotproduct \shiftmatrix{\degreevector}{\impulsevariable} ) \cdot \unitvector{1}) \cdot \unitvector{2} )
	\nonumber
	\\
	&
	+
	(((\amplitudematrix \matrixdotproduct \shiftmatrix{\degreevector}{\impulsevariable}  )\cdot \unitvector{2}) \cdot \unitvector{1}) \ast (((\amplitudematrix \matrixdotproduct \shiftmatrix{\degreevector}{\impulsevariable}  )\cdot \unitvector{2}) \cdot \unitvector{2})
	\\
	&
	=
	\left( \left( 
	\begin{pmatrix}
		\begin{pmatrix}
			2 \\ 4 \\ 5
		\end{pmatrix} 
		\matrixdotproduct
		\begin{pmatrix}
			\diracdelta(\impulsevariable) \\	\diracdelta(\impulsevariable - 5) \\ \diracdelta(\impulsevariable - 10)
		\end{pmatrix}
		& 
		\begin{pmatrix}
			3 \\ 1 \\ 8
		\end{pmatrix}
		\matrixdotproduct
		\begin{pmatrix}
			\diracdelta(\impulsevariable) \\	\diracdelta(\impulsevariable - 3) \\ \diracdelta(\impulsevariable - 6)
		\end{pmatrix}
		\\
		\begin{pmatrix}
			6 \\ 2 \\ 3
		\end{pmatrix}
		\matrixdotproduct
		\begin{pmatrix}
			\diracdelta(\impulsevariable) \\	\diracdelta(\impulsevariable - 7) \\ \diracdelta(\impulsevariable - 14)
		\end{pmatrix}
		&
		\begin{pmatrix}
			9 \\ 4 \\ 2
		\end{pmatrix}
		\matrixdotproduct
		\begin{pmatrix}
			\diracdelta(\impulsevariable) \\	\diracdelta(\impulsevariable - 4) \\	\diracdelta(\impulsevariable - 8)
		\end{pmatrix}
	\end{pmatrix} 
	\cdot 
	\begin{pmatrix}
		1 \\ 0
	\end{pmatrix}
	\right)
	\cdot 
	\begin{pmatrix}
		1 \\ 0
	\end{pmatrix}
	\right)
	\nonumber
	\\
	&
	\ast 
	\left( \left(	
	\begin{pmatrix}
	\begin{pmatrix}
		2 \\ 4 \\ 5
	\end{pmatrix} 
	\matrixdotproduct
	\begin{pmatrix}
		\diracdelta(\impulsevariable) \\	\diracdelta(\impulsevariable - 5) \\ \diracdelta(\impulsevariable - 10)
	\end{pmatrix}
	& 
	\begin{pmatrix}
		3 \\ 1 \\ 8
	\end{pmatrix}
	\matrixdotproduct
	\begin{pmatrix}
		\diracdelta(\impulsevariable) \\	\diracdelta(\impulsevariable - 3) \\ \diracdelta(\impulsevariable - 6)
	\end{pmatrix}
	\\
	\begin{pmatrix}
		6 \\ 2 \\ 3
	\end{pmatrix}
	\matrixdotproduct
	\begin{pmatrix}
		\diracdelta(\impulsevariable) \\	\diracdelta(\impulsevariable - 7) \\ \diracdelta(\impulsevariable - 14)
	\end{pmatrix}
	&
	\begin{pmatrix}
		9 \\ 4 \\ 2
	\end{pmatrix}
	\matrixdotproduct
	\begin{pmatrix}
		\diracdelta(\impulsevariable) \\	\diracdelta(\impulsevariable - 4) \\	\diracdelta(\impulsevariable - 8)
	\end{pmatrix}
\end{pmatrix} 
	\cdot 
	\begin{pmatrix}
		0 \\ 1
	\end{pmatrix}
	\right) 
	\cdot 
	\begin{pmatrix}
		1 \\ 0
	\end{pmatrix}
	\right)
	\nonumber
	\\
	&
	+ \,
	\left(\left( 	
	\begin{pmatrix}
		\begin{pmatrix}
			2 \\ 4 \\ 5
		\end{pmatrix} 
		\matrixdotproduct
		\begin{pmatrix}
			\diracdelta(\impulsevariable) \\	\diracdelta(\impulsevariable - 5) \\ \diracdelta(\impulsevariable - 10)
		\end{pmatrix}
		& 
		\begin{pmatrix}
			3 \\ 1 \\ 8
		\end{pmatrix}
		\matrixdotproduct
		\begin{pmatrix}
			\diracdelta(\impulsevariable) \\	\diracdelta(\impulsevariable - 3) \\ \diracdelta(\impulsevariable - 6)
		\end{pmatrix}
		\\
		\begin{pmatrix}
			6 \\ 2 \\ 3
		\end{pmatrix}
		\matrixdotproduct
		\begin{pmatrix}
			\diracdelta(\impulsevariable) \\	\diracdelta(\impulsevariable - 7) \\ \diracdelta(\impulsevariable - 14)
		\end{pmatrix}
		&
		\begin{pmatrix}
			9 \\ 4 \\ 2
		\end{pmatrix}
		\matrixdotproduct
		\begin{pmatrix}
			\diracdelta(\impulsevariable) \\	\diracdelta(\impulsevariable - 4) \\	\diracdelta(\impulsevariable - 8)
		\end{pmatrix}
	\end{pmatrix} 
	\cdot
	\begin{pmatrix}
		1 \\ 0
	\end{pmatrix}
	\right) 
	\cdot
	\begin{pmatrix}
		0 \\ 1
	\end{pmatrix}
	\right) 
	\nonumber
	\\
	&
	\ast 
	\left(\left( 	
	\begin{pmatrix}
	\begin{pmatrix}
		2 \\ 4 \\ 5
	\end{pmatrix} 
	\matrixdotproduct
	\begin{pmatrix}
		\diracdelta(\impulsevariable) \\	\diracdelta(\impulsevariable - 5) \\ \diracdelta(\impulsevariable - 10)
	\end{pmatrix}
	& 
	\begin{pmatrix}
		3 \\ 1 \\ 8
	\end{pmatrix}
	\matrixdotproduct
	\begin{pmatrix}
		\diracdelta(\impulsevariable) \\	\diracdelta(\impulsevariable - 3) \\ \diracdelta(\impulsevariable - 6)
	\end{pmatrix}
	\\
	\begin{pmatrix}
		6 \\ 2 \\ 3
	\end{pmatrix}
	\matrixdotproduct
	\begin{pmatrix}
		\diracdelta(\impulsevariable) \\	\diracdelta(\impulsevariable - 7) \\ \diracdelta(\impulsevariable - 14)
	\end{pmatrix}
	&
	\begin{pmatrix}
		9 \\ 4 \\ 2
	\end{pmatrix}
	\matrixdotproduct
	\begin{pmatrix}
		\diracdelta(\impulsevariable) \\	\diracdelta(\impulsevariable - 4) \\	\diracdelta(\impulsevariable - 8)
	\end{pmatrix}
	\end{pmatrix} 
	\cdot 
	\begin{pmatrix}
		0 \\ 1
	\end{pmatrix}
	\right) 
	\cdot 
	\begin{pmatrix}
		0 \\ 1
	\end{pmatrix}
	\right)
	\\
	&
	=
	\left(
	\begin{pmatrix}
		2 \\ 4 \\ 5
	\end{pmatrix} 
	\matrixdotproduct
	\begin{pmatrix}
		\diracdelta(\impulsevariable) \\	\diracdelta(\impulsevariable - 5) \\ \diracdelta(\impulsevariable - 10)
	\end{pmatrix}
	\right)
	\ast 
	\left(
	\begin{pmatrix}
		3 \\ 1 \\ 8
	\end{pmatrix}
	\matrixdotproduct
	\begin{pmatrix}
		\diracdelta(\impulsevariable) \\	\diracdelta(\impulsevariable - 3) \\ \diracdelta(\impulsevariable - 6)
	\end{pmatrix}
	\right)
	\nonumber
	\\
	&
	+
	\left(
		\begin{pmatrix}
		6 \\ 2 \\ 3
	\end{pmatrix}
	\matrixdotproduct
	\begin{pmatrix}
		\diracdelta(\impulsevariable) \\	\diracdelta(\impulsevariable - 7) \\ \diracdelta(\impulsevariable - 14)
	\end{pmatrix}
	\right)
	\ast
	\left(
	\begin{pmatrix}
		9 \\ 4 \\ 2
	\end{pmatrix}
	\matrixdotproduct
	\begin{pmatrix}
		\diracdelta(\impulsevariable) \\	\diracdelta(\impulsevariable - 4) \\	\diracdelta(\impulsevariable - 8)
	\end{pmatrix}
	\right)
	\\
	&
	=
	\left(
	\begin{pmatrix}
		2 \impulsespectralspacemult \diracdelta(\impulsevariable)  \\	4 \impulsespectralspacemult \diracdelta(\impulsevariable - 5) \\	5 \impulsespectralspacemult\diracdelta(\impulsevariable - 10)
	\end{pmatrix}
	\cdot 
	\begin{pmatrix}
		1 \\ 1 \\ 1
	\end{pmatrix}
	\right)
	\ast 
	\left(
	\begin{pmatrix}
		3 \impulsespectralspacemult \diracdelta(\impulsevariable)  \\	1 \impulsespectralspacemult \diracdelta(\impulsevariable - 3) \\	8 \impulsespectralspacemult\diracdelta(\impulsevariable - 6)
	\end{pmatrix}
	\cdot 
	\begin{pmatrix}
		1 \\ 1 \\ 1
	\end{pmatrix}
	\right)
	\nonumber
	\\
	&
	+
	\left(
	\begin{pmatrix}
		6 \impulsespectralspacemult \diracdelta(\impulsevariable)  \\	2 \impulsespectralspacemult \diracdelta(\impulsevariable - 7) \\	3 \impulsespectralspacemult\diracdelta(\impulsevariable - 14)
	\end{pmatrix}
	\cdot 
	\begin{pmatrix}
		1 \\ 1 \\ 1
	\end{pmatrix}
	\right)
	\ast
	\left(
	\begin{pmatrix}
		9 \impulsespectralspacemult \diracdelta(\impulsevariable)  \\	4 \impulsespectralspacemult \diracdelta(\impulsevariable - 4) \\	2 \impulsespectralspacemult\diracdelta(\impulsevariable - 8)
	\end{pmatrix}
	\cdot 
	\begin{pmatrix}
		1 \\ 1 \\ 1
	\end{pmatrix}
	\right)
	\\
	&
	=
	\left(
	2 \, \diracdelta(\impulsevariable) + 4   \, \diracdelta(\impulsevariable - 5) + 5  \, \diracdelta(\impulsevariable - 10)
	\right)
	\ast 
	\left(
	3  \, \diracdelta(\impulsevariable) + \diracdelta(\impulsevariable - 3) + 8 \,  \diracdelta(\impulsevariable - 6)
	\right)
	\nonumber
	\\
	&
	+
	\left(
	6 \,  \diracdelta(\impulsevariable) + 2  \,  \diracdelta(\impulsevariable - 7) + 3 \,  \diracdelta(\impulsevariable - 14) 
	\right)
	\ast
	\left(
	9  \, \diracdelta(\impulsevariable) + 4  \,  \diracdelta(\impulsevariable - 4) + 2  \, \diracdelta(\impulsevariable - 8)
	\right)
	\nonumber
	\\
	&
	=
	6   \, \diracdelta(\impulsevariable) + 12  \,  \diracdelta(\impulsevariable - 5) + 15  \,  \diracdelta(\impulsevariable - 10) 
	\nonumber
	\\
	&
	+
	2  \,  \diracdelta(\impulsevariable - 3) +  4   \, \diracdelta(\impulsevariable - 8) +  5  \,  \diracdelta(\impulsevariable - 13) 
	\nonumber
	\\
	&
	+
	16  \,  \diracdelta(\impulsevariable - 6) +  32  \,  \diracdelta(\impulsevariable - 11) +  40   \,  \diracdelta(\impulsevariable - 16)
	\nonumber
	\\
	&
	+
	54  \,  \diracdelta(\impulsevariable) + 18  \,  \diracdelta(\impulsevariable - 7) + 27  \,  \diracdelta(\impulsevariable - 14) 
	\nonumber
	\\
	&
	+
	24  \,  \diracdelta(\impulsevariable - 4) + 8  \,  \diracdelta(\impulsevariable - 11) + 12   \, \diracdelta(\impulsevariable - 18) 
	\nonumber
	\\
	&
	+
	12  \,  \diracdelta(\impulsevariable - 8) + 4  \,  \diracdelta(\impulsevariable - 15) + 6   \, \diracdelta(\impulsevariable - 22)
	\\
	&
	= 
	\sum_{\impulseindex_{1,1} = 1}^{3} \sum_{\impulseindex_{1,2} = 1}^{3} \amplitude^{\impulseindex_{1,1}} \amplitude^{\impulseindex_{1,2}} \diracdelta(\impulsevariable - (\impulseindex_{1,1}  - 1) \phasevectorcomponent{1,1} - (\impulseindex_{1,2}  - 1) \phasevectorcomponent{1,2})
	\nonumber
	\\
	&
	+
	\sum_{\impulseindex_{2,1} = 1}^{3} \sum_{\impulseindex_{2,2} = 1}^{3} \amplitude^{\impulseindex_{2,1}} \amplitude^{\impulseindex_{2,2}} \diracdelta(\impulsevariable - (\impulseindex_{2,1}  - 1) \phasevectorcomponent{2,1} - (\impulseindex_{2,2}  - 1) \phasevectorcomponent{2,2})
	\\
	&
	=
	\sum_{\impulseinterferenceindexone = 1}^{2}
	\sum_{\impulseindex_{\impulseinterferenceindexone,1} = 1}^{3} \sum_{\impulseindex_{\impulseinterferenceindexone,2} = 1}^{3} \amplitude^{\impulseindex_{\impulseinterferenceindexone,1}} \amplitude^{\impulseindex_{\impulseinterferenceindexone,2}} \diracdelta(\impulsevariable - (\impulseindex_{\impulseinterferenceindexone,1}  - 1) \phasevectorcomponent{\impulseinterferenceindexone,1} - (\impulseindex_{\impulseinterferenceindexone,2}  - 1) \phasevectorcomponent{\impulseinterferenceindexone,2})
	\\
	&
	=
	\impulseinterference{\impulseshift}{\impulsedegree}{\impulsevariable} 
\end{alignat}
\section{Impulse bounds}\label{section:operations}
%\begin{figure}
%	\centering
%	\includegraphics[scale=0.7]{img/heavisideapprox.pdf}
%	\caption{Family of functions $h_{k}(x)$ as defined by Equation \eqref{eq:heavisideapprox} for $k = 0.1, 0.2, \dots, 1$}
%	\label{fig:heavisideapprox}
%\end{figure}
The calculus of the impulse algebra originates from the unified event bound (UEB) model of \cite{slomka2021beyond}. 
We show how a Heaviside mask can be sampled by a Dirac impulse based on which we can discretize the analysis functions of the UEB model for fast computation.
\begin{definition}[Heaviside mask \cite{slomka2021beyond}]
	Let $a,b,\impulsevariable \in \realnumbers$. The \textbf{Heaviside mask} is defined by $\heavisidemaskone{\impulsevariable}{a}{b} = \heaviside(\impulsevariable - a) \cdot (b - \impulsevariable)$. In distributional short-form notation, we write $\heavisidemaskonedistributional{a}{b} \coloneqq \heavisidemaskone{\impulsevariable}{a}{b}$.
\end{definition}
\subsection{Unified bound discretization}
The UEB function of \cite{slomka2021beyond} applied to a \impulseinterferencename{}
\begin{alignat}{1}
	& \int_{-\infty}^{\infty} \impulseinterference{\impulseshift}{\impulsedegree}{\impulsevariable} \cdot \heaviside(b - \impulsevariable) \cdot \heaviside(\impulsevariable - a) \, d\impulsevariable
	\\
	=
	&
	\int_{-\infty}^{\infty} 
	\sum_{\impulseinterferenceindexone = 1}^{\impulseinterferencedegreeone}
	\sum_{\impulseindex_{\impulseinterferenceindexone,1} = 1}^{\degreevectorcomponent{\impulseinterferenceindexone,1}} \ldots \sum_{\impulseindex_{\impulseinterferenceindexone,\multiperiodicimpulsedimension} = 1}^{\degreevectorcomponent{\impulseinterferenceindexone,\multiperiodicimpulsedimension}}
	\amplitudevector{\impulseindex_{\impulseinterferenceindexone,1},\dots,\impulseindex_{\impulseinterferenceindexone,\multiperiodicimpulsedimension}} 
	\cdot \diracdelta(\impulsevariable - (\impulseindex_{\impulseinterferenceindexone,1} - 1) \phasevectorcomponent{\impulseinterferenceindexone,1}  - \ldots - (\impulseindex_{\impulseinterferenceindexone,\multiperiodicimpulsedimension} - 1) \phasevectorcomponent{\impulseinterferenceindexone,\multiperiodicimpulsedimension}) 
	\nonumber
	\\
	& ~~~~~~~~~~~~~~~~~~~~~~~~~~~~~~~~~~~~~~~~~~~~~~~~
	\cdot   \heaviside(\impulsevariable - a) \cdot \heaviside(b - \impulsevariable) \, d\impulsevariable \label{eq:uebtomist}
\end{alignat}
counts the number of impulses of the \impulseinterferencenameshort{} in the given Heaviside mask $\heavisidemaskonedistributional{a}{b}$.
We observe in Equation \ref{eq:uebtomist} that the Heaviside mask is multiplied by a Dirac impulse within an integral.  
From Lemma \ref{lemma:translationdirac}, we know if a test function is multiplied by a Dirac impulse, then the test function is sampled at the shift of the impulse.
Considering the Heaviside function, we know that it does not have a compact support since its value is 1 for $x > 0$, so that it does not fulfill the properties of a test function.
However, the Dirac impulse has the compact support $\{0\}$, which allows us to apply the Dirac impulse on any infinitely continuous differentiable function \cite{strichartz2003guide} (page 81).
Therefore, if we approximate the Heaviside mask by an infinitely continuous differentiable function, we can apply the sampling property of the Dirac impulse to the Heaviside mask in the integral of Equation \eqref{eq:uebtomist}. 
In other words, the computation of the UEB function is reduced from the evaluation of an integral to the evaluation of a single Heaviside mask. 
To this end, we apply the sequence $(h_{k})$ of functions from Equation \eqref{eq:heavisideapproxproperties} and \eqref{eq:heavisideapprox} that approximates the Heaviside function for $k \to 0$.
% with the properties
%\begin{alignat}{1}\label{eq:heavisideapproxproperties}
%	h_{k}(x) &= 1,~~~ x > k \\
%	h_{k}(x) &= 0, ~~~ x < -k
%\end{alignat}
%and $h_{k}(x) \in C^{\infty}$ for all $k \in \naturalnumbers$. An exemplary approximation is the family of functions
%\begin{equation}\label{eq:heavisideapprox}
%	h_{k}(x) = \left\{\begin{array}{lll}
%		\frac{1}{1 + \operatorname{e}^{\frac{4kx}{x^{2}-k^{2}}}}   &,|x| &< k 
%		\\
%		0  &,x &< - k 
%		\\
%		1  &,x &> k
%	\end{array}
%	\right.
%\end{equation}
%that is depicted in Figure \ref{fig:heavisideapprox}.
Since $h_{k}(x) \in C^{\infty}$, it follows that
% the product of $h_{k}(x)$ and a test function is a test function by theorem 2.6 of \cite{grubb2008distributions} 
$h_{k}(x)$ can  be sampled by the Dirac impulse 
based on which the translation (sampling) of a Heaviside function by a Dirac impulse can be defined.
\begin{lemma}[Translation of Heaviside product]\label{lemma:translationheaviside}
	Let $a,b, \epsilon \in \realnumbers$. 
	Let $(f_{n})$ be a sequence of test functions approximating the Dirac impulse with the properties $	\int_{-1/n}^{1/n}f_n(x) \, dx = 1 $ and $	f_n(x) = 0, x \geq |1/n|$.
	Let $(h_{k})$ be a sequence of $C^{\infty}$-functions that approximate the heaviside function, i.e. $\heaviside(x) = \lim_{k \to 0} h_{k}(x)$ and the $h_{k}(x)$ have the properties described in Equation \eqref{eq:heavisideapproxproperties}. 
	Then, the Heaviside mask is translated by the Dirac impulse as follows:
	\begin{equation}
		\langle T_{-\epsilon} \diracdelta,  \heavisidemaskonedistributional{a}{b} \rangle = \heavisidemaskone{\epsilon}{a}{b}
	\end{equation}
	which means that
	\begin{equation}
		\int_{-\infty}^{\infty} \diracdelta(\impulsevariable - \epsilon) \cdot \heavisidemaskone{\impulsevariable}{a}{b} \, dx = \heavisidemaskone{\epsilon}{a}{b}
	\end{equation}
\end{lemma}
\begin{proof}
	\begin{alignat}{1}
		\langle T_{-\epsilon} \diracdelta,  \heavisidemaskonedistributional{a}{b} \rangle
		&
		=
		\lim\limits_{\substack{n \, \to \infty \\ k \to 0}}
		\int_{-\infty}^{\infty} 
		\underbrace{f_{n}(x - \epsilon)}_{u^\prime} \cdot  \underbrace{ h_{k}(x-a) \cdot h_{k}(b-x) }_{v} 
		\, dx 
		\\
		&
		=
		\lim\limits_{\substack{n \, \to \infty \\ k \to 0}}
		[ F_{n}(x- \epsilon) \cdot h_{k}(x-a) \cdot h_{k}(b-x) ]_{- \infty}^{\infty}
		\nonumber
		\\
		&
		-
		\int_{-\infty}^{\infty} 
		F_{n}(x - \epsilon) \cdot (h_{k}(x-a) \cdot h_{k}(b-x) )'
		\, dx 
		\label{eq:heavisidemaskintegral1}
		\\
		&
		=
		\lim\limits_{\substack{n \, \to \infty \\ k \to 0}}
		1 \cdot 0 \cdot 1 - 0 \cdot 1 \cdot 0 
		-
		\int_{-\infty}^{\infty} 
		F_{n}(x - \epsilon) \, ( h_{k}(x-a) \cdot h_{k}(b-x) )'
		\, dx 
		\\
		&
		=
		\lim\limits_{\substack{n \, \to \infty \\ k \to 0}}
		-
		\left(
		\int_{-\infty}^{- 1/n + \epsilon} 
		\underbrace{F_{n}(x - \epsilon)}_{ = \, 0} \, ( h_{k}(x-a) \, h_{k}(b-x) )'
		\, dx \right.
		\\
		&
		\left.
		+
			\int_{- 1/n + \epsilon}^{ 1/n + \epsilon} 
		F_{n}(x - \epsilon)\, ( h_{k}(x-a) \, h_{k}(b-x) )'
		\, dx 
		\right.
		\\
		&
		\left.
		+
		\int_{1/n + \epsilon}^{ \infty} 
		\underbrace{F_{n}(x - \epsilon)}_{ = \, 1} \, ( h_{k}(x-a) \, h_{k}(b-x) )'
		\, dx 
		\right)
		\\
		&
		=
		\lim\limits_{\substack{n \, \to \infty \\ k \to 0}}
		-
		\left(
			\int_{- 1/n + \epsilon}^{ 1/n + \epsilon} 
		F_{n}(x - \epsilon)\, (h_{k}(x-a) \, h_{k}(b-x)  )'
		\, dx 
		\right.
		\\
		& ~~~~~~~~~~~~
		\left.
		+
		\int_{1/n + \epsilon}^{ \infty} 
		 ( h_{k}(x-a) \, h_{k}(b-x) )'
		\, dx 
		\right)
		\\
		&
		=
		\lim\limits_{k \to 0}
		-
		\int_{\epsilon}^{ \infty} 
		( h_{k}(x-a) \, h_{k}(b-x) )'
		\, dx 
		\\
		&
		=
		\lim\limits_{k \to 0}
		-
		[ h_{k}(x-a) \, h_{k}(b-x) ]_{\epsilon}^{\infty}
		\label{eq:heavisidemaskintegral2}
		\\
		&
		=
		\lim\limits_{k \to 0}
		- ( h_{k}(\infty-a) \, \underbrace{h_{k}(b-\infty)}_{= \, 0} -  h_{k}(\epsilon-a) \, h_{k}(b-\epsilon) )
		\\
		&
		=
		\lim\limits_{k \to 0}
		- ( - \, h_{k}(\epsilon-a) \, h_{k}(b-\epsilon)  )
		\\
		&
		=
		\heaviside(\epsilon - a) \, \heaviside(b - \epsilon) 
	\end{alignat}
%	\begin{alignat}{1}
%	\langle \heavisidemaskonedistributional{a}{b} T_{\epsilon} \diracdelta, \phi \rangle 
%		&
%		=
%		\lim_{k \to \infty}
%		\int_{-\infty}^{\infty} h_{k}(x-a) \cdot h_{k}(b-x) \cdot \delta(x- \epsilon) \cdot \phi(x) \, dx 
%		\\
%		&
%		=
%		\lim_{k \to \infty}
%		\int_{-\infty}^{\infty}  \delta(x-\epsilon) \cdot \underbrace{h_{k}(x-a) \cdot h_{k}(b-x) \cdot \phi(x)}_{\in \mathcal{D}} \, dx 
%		\\
%		&
%		\oset[1.5ex]{\mathclap{\mathrm{Le.} \ref{lemma:translationdirac}}}{=}
%		\lim_{k \to \infty} h_{k}(\epsilon - a) \cdot h_{k}(b - \epsilon) \cdot \phi(\epsilon)
%		\\
%		&
%		=
%		\heaviside(\epsilon - a) \cdot \heaviside(b - \epsilon) \cdot  \phi(\epsilon)
%		\\
%		&
%		= \langle \diracdelta, T_{\epsilon} \heavisidemaskonedistributional{a}{b} \phi \rangle
%	\end{alignat}
%	We note that $h_{k}(x-a) \cdot h_{k}(b-x) \cdot \phi(x) \in \mathcal{D}$ follows from theorem 2.6 of \cite{grubb2008distributions} stating that the product of a $C^{\infty}$-function and a test function is a test function.
\end{proof}
	Dependent on the definition of the Heaviside approximation $h_{k}(x)$, we can replace $\heaviside(\epsilon - a) \cdot \heaviside(b - \epsilon)$ with upper and lower Heaviside functions and hence obtain the Heaviside masks for all combinations of open and closed interval boundaries as this replacement does not change the result of the integral in Equation \ref{eq:heavisidemaskintegral1} and \ref{eq:heavisidemaskintegral2}.
	
	We apply a Heaviside mask to any \impulseinterferencename{} to mask out the amplitudes of impulses that shall not be considered in further analysis. This masking procedure can be computed by an integral as known from the UEB model but there exists also a discrete computation based on the sifting property of the Dirac impulse. We call this discretized version the \heavisidedurationname{} for which we introduce a short-form notation.
	\begin{definition}[\heavisidedurationnamecapital]\label{def:heavisideduration}
		The \textbf{
			\heavisidedurationname} is defined as the discretized unified request bound function of \cite{slomka2021beyond} as follows:
		\begin{equation}
			\heavisideduration{\Sh}{a,b}{} \coloneqq 	
			\amplitudevector{\impulseindex_{\impulseinterferenceindexone,1} \dots \impulseindex_{\impulseinterferenceindexone,\multiperiodicimpulsedimension}} \overline{| \langle \phasevectorcomponent{\impulseindex_{\impulseinterferenceindexone,1}} \dots \phasevectorcomponent{\impulseindex_{\impulseinterferenceindexone,\multiperiodicimpulsedimension}} \rangle  |}_{a,b}
		\end{equation}
		where
		\begin{alignat}{1}
			&
			\amplitudevector{\impulseindex_{\impulseinterferenceindexone,1} \dots \impulseindex_{\impulseinterferenceindexone,\multiperiodicimpulsedimension}} \overline{| \langle \phasevectorcomponent{\impulseindex_{\impulseinterferenceindexone,1}} \dots \phasevectorcomponent{\impulseindex_{\impulseinterferenceindexone,\multiperiodicimpulsedimension}} \rangle  |}_{a,b}
			\\
			\coloneqq 
			&
			\sum_{\impulseinterferenceindexone = 1}^{\impulseinterferencedegreeone}
			\sum_{\impulseindex_{\impulseinterferenceindexone,1} = 1}^{\degreevectorcomponent{\impulseinterferenceindexone,1}} \dots \sum_{\impulseindex_{\impulseinterferenceindexone,\multiperiodicimpulsedimension} = 1}^{\degreevectorcomponent{\impulseinterferenceindexone,\multiperiodicimpulsedimension}}
			\amplitudevector{\impulseindex_{\impulseinterferenceindexone,1},\dots,\impulseindex_{\impulseinterferenceindexone,\multiperiodicimpulsedimension}} 
			\cdot \heaviside(b - (\impulseindex_{\impulseinterferenceindexone,1} - 1) \phasevectorcomponent{\impulseinterferenceindexone,1}  - \dots - (\impulseindex_{\impulseinterferenceindexone,\multiperiodicimpulsedimension} - 1) \phasevectorcomponent{\impulseinterferenceindexone,\multiperiodicimpulsedimension}) 
			\nonumber
			\\
			&
			\cdot \heaviside((\impulseindex_{\impulseinterferenceindexone,1} - 1) \phasevectorcomponent{\impulseinterferenceindexone,1}  + \dots + (\impulseindex_{\impulseinterferenceindexone,\multiperiodicimpulsedimension} - 1) \phasevectorcomponent{\impulseinterferenceindexone,\multiperiodicimpulsedimension} - a) 		
		\end{alignat}
	\end{definition}
	Based on Lemma \ref{lemma:translationheaviside}, we derive in the following the discretized UEB function which is equal to a corresponding \heavisidedurationname{}.
	
\begin{theorem}[Unified event bound function discretization]\label{theorem:discretization}
	Let $\impulseinterference{\impulseshift}{\impulsedegree}{\impulsevariable}$ be a \impulseinterferencename{}.
	The discretized unified event bound function given by theorem 1 in \cite{slomka2021beyond} is equal to a \heavisidedurationname{}:
	\begin{alignat}{1}
		\int_{-\infty}^{\infty} \impulseinterference{\impulseshift}{\impulsedegree}{\impulsevariable} \cdot \heaviside(b - x) \cdot \heaviside(x - a) \, dx
		=
		\heavisideduration{\Sh}{a,b}{} 
	\end{alignat}
\end{theorem}
\begin{proof}
	\begin{alignat}{1}
		&
		\int_{-\infty}^{\infty} \impulseinterference{\impulseshift}{\impulsedegree}{\impulsevariable} \cdot \heaviside(b - \impulsevariable) \cdot \heaviside(\impulsevariable - a) \, d\impulsevariable
		\\
		=
		&
		\int_{-\infty}^{\infty} 
		\sum_{\impulseinterferenceindexone = 1}^{\impulseinterferencedegreeone}
		\sum_{\impulseindex_{\impulseinterferenceindexone,1} = 1}^{\degreevectorcomponent{\impulseinterferenceindexone,1}} \ldots \sum_{\impulseindex_{\impulseinterferenceindexone,\multiperiodicimpulsedimension} = 1}^{\degreevectorcomponent{\impulseinterferenceindexone,\multiperiodicimpulsedimension}}
		\amplitudevector{\impulseindex_{\impulseinterferenceindexone,1},\dots,\impulseindex_{\impulseinterferenceindexone,\multiperiodicimpulsedimension}} 
		\cdot \diracdelta(\impulsevariable - (\impulseindex_{\impulseinterferenceindexone,1} - 1) \phasevectorcomponent{\impulseinterferenceindexone,1}  - \ldots - (\impulseindex_{\impulseinterferenceindexone,\multiperiodicimpulsedimension} - 1) \phasevectorcomponent{\impulseinterferenceindexone,\multiperiodicimpulsedimension}) 
		\nonumber
		\\
		& ~~~~~~~~~~~~~~~~~~~~~~~~~~~~~~~~~~~~~~~~~~~~~~~~
		 \cdot \heaviside(b - \impulsevariable) \cdot \heaviside(\impulsevariable - a) \, d\impulsevariable
		\label{eq:discretization1}
		\\
		=
		&
		\sum_{\impulseinterferenceindexone = 1}^{\impulseinterferencedegreeone}
		\sum_{\impulseindex_{\impulseinterferenceindexone,1} = 1}^{\degreevectorcomponent{\impulseinterferenceindexone,1}} \ldots \sum_{\impulseindex_{\impulseinterferenceindexone,\multiperiodicimpulsedimension} = 1}^{\degreevectorcomponent{\impulseinterferenceindexone,\multiperiodicimpulsedimension}}
				\int_{-\infty}^{\infty} 
	\amplitudevector{\impulseindex_{\impulseinterferenceindexone,1},\dots,\impulseindex_{\impulseinterferenceindexone,\multiperiodicimpulsedimension}} 
		\cdot \diracdelta(\impulsevariable - (\impulseindex_{\impulseinterferenceindexone,1} - 1) \phasevectorcomponent{\impulseinterferenceindexone,1}  - \ldots - (\impulseindex_{\impulseinterferenceindexone,\multiperiodicimpulsedimension} - 1) \phasevectorcomponent{\impulseinterferenceindexone,\multiperiodicimpulsedimension}) 
		\nonumber
		\\
		& ~~~~~~~~~~~~~~~~~~~~~~~~~~~~~~~~~~~~~~~~~~~~~~~~
		\cdot \heaviside(b - \impulsevariable) \cdot \heaviside(\impulsevariable - a) \, d\impulsevariable
		\label{eq:discretization2}
		\\
		=
		&
		\sum_{\impulseinterferenceindexone = 1}^{\impulseinterferencedegreeone}
		\sum_{\impulseindex_{\impulseinterferenceindexone,1} = 1}^{\degreevectorcomponent{\impulseinterferenceindexone,1}} \ldots \sum_{\impulseindex_{\impulseinterferenceindexone,\multiperiodicimpulsedimension} = 1}^{\degreevectorcomponent{\impulseinterferenceindexone,\multiperiodicimpulsedimension}}
		\amplitudevector{\impulseindex_{\impulseinterferenceindexone,1},\dots,\impulseindex_{\impulseinterferenceindexone,\multiperiodicimpulsedimension}} 
		\cdot \heaviside(b - (\impulseindex_{\impulseinterferenceindexone,1} - 1) \phasevectorcomponent{\impulseinterferenceindexone,1}  - \ldots - (\impulseindex_{\impulseinterferenceindexone,\multiperiodicimpulsedimension} - 1) \phasevectorcomponent{\impulseinterferenceindexone,\multiperiodicimpulsedimension}) 
		\nonumber
		\\
		&		~~\,~~~~~~~~~~~~~~~~~~~~~~~~~~~~~~~~~~~~~~
		\cdot \heaviside((\impulseindex_{\impulseinterferenceindexone,1} - 1) \phasevectorcomponent{\impulseinterferenceindexone,1}  + \ldots + (\impulseindex_{\impulseinterferenceindexone,\multiperiodicimpulsedimension} - 1) \phasevectorcomponent{\impulseinterferenceindexone,\multiperiodicimpulsedimension} - a) 		
		\label{eq:discretization3}
		\\
		=
		&~
		\amplitudevector{\impulseindex_{\impulseinterferenceindexone,1} \ldots \impulseindex_{\impulseinterferenceindexone,\multiperiodicimpulsedimension}} \overline{| \langle \phasevectorcomponent{\impulseindex_{\impulseinterferenceindexone,1}} \ldots \phasevectorcomponent{\impulseindex_{\impulseinterferenceindexone,\multiperiodicimpulsedimension}} \rangle  |}_{a,b}
		\\
		=
		&
		\heavisideduration{\Sh}{a,b}{} 
	\end{alignat}
\end{proof}
From Equation \eqref{eq:discretization1} to \eqref{eq:discretization2} we apply Fubini's theorem \cite{tao2009analysis} to switch the summation and the integral as the integral is bounded from above which is discussed in details in lemma 1 of \cite{slomka2021beyond}. 
From Equation \eqref{eq:discretization2} to \eqref{eq:discretization3} we apply the sampling property of the Dirac impulse to the Heaviside mask which is shown in Lemma \ref{lemma:translationheaviside}.
We note that discretization means that the integral over the variable $\impulsevariable$ is substituted by a summation of Heaviside masks as demonstrated in the following.
\begin{example}[Unified bound discretization]\label{example:unifiedbounddiscretization}
	Let $4 \, \diracdelta(\impulsevariable - 1) \impulsegroupoperation 5 \, \diracdelta(\impulsevariable - 2) \impulsegroupoperation  6 \,\diracdelta(\impulsevariable - 3)$ be a Dirac train. 
	Then, the sum of the amplitudes of impulses occuring in the interval $[2,4]$ is computed by the UEB function as follows
	\begin{alignat}{1}
		&
		\int_{-\infty}^{\infty} (4 \, \diracdelta(\impulsevariable - 1) \impulsegroupoperation 5 \, \diracdelta(\impulsevariable - 2) \impulsegroupoperation  6 \,\diracdelta(\impulsevariable - 3))\cdot \heaviside(\impulsevariable - 2) \cdot \heaviside(4 - \impulsevariable) \, d\impulsevariable
		\label{eq:exampleunifiedbounddiscretization1}
		\\
		=
		&
		\int_{-\infty}^{\infty} 4 \, \diracdelta(\impulsevariable - 1) \, \heaviside(\impulsevariable - 2) \, \heaviside(3 - \impulsevariable)  \impulsegroupoperation 5 \, \diracdelta(\impulsevariable - 2) \, \heaviside(\impulsevariable - 2) \, \heaviside(4 - \impulsevariable)
		\nonumber
		\\
		&~~~~
		\impulsegroupoperation  6 \, \diracdelta(\impulsevariable - 3) \,  \heaviside(\impulsevariable - 2) \, \heaviside(4 - \impulsevariable) \, d\impulsevariable
		\label{eq:exampleunifiedbounddiscretization2}
		\\
		=
		&
		\int_{-\infty}^{\infty} 4 \, \diracdelta(\impulsevariable - 1) \, \heaviside(\impulsevariable - 2) \, \heaviside(4 - \impulsevariable) \, d\impulsevariable + \int_{-\infty}^{\infty} 5 \, \diracdelta(\impulsevariable - 2) \, \heaviside(\impulsevariable - 2) \, \heaviside(4 - \impulsevariable) \, d\impulsevariable
		\nonumber
		\\
		+ 
		&
		\int_{-\infty}^{\infty} 6 \, \diracdelta(\impulsevariable - 3) \,  \heaviside(\impulsevariable - 2) \, \heaviside(4 - \impulsevariable) \, d\impulsevariable
		\label{eq:exampleunifiedbounddiscretization3}
		\\
		\oset[1.5ex]{\mathclap{\mathrm{Le.} \, \ref{lemma:translationheaviside}}}{=}
		& ~
		4 \, \heaviside(1 - 2) \, \heaviside(4 - 1) + 5 \, \heaviside(2 - 2) \, \heaviside(4 - 2) + 6 \, \heaviside(3 - 2) \, \heaviside(4 - 3)
		\label{eq:exampleunifiedbounddiscretization4}
		\\
		= 
		&~
		4 \cdot 0 \cdot 1 + 5 \cdot 1 \cdot 1 + 6 \cdot 1 \cdot 1 = 11
	\end{alignat}
	We observe that the integral in Equation \ref{eq:exampleunifiedbounddiscretization1} is solved by distributing the integral to each Dirac impulse in the series in Equation \ref{eq:exampleunifiedbounddiscretization3}. Then, the sampling property of the Dirac impulse is applied to sample the Heaviside mask in Equation \ref{eq:exampleunifiedbounddiscretization4}.
	Note that the discretization of the UEB function to the \heavisidedurationname{} occurs from Equation \ref{eq:exampleunifiedbounddiscretization3} to \ref{eq:exampleunifiedbounddiscretization4} by applying Lemma \ref{lemma:translationheaviside}.
	Furthermore, all the parameters to compute Equation \ref{eq:exampleunifiedbounddiscretization4} are already known: the amplitudes and the shifts of the impulses and the interval $[2,3]$ of the Heaviside mask. Therefore, we can skip the computation steps from Equation \ref{eq:exampleunifiedbounddiscretization1} to \ref{eq:exampleunifiedbounddiscretization3} and directly compute the \heavisidedurationname{} in Equation \ref{eq:exampleunifiedbounddiscretization4}.
\end{example}

\subsection{Maximum and minimum bound on the Heaviside distribution}
To find the maximal or minimal density of impulses, we formulate an extreme value problem on \impulseinterferencename s. 
This allows us to compute worst- and best-cases in a performance analysis of a computing system or a network which can be e.g. the maximum or minimum computational (execution time) or network load (bits). 
By modeling computational or network load as the amplitude of the impulse, we can represent an extreme value problem of computational or network load by an extreme value problem of the \heavisidedurationname{}. Therefore, we assume that the amplitudes of all impulses are positive real numbers.
\begin{problem}[Local extrema of \heavisidedurationname]\label{problem:localextrema}
	Given is an 
	\impulseinterferencename{} $\impulseinterference{\impulseshift}{\impulsedegree}{\impulsevariable}$
	 and a distance $\intervalduration \in \realnumbers$. The problem is to find values for $\impulsevariable^{+},\impulsevariable^{-} \in [\localextremevalueintervalleft,\localextremevalueintervalright - \intervalduration]$ with $\localextremevalueintervalleft,\localextremevalueintervalright \in \realnumbers, \localextremevalueintervalleft \leq \localextremevalueintervalright$, such that
	\begin{alignat}{1}
		\heavisideduration{\Sh}{\impulsevariable^{+},\impulsevariable^{+} + \intervalduration}{} \geq \heavisideduration{\Sh}{\impulsevariable, \impulsevariable + \intervalduration}{} \geq \heavisideduration{\Sh}{\impulsevariable^{-}, \impulsevariable^{-} + \intervalduration}{} ~~~~~~\forall \impulsevariable \in [\localextremevalueintervalleft,\localextremevalueintervalright-\intervalduration]
	\end{alignat}
\end{problem}
To solve Problem \ref{problem:localextrema}, let us consider the extreme value theorem by Bolzano \cite{keisler2013elementary}. It states that a continuous real-valued function attains a maximum and minimum in a closed interval $[\localextremevalueintervalleft,\localextremevalueintervalright - \intervalduration]$.
However, the \heavisidedurationname{} is not a continuous but a discontinuous function, so that we cannot apply Bolzano's theorem to ensure the existence of extreme values of the \heavisidedurationname{}. Hence, we need to show the existence of a maximum and minimum in $[\localextremevalueintervalleft,\localextremevalueintervalright - \intervalduration]$.
%TODO hier die s_m ersetzen druch shiftmatrixentry. s also grenze ersetzen.
\begin{lemma}[Existence of extreme values]\label{lemma:existenceextremevalues}
	Let $\localextremevalueintervalleft,\localextremevalueintervalright \in \realnumbers$ with $\localextremevalueintervalleft \leq \localextremevalueintervalright$, $\intervalduration \in \nonnegativerealnumbers$ and $\impulseinterference{\impulseshift}{\impulsedegree}{\impulsevariable}$ be a \impulseinterferencename{}.
	If 
	\begin{equation}
		\forall \impulseinterferenceindexone \in \naturalnumbersuntil{\impulseinterferencedegreeone}, \impulseinterferenceindextwo \in \naturalnumbersuntil{\impulseinterferencedegreetwo} \colon \degreevectorcomponent{\impulseinterferenceindexone,\impulseinterferenceindextwo} \in \naturalnumbers \setminus \{\infty\} \lor \phasevectorcomponent{\impulseinterferenceindexone,\impulseinterferenceindextwo} \in \realnumbers \setminus \{0\} \label{eq:extremevalueassumption}
	\end{equation}
	then, there exist values $\impulsevariable^{+},\impulsevariable^{-} \in [\localextremevalueintervalleft,\localextremevalueintervalright]$, such that 
	\begin{alignat}{1}
		\heavisideduration{\Sh}{\impulsevariable^{+},\impulsevariable^{+} + \intervalduration}{} \geq \heavisideduration{\Sh}{\impulsevariable, \impulsevariable + \intervalduration}{} \geq \heavisideduration{\Sh}{\impulsevariable^{-}, \impulsevariable^{-} + \intervalduration}{} ~~~~~~\forall \impulsevariable \in [\localextremevalueintervalleft,\localextremevalueintervalright - \intervalduration]
	\end{alignat}
holds.
\end{lemma}
\begin{proof}	
	Assume that 	$\exists \impulseinterferenceindexone \in \naturalnumbersuntil{\impulseinterferencedegreeone}, \impulseinterferenceindextwo \in \naturalnumbersuntil{\impulseinterferencedegreetwo} \colon \degreevectorcomponent{\impulseinterferenceindexone,\impulseinterferenceindextwo} = \infty \land \phasevectorcomponent{\impulseinterferenceindexone,\impulseinterferenceindextwo} = 0$. This means that infinitely many impulses occur at a certain point. The \heavisidedurationname{} then sums up the positive amplitudes of infinitely many impulses which is infinite. Hence, a maximum does not exists.
	
	Now, assume the opposite statement shown in Equation \eqref{eq:extremevalueassumption}.
	If $ \phasevectorcomponent{\impulseinterferenceindexone,\impulseinterferenceindextwo} \in \realnumbers \setminus \{0\}$, then the $(\impulseinterferenceindexone, \impulseinterferenceindextwo)$-th \impulsespectraltrainname{} generates finitely many impulses in the interval $[\localextremevalueintervalleft ,\localextremevalueintervalright]$ as they are separated by a non-zero phase $\phasevectorcomponent{\impulseinterferenceindexone,\impulseinterferenceindextwo}$. 
	If $\degreevectorcomponent{\impulseinterferenceindexone,\impulseinterferenceindextwo} \in \naturalnumbers \setminus \{\infty\} $, then the $(\impulseinterferenceindexone,\impulseinterferenceindextwo)$-th \impulsespectraltrainname{} generates finitely many impulses over time, hence, also finitely many impulses in $[\localextremevalueintervalleft ,\localextremevalueintervalright ]$. 
	As there are $\impulseinterferencedegreeone \cdot \impulseinterferencedegreetwo \in \naturalnumbers$ many \impulsespectraltrainname s, the \impulseinterferencename{} generates finitely many impulses in $[\localextremevalueintervalleft ,\localextremevalueintervalright ]$. 
	Furthermore, the impulses have a finite and positive amplitude. 
	%The sum of the amplitudes of finitely many impulses of some interval $[s ,\localextremevalueintervalright + \intervalduration]$ is finite. Hence,  $\heavisideduration{\Sh}{\impulsevariable, \impulsevariable + \intervalduration}{} $ is finite.
	If $k \in \naturalnumbers$ is the number of impulses in $[\localextremevalueintervalleft ,\localextremevalueintervalright ]$ and $a^{+}$ is the maximum amplitude of the impulses in $[\localextremevalueintervalleft ,\localextremevalueintervalright ]$, then the \heavisidedurationname{} is bounded by 
	$$0 \leq \heavisideduration{\Sh}{\impulsevariable, \impulsevariable + \intervalduration}{} \leq k\cdot a^{+}$$ 
	since the minimum amplitude is 0. %and their may be no impulses counted by the \heavisidedurationname.
	For increasing $\impulsevariable$ from $\localextremevalueintervalleft$ to $\localextremevalueintervalright - \intervalduration$, an impulse occurring at $\impulsevariable_{1} \in [\localextremevalueintervalleft,\localextremevalueintervalright - \intervalduration]$ changes the value of the \heavisidedurationname{} at most twice: When it is included for $ \impulsevariable < \impulsevariable_{1} < \impulsevariable + \intervalduration $ and when it is excluded from the \heavisidedurationname{} for $ \impulsevariable_{1} < \impulsevariable $.
	As the \heavisidedurationname{} is otherwise constant,
	it follows that the cardinality of its image in $[\localextremevalueintervalleft, \localextremevalueintervalright]$ is upper bounded by $2k$.  A finite image of bounded values implies that the  \heavisidedurationname{} $ \heavisideduration{\Sh}{\impulsevariable, \impulsevariable + \intervalduration}{} $ has a maximum and minimum in $[\localextremevalueintervalleft,\localextremevalueintervalright]$.
\end{proof}
After showing the existence of the extreme values, we formalize the maximum and minimum \heavisidedurationname{}.
\begin{definition}[Maximum and minimum Heaviside duration]\label{def:maxminheavisideduration}
	Let $\localextremevalueintervalleft,\localextremevalueintervalright \in \realnumbers$ with $\localextremevalueintervalleft \leq \localextremevalueintervalright$, $\intervalduration \in \nonnegativerealnumbers$ and  $\impulseinterference{\impulseshift}{\impulsedegree}{\impulsevariable}$ be a \impulseinterferencename{}.
	Then,
	\begin{alignat}{1}
		\heavisideduration{\Sh}{\localextremevalueintervalleft, \localextremevalueintervalright,\intervalduration}{+} \coloneqq \max_{\impulsevariable \in [\localextremevalueintervalleft, \localextremevalueintervalright- \intervalduration]}\{ \heavisideduration{\Sh}{\impulsevariable, \impulsevariable + \intervalduration}{} \} 
		=
		\max_{\impulsevariable \in [\localextremevalueintervalleft, \localextremevalueintervalright - \intervalduration]}\{
					\amplitudevector{\impulseindex_{\impulseinterferenceindexone,1} \dots \impulseindex_{\impulseinterferenceindexone,\multiperiodicimpulsedimension}} \overline{| \langle \phasevectorcomponent{\impulseindex_{\impulseinterferenceindexone,1}} \dots \phasevectorcomponent{\impulseindex_{\impulseinterferenceindexone,\multiperiodicimpulsedimension}} \rangle  |}_{\impulsevariable,\impulsevariable + \intervalduration} \}
				\\
		\heavisideduration{\Sh}{\localextremevalueintervalleft,\localextremevalueintervalright,\intervalduration}{-} \coloneqq \min_{\impulsevariable \in [\localextremevalueintervalleft, \localextremevalueintervalright- \intervalduration]}\{ \heavisideduration{\Sh}{\impulsevariable, \impulsevariable + \intervalduration}{} \} 
				=
		\min_{\impulsevariable \in [\localextremevalueintervalleft, \localextremevalueintervalright- \intervalduration]}\{
		\amplitudevector{\impulseindex_{\impulseinterferenceindexone,1} \dots \impulseindex_{\impulseinterferenceindexone,\multiperiodicimpulsedimension}} \overline{| \langle \phasevectorcomponent{\impulseindex_{\impulseinterferenceindexone,1}} \dots \phasevectorcomponent{\impulseindex_{\impulseinterferenceindexone,\multiperiodicimpulsedimension}} \rangle  |}_{\impulsevariable,\impulsevariable + \intervalduration} \}
	\end{alignat}
	are called \textbf{\maximumheavisidedurationname} and \textbf{\minimumheavisidedurationname}.
\end{definition}
As the maximum and minimum \heavisidedurationname{} are local extrema of the \heavisidedurationname{} on intervals of length $\intervalduration$, we call them the \textbf{interval transformation} of the \heavisidedurationname.
\subsection{Maximum and minimum bounds as functions}\label{section:analysisofextremevalues}
The \heavisidedurationname{}  based on the Heaviside distribution abstracts from the concrete definition of its values at jump discontinuities. Indeed, the left- and right-continuous versions of the \heavisidedurationname{} belong to the same distribution \cite{strichartz2003guide}, as discussed in Equation \eqref{eq:heavisideintegralzero} of Section \ref{section:heavisidefunctionanddiracdelta}. 
Therefore, we can work on this abstraction level as long as concrete function values at the jump discontinuities are not considered.
If we want to compute the extrema of the \heavisidedurationname{} in an interval $[\localextremevalueintervalleft,\localextremevalueintervalright]$, and if e.g. a jump occurs at $\localextremevalueintervalleft$ or $\localextremevalueintervalright$, then the left- or right-continuity may decide whether the extremum occurs at the lower or upper point of the jump. The continuity also impacts the analysis to compute the extrema.
To describe these analyses, we present the upper and lower Heaviside function of \cite{slomka2021beyond} that define whether the upper or lower point of the jump is the function value.
\begin{definition}[Upper and lower Heaviside functions]
	The \textbf{upper Heaviside function} and \textbf{lower Heaviside function} are defined by
	\begin{alignat}{1}
		\upperheaviside(\impulsevariable) \coloneqq 
		\left\{ 
		\begin{array}{ll}
			1,& \impulsevariable \geq 0 \\
			0,& \impulsevariable < 0
		\end{array}
		\right.
		\\
		\lowerheaviside(\impulsevariable) \coloneqq 
		\left\{ 
		\begin{array}{ll}
			1,& \impulsevariable > 0 \\
			0,& \impulsevariable \leq 0
		\end{array}
		\right.
	\end{alignat}
\end{definition}
Based on these two functions, we concretize the \heavisidedurationname{} and its maxima and minima in Table \ref{table:listofsymbols}.
The analysis of the extrema is presented. 
\begin{theorem}[Analysis of maximum \heavisidedurationname]\label{theorem:analysisofmaximumheavisideduration} 
	Let $\localextremevalueintervalleft,\localextremevalueintervalright \in \realnumbers$ with $\localextremevalueintervalleft \leq \localextremevalueintervalright$, $\intervalduration \in \nonnegativerealnumbers$ and  $\impulseinterference{\impulseshift}{\impulsedegree}{\impulsevariable}$ be a \impulseinterferencename{}.
	Let $\setofimpulsetimepoint$ be the set of points at which impulses of $\impulseinterference{\impulseshift}{\impulsedegree}{\impulsevariable}$  occur and let $\setofimpulsetimepointminusintervalduration \coloneqq \{ a - \intervalduration \, | \, a \in \setofimpulsetimepoint \}$ be the set of impulse points subtracted by the interval duration $\intervalduration \in \nonnegativerealnumbers$.
	Let
	\begin{alignat}{1}\label{eq:testpointsmax1}
		\testsetleft &=  ([\localextremevalueintervalleft, \localextremevalueintervalright] \cap \setofimpulsetimepoint) \cup \{\localextremevalueintervalright\}
		\\
		\testsetright &= ([\localextremevalueintervalleft + \intervalduration, \localextremevalueintervalright + \intervalduration] \cap \setofimpulsetimepoint) \cup \{\localextremevalueintervalleft + \intervalduration\}\label{eq:testpointsmax2}
	\end{alignat}
	Then, the \maximumheavisidedurationname{} is computed by
	\begin{alignat}{1}
		\heavisidedurationupperupper{\Sh}{\intervalduration}{+} 
		&
		=
		 \max_{\impulsevariable \in \testsetleft}\{ \heavisidedurationupperupper{\Sh}{\impulsevariable, \impulsevariable + \intervalduration}{} \} 
		=
		 \max_{\impulsevariable \in \testsetleft}\{
		\amplitudevector{\impulseindex_{\impulseinterferenceindexone,1} \dots \impulseindex_{\impulseinterferenceindexone,\multiperiodicimpulsedimension}} \overline{| \langle \phasevectorcomponent{\impulseindex_{\impulseinterferenceindexone,1}} \dots \phasevectorcomponent{\impulseindex_{\impulseinterferenceindexone,\multiperiodicimpulsedimension}} \rangle  |}_{\overline{\impulsevariable},\overline{\impulsevariable + \intervalduration}} \}
		\\
		\heavisidedurationupperlower{\Sh}{\intervalduration}{+} 
		&
		=
		\max_{\impulsevariable \in \testsetleft}\{ \heavisidedurationupperlower{\Sh}{\impulsevariable, \impulsevariable + \intervalduration}{} \} 
		=
		\max_{\impulsevariable \in \testsetleft}\{
	\amplitudevector{\impulseindex_{\impulseinterferenceindexone,1} \dots \impulseindex_{\impulseinterferenceindexone,\multiperiodicimpulsedimension}} \overline{| \langle \phasevectorcomponent{\impulseindex_{\impulseinterferenceindexone,1}} \dots \phasevectorcomponent{\impulseindex_{\impulseinterferenceindexone,\multiperiodicimpulsedimension}} \rangle  |}_{\overline{\impulsevariable},\underline{\impulsevariable + \intervalduration}} \}
		\\
		\heavisidedurationlowerupper{\Sh}{\intervalduration}{+} 
		&
		=
		\max_{\impulsevariable \in \testsetright}\{ \heavisidedurationlowerupper{\Sh}{\impulsevariable - \intervalduration, \impulsevariable }{} \} 
		=
		\max_{\impulsevariable \in \testsetright }\{
	\amplitudevector{\impulseindex_{\impulseinterferenceindexone,1} \dots \impulseindex_{\impulseinterferenceindexone,\multiperiodicimpulsedimension}} \overline{| \langle \phasevectorcomponent{\impulseindex_{\impulseinterferenceindexone,1}} \dots \phasevectorcomponent{\impulseindex_{\impulseinterferenceindexone,\multiperiodicimpulsedimension}} \rangle  |}_{\underline{\impulsevariable - \intervalduration},\overline{\impulsevariable }} \}
	\end{alignat}
\end{theorem}
\begin{proof}
	Let $\impulsevariable \in \realnumbers$ be a point.
	Let $\impulsevariable^{*} = \min \{ \impulsevariable' \in \testsetleft \, | \, \impulsevariable' \geq \impulsevariable \}$. By definition of $\impulsevariable^{*}$, there does not exist an impulse in the interval $(\impulsevariable, \impulsevariable^{*})$. It follows that no impulse is excluded when shifting the \heavisidedurationname s $\heavisidedurationupperupper{\Sh}{\impulsevariable, \impulsevariable + \intervalduration}{} $ and $\heavisidedurationupperlower{\Sh}{\impulsevariable, \impulsevariable + \intervalduration}{} $ to $\heavisidedurationupperupper{\Sh}{\impulsevariable^{*}, \impulsevariable^{*} + \intervalduration}{} $ and $\heavisidedurationupperlower{\Sh}{\impulsevariable^{*}, \impulsevariable^{*} + \intervalduration}{} $, since the impulse at $\impulsevariable^{*}$ is included by these \heavisidedurationname s.
	This means the \heavisidedurationname{} can only increase from $\impulsevariable$ to $\impulsevariable^{*}$. Formally,
	\begin{alignat}{1}
		\heavisidedurationupperupper{\Sh}{\impulsevariable, \impulsevariable + \intervalduration}{} \leq \heavisidedurationupperupper{\Sh}{\impulsevariable^{*}, \impulsevariable^{*} + \intervalduration}{}
		\\
		\heavisidedurationupperlower{\Sh}{\impulsevariable, \impulsevariable + \intervalduration}{} \leq \heavisidedurationupperlower{\Sh}{\impulsevariable^{*}, \impulsevariable^{*} + \intervalduration}{}
	\end{alignat}
	which means the \heavisidedurationnameplural{} in the intervals $[\impulsevariable, \impulsevariable + \intervalduration]$ and $[\impulsevariable, \impulsevariable + \intervalduration)$ are respectively smaller or equal to the \heavisidedurationnameplural{} in the intervals $[\impulsevariable^{*} + \intervalduration, \impulsevariable^{*} + \intervalduration]$ and $[\impulsevariable^{*} + \intervalduration, \impulsevariable^{*} + \intervalduration)$.
	Therefore, $		 \max_{\impulsevariable \in [\localextremevalueintervalleft, \localextremevalueintervalright] \cap \setofimpulsetimepoint}\{ \heavisidedurationupperupper{\Sh}{\impulsevariable, \impulsevariable + \intervalduration}{} \}$ and respectively $		 \max_{\impulsevariable \in [\localextremevalueintervalleft, \localextremevalueintervalright] \cap \setofimpulsetimepoint}\{ \heavisidedurationupperlower{\Sh}{\impulsevariable, \impulsevariable + \intervalduration}{} \} $ is found at some $\impulsevariable = \impulsevariable^{*} \in \testsetleft$.
	
	Now, consider $\impulsevariable^{*} = \max \{ \impulsevariable' \in \testsetright \, | \, \impulsevariable' \leq \impulsevariable + \intervalduration \}$.
	By definition of $\impulsevariable^{*}$, there does not exist an impulse in the interval $(\impulsevariable^{*} + \intervalduration, \impulsevariable + \intervalduration)$. It follows that no impulse is excluded when shifting the \heavisidedurationname{} $\heavisidedurationlowerupper{\Sh}{\impulsevariable - \intervalduration, \impulsevariable }{} $ to $\heavisidedurationlowerupper{\Sh}{\impulsevariable^{*} - \intervalduration, \impulsevariable^{*}}{} $ since the impulse at $\impulsevariable^{*}$ is included by the \heavisidedurationname{}.
	This means the \heavisidedurationname{} can only increase from $\impulsevariable + \intervalduration$ to $\impulsevariable^{*}$. Formally,
	\begin{alignat}{1}
		\heavisidedurationlowerupper{\Sh}{\impulsevariable - \intervalduration, \impulsevariable }{} \leq \heavisidedurationlowerupper{\Sh}{\impulsevariable^{*} - \intervalduration, \impulsevariable^{*}}{}
	\end{alignat}
	Therefore, $		 \max_{\impulsevariable \in [\localextremevalueintervalleft, \localextremevalueintervalright] \cap \setofimpulsetimepoint}\{ \heavisidedurationlowerupper{\Sh}{\impulsevariable - \intervalduration, \impulsevariable}{} \}$  is found at some $\impulsevariable = \impulsevariable^{*} \in \testsetright$.
	\end{proof}

\begin{theorem}[Analysis of minimum \heavisidedurationname]\label{theorem:analysisofminimumheavisideduration} 
	Let $\localextremevalueintervalleft,\localextremevalueintervalright \in \realnumbers$ with $\localextremevalueintervalleft \leq \localextremevalueintervalright$, $\intervalduration \in \nonnegativerealnumbers$ and  $\impulseinterference{\impulseshift}{\impulsedegree}{\impulsevariable}$ be a \impulseinterferencename{}.
	Let $\setofimpulsetimepoint$ be the set of values at which impulses of $\impulseinterference{\impulseshift}{\impulsedegree}{\impulsevariable}$ occur and let $\setofimpulsetimepointminusintervalduration \coloneqq \{ a - \intervalduration \, | \, a \in \setofimpulsetimepoint \}$ be the set of impulse points subtracted by the interval duration $\intervalduration \in \nonnegativerealnumbers$.
	Let 
	\begin{alignat}{1}
		\testsetleft &=  ([\localextremevalueintervalleft, \localextremevalueintervalright] \cap \setofimpulsetimepoint) \cup \{\localextremevalueintervalleft\}
		\\
		\testsetright &= ([\localextremevalueintervalleft + \intervalduration, \localextremevalueintervalright + \intervalduration] \cap \setofimpulsetimepoint) \cup \{\localextremevalueintervalright + \intervalduration\}
	\end{alignat}
Then, the \minimumheavisidedurationname{} is computed by
\begin{alignat}{1}
	\heavisidedurationlowerlower{\Sh}{\intervalduration}{-} 
	&
	=
	\min_{\impulsevariable \in \testsetleft}\{ \heavisidedurationlowerlower{\Sh}{\impulsevariable, \impulsevariable + \intervalduration}{} \} 
	=
	\min_{\impulsevariable \in \testsetleft}\{
	\amplitudevector{\impulseindex_{\impulseinterferenceindexone,1} \dots \impulseindex_{\impulseinterferenceindexone,\multiperiodicimpulsedimension}} \overline{| \langle \phasevectorcomponent{\impulseindex_{\impulseinterferenceindexone,1}} \dots \phasevectorcomponent{\impulseindex_{\impulseinterferenceindexone,\multiperiodicimpulsedimension}} \rangle  |}_{\underline{\impulsevariable},\underline{\impulsevariable + \intervalduration}} \}
	\\
	\heavisidedurationlowerupper{\Sh}{\intervalduration}{-} 
	&
	=
	\min_{\impulsevariable \in \testsetleft}\{ \heavisidedurationlowerupper{\Sh}{\impulsevariable, \impulsevariable + \intervalduration}{} \} 
	=
	\min_{\impulsevariable \in \testsetleft}\{
	\amplitudevector{\impulseindex_{\impulseinterferenceindexone,1} \dots \impulseindex_{\impulseinterferenceindexone,\multiperiodicimpulsedimension}} \overline{| \langle \phasevectorcomponent{\impulseindex_{\impulseinterferenceindexone,1}} \dots \phasevectorcomponent{\impulseindex_{\impulseinterferenceindexone,\multiperiodicimpulsedimension}} \rangle  |}_{\underline{\impulsevariable},\overline{\impulsevariable + \intervalduration}} \}
	\\
	\heavisidedurationupperlower{\Sh}{\intervalduration}{-} 
	&
	=
	\min_{\impulsevariable \in \testsetright}\{ \heavisidedurationlowerupper{\Sh}{\impulsevariable - \intervalduration, \impulsevariable }{} \} 
	=
	\min_{\impulsevariable \in \testsetright}\{
	\amplitudevector{\impulseindex_{\impulseinterferenceindexone,1} \dots \impulseindex_{\impulseinterferenceindexone,\multiperiodicimpulsedimension}} \overline{| \langle \phasevectorcomponent{\impulseindex_{\impulseinterferenceindexone,1}} \dots \phasevectorcomponent{\impulseindex_{\impulseinterferenceindexone,\multiperiodicimpulsedimension}} \rangle  |}_{\overline{\impulsevariable - \intervalduration},\underline{\impulsevariable }} \}
\end{alignat}
\end{theorem}
\begin{proof}
	Let $\impulsevariable \in \realnumbers$ be a value.
	Let $\impulsevariable^{*} = \max \{ \impulsevariable' \in \testsetleft \, | \, \impulsevariable' \leq \impulsevariable \}$. By definition of $\impulsevariable^{*}$, there does not exist an impulse in the interval $(\impulsevariable^{*}, \impulsevariable)$. It follows that no impulse is included when shifting the \heavisidedurationname s $\heavisidedurationlowerlower{\Sh}{\impulsevariable, \impulsevariable + \intervalduration}{} $ and $\heavisidedurationlowerupper{\Sh}{\impulsevariable, \impulsevariable + \intervalduration}{} $ to $\heavisidedurationlowerlower{\Sh}{\impulsevariable^{*}, \impulsevariable^{*} + \intervalduration}{} $ and $\heavisidedurationlowerupper{\Sh}{\impulsevariable^{*}, \impulsevariable^{*} + \intervalduration}{} $, since the impulse at $\impulsevariable^{*}$ is not included by these \heavisidedurationname s.
	This means the \heavisidedurationname{} can only decrease from $\impulsevariable$ to $\impulsevariable^{*}$. Formally,
	\begin{alignat}{1}
		\heavisidedurationlowerlower{\Sh}{\impulsevariable, \impulsevariable + \intervalduration}{} \geq \heavisidedurationlowerlower{\Sh}{\impulsevariable^{*}, \impulsevariable^{*} + \intervalduration}{}
		\\
		\heavisidedurationlowerupper{\Sh}{\impulsevariable, \impulsevariable + \intervalduration}{} \geq \heavisidedurationlowerupper{\Sh}{\impulsevariable^{*}, \impulsevariable^{*} + \intervalduration}{}
	\end{alignat}
	Therefore, $\min_{\impulsevariable \in [\localextremevalueintervalleft,\localextremevalueintervalright]}\{ \heavisidedurationlowerlower{\Sh}{\impulsevariable, \impulsevariable + \intervalduration}{} \} $ and respectively $\min_{\impulsevariable \in [\localextremevalueintervalleft,\localextremevalueintervalright]}\{ \heavisidedurationlowerupper{\Sh}{\impulsevariable, \impulsevariable + \intervalduration}{} \}$ is found at some $\impulsevariable = \impulsevariable^{*} \in \testsetleft$.
	
	Now, consider $\impulsevariable^{*} = \min \{ a \in \testsetright \, | \, a \geq \impulsevariable + \intervalduration \}$.
	By definition of $\impulsevariable^{*}$, there does not exist an impulse in the interval $(\impulsevariable + \intervalduration, \impulsevariable^{*})$. It follows that no impulse is included when shifitng the \heavisidedurationname{} $\heavisidedurationupperlower{\Sh}{\impulsevariable, \impulsevariable + \intervalduration}{} $ to $\heavisidedurationupperlower{\Sh}{\impulsevariable^{*} - \intervalduration, \impulsevariable^{*}}{} $ since the impulse at $\impulsevariable^{*}$ is excluded by the \heavisidedurationname{}.
	This means the \heavisidedurationname{} can only decrease from $\impulsevariable + \intervalduration$ to $\impulsevariable^{*}$. Formally,
	\begin{alignat}{1}
		\heavisidedurationupperlower{\Sh}{\impulsevariable, \impulsevariable + \intervalduration}{} \geq \heavisidedurationupperlower{\Sh}{\impulsevariable^{*} - \intervalduration, \impulsevariable^{*}}{}
	\end{alignat}
	Therefore, $\min_{\impulsevariable \in [\localextremevalueintervalleft,\localextremevalueintervalright]}\{ \heavisidedurationupperlower{\Sh}{\impulsevariable - \intervalduration, \impulsevariable}{} \}$ is found at some $\impulsevariable = \impulsevariable^{*} \in \testsetright$.
\end{proof}

\subsection{Graph of the maximum bound}
The interval transformation describes the \maximumheavisidedurationname{} for a single distance $\intervalduration$. 
We showed that the transformation is computed by a maximum over a finite set of \heavisidedurationnameplural{}.
However, if we want to compute the graph of the \maximumheavisidedurationname{} of the interval $[\localextremevalueintervalleft,\localextremevalueintervalright]$, then we have to find all intervals $[x_{1},x_{2}] \subseteq [\localextremevalueintervalleft,\localextremevalueintervalright]$ that result in the \maximumheavisidedurationname{} and \minimumheavisidedurationname{}.

The approach is to determine all interval length $x_{2} - x_{1}$ that are required to find all function values of the interval transformation and then to apply Theorem \ref{theorem:analysisofmaximumheavisideduration} and \ref{theorem:analysisofminimumheavisideduration} to compute the graph of the interval transformation. We begin by defining the set of \distancesetname{} in the interval $[a,b]$.
\begin{definition}[\distancesetnamecapital]
	Let
	 $\impulseinterference{\impulseshift}{\impulsedegree}{\impulsevariable}$
	be a \impulseinterferencenameshort{} and $a,b \in \realnumbers$ with $a \leq b$. 
	Let $\setofimpulsetimepoint$ be the set of values at which impulses of $\impulseinterference{\impulseshift}{\impulsedegree}{\impulsevariable}$ occur. Then,
	\begin{equation}
		\distanceset{\setofimpulsetimepoint}{a}{b} = \big\{ |x_{1} - x_{2}|  ~\big| ~  x_{1},x_{2} \in \setofimpulsetimepoint \cup \{a,b\} ~ \land ~ x_{1},x_{2} \in [a,b] \big\}
	\end{equation}
	describes the set of absolute differences of \shiftnameplural{} of $\impulseinterference{\impulseshift}{\impulsedegree}{\impulsevariable}$ in the interval $[a,b]$ and their absolute differences to the interval boundaries $a$ and $b$. $\distanceset{\setofimpulsetimepoint}{a}{b} $ is called  \textbf{\distancesetname} of $[a,b]$.
\end{definition}
In Section \ref{section:analysisofextremevalues}, we derived the finite set of test values to compute the local extrema of a given interval $[a,b]$ and some $\intervalduration$. To complete the interval transformation, we show that the required $\intervalduration$ to find the local extrema in $[a,b]$ are equal to the \distancesetname{} in $[a,b]$.
\begin{lemma}[Piece-wise constant \heavisidedurationname]\label{lemma:piecewiseconstant}
	Let
	$\impulseinterference{\impulseshift}{\impulsedegree}{\impulsevariable}$
	be a \impulseinterferencenameshort{} and $a,b \in \realnumbers$ with $a \leq b$. 
	Let $\setofimpulsetimepoint$ be the set of values at which impulses of $\impulseinterference{\impulseshift}{\impulsedegree}{\impulsevariable}$ occur. 
	Let $\distanceset{\setofimpulsetimepoint}{a}{b} $ be the  \distancesetname{} of $[a,b]$.
	Let $x^{*} \in \setofimpulsetimepoint \cap [a,b]$ and $\intervalduration_{1}, \intervalduration_{2} \in \distanceset{\setofimpulsetimepoint}{a}{b} $ with $\intervalduration_{1} < \intervalduration_{2}$, such that $x^{*}  + \intervalduration_{1} \in [a,b]$ and $x^{*} + \intervalduration_{2} \in [a,b]$ are two consecutive impulse shifts, i.e. an impulse occurs at $x^{*}  + \intervalduration_{1}$ and another impulse occurs at  $x^{*} + \intervalduration_{2}$ and there does not exist an impulse occurring in the interval $(x^{*} + \intervalduration_{1}, x^{*} + \intervalduration_{2})$.
	Then,
	\begin{alignat}{1}
		\forall \intervalduration \in (\intervalduration_{1}, \intervalduration_{2}) \colon
		\heavisidedurationupperupper{\Sh}{x^{*}, x^{*} + \intervalduration_{1}}{} 
		&
		=
		\heavisidedurationupperupper{\Sh}{x^{*}, x^{*} + \intervalduration}{} 
		\nonumber
		\\
		&
		<
		\heavisidedurationupperupper{\Sh}{x^{*}, x^{*}+ \intervalduration_{2}}{} 
		\\
		\forall \intervalduration \in ( \intervalduration_{1}, \intervalduration_{2}) \colon
		\heavisidedurationlowerupper{\Sh}{x^{*}, x^{*} + \intervalduration_{1}}{} 
		&
		=
		\heavisidedurationlowerupper{\Sh}{x^{*}, x^{*} + \intervalduration}{} 
		\nonumber
		\\
		&
		<
		\heavisidedurationupperlower{\Sh}{x^{*}, x^{*}+ \intervalduration_{2}}{} 
		\\
		\forall \intervalduration \in (\intervalduration_{1}, \intervalduration_{2}) \colon
		\heavisidedurationupperlower{\Sh}{x^{*}, x^{*}+ \intervalduration_{1}}{} 
		&
		<
		\heavisidedurationupperlower{\Sh}{x^{*}, x^{*}+ \intervalduration}{} 
		\nonumber
		\\
		&
		=
		\heavisidedurationupperlower{\Sh}{x^{*}, x^{*}+ \intervalduration_{2}}{} 
	\end{alignat}
	which means that the \heavisidedurationname{} is a piece-wise constant and monotonically increasing function. 
\end{lemma}
\begin{proof}
	Consider the intervals $[x^{*},x^{*} + \intervalduration_{1}]$, $(x^{*},x^{*} + \intervalduration_{1}]$ and $[x^{*},x^{*} + \intervalduration_{1})$.
	The sum of the amplitudes of impulses occurring in these intervals are computed by the \heavisidedurationnameplural{} $\heavisidedurationupperupper{\Sh}{x^{*}, x^{*}+ \intervalduration_{1}}{} $, $\heavisidedurationlowerupper{\Sh}{x^{*}, x^{*}+ \intervalduration_{1}}{} $ and $\heavisidedurationupperlower{\Sh}{x^{*}, x^{*}+ \intervalduration_{1}}{} $, called the first, second and third \heavisidedurationname{} in this proof.
	Note that the impulse at $x^{*} + \intervalduration_{1}$ is included by the first and second and excluded by the third \heavisidedurationname.
	This means that
	\begin{equation}
		\forall \intervalduration \in (\intervalduration_{1}, \intervalduration_{2}) \colon
		\heavisidedurationupperlower{\Sh}{x^{*}, x^{*}+ \intervalduration_{1}}{} 
		<
		\heavisidedurationupperlower{\Sh}{x^{*}, x^{*}+ \intervalduration}{} 
	\end{equation}
	since the impulse is included as soon as $\intervalduration > \intervalduration_{1}$.
	However, by assumption, there does not occur an impulse in the interval $(x^{*} + \intervalduration_{1}, x^{*} + \intervalduration_{2})$. 
	Hence, the \heavisidedurationnameplural{} $\heavisidedurationupperupper{\Sh}{x^{*}, x^{*}+ \intervalduration}{} $, $\heavisidedurationlowerupper{\Sh}{x^{*}, x^{*}+ \intervalduration}{} $ and $\heavisidedurationupperlower{\Sh}{x^{*}, x^{*}+ \intervalduration}{}$  are constant $\forall \intervalduration \in (\intervalduration_{1}, \intervalduration_{2})$. 
	This implies for the third \heavisidedurationname{} that
	\begin{equation}\label{eq:thirdheavisideduration}
		\forall \intervalduration \in (\intervalduration_{1}, \intervalduration_{2}) \colon
		\heavisidedurationupperlower{\Sh}{x^{*}, x^{*}+ \intervalduration_{1}}{} 
		<
		\heavisidedurationupperlower{\Sh}{x^{*}, x^{*}+ \intervalduration}{} 
		= 
				\heavisidedurationupperlower{\Sh}{x^{*}, x^{*}+ \intervalduration_{2}}{} 
	\end{equation}
	Furthermore, as there occurs an impulse at $x^{*} + \intervalduration_{2}$, it follows that the first and second \heavisidedurationname{} increase by the amplitude of that impulse, formally, 
	\begin{alignat}{1}\label{eq:firstheavisideduration}
		\heavisidedurationupperupper{\Sh}{x^{*}, x^{*} + \intervalduration}{} 
		&
		<
		\heavisidedurationupperupper{\Sh}{x^{*}, x^{*}+ \intervalduration_{2}}{} 
		\\
		\heavisidedurationlowerupper{\Sh}{x^{*}, x^{*} + \intervalduration}{} 
		&
		<
		\heavisidedurationupperlower{\Sh}{x^{*}, x^{*}+ \intervalduration_{2}}{}  \label{eq:secondheavisideduration}
	\end{alignat}
	From \eqref{eq:thirdheavisideduration}, \eqref{eq:firstheavisideduration} and \eqref{eq:secondheavisideduration}, it follows that the \heavisidedurationname{} is constant $\forall \intervalduration \in (\intervalduration_{1}, \intervalduration_{2})$ and its function value discontinuously increases at \shiftnameplural{} of impulses such as $x^{*}+ \intervalduration_{1}$ and $x^{*}+ \intervalduration_{2}$.
\end{proof}
The left- or right-continuity at jump discontinuities depends on the choice of the Heaviside mask.
Based on the piece-wise constant property of the \heavisidedurationname{}, we present the computation of the graph of the interval transformation for an interval $[a,b]$.
\begin{theorem}[Graph of interval transformation]\label{theorem:graphofintervaltransformation}
	Let
	$\impulseinterference{\impulseshift}{\impulsedegree}{\impulsevariable}$
	be a \impulseinterferencenameshort{} and $a,b \in \realnumbers$ with $a \leq b$. 
	Let $\setofimpulsetimepoint$ be the set of \shiftnameplural{} of $\impulseinterference{\impulseshift}{\impulsedegree}{\impulsevariable}$ occur. 
	Let $\distanceset{\setofimpulsetimepoint}{a}{b} $ be the  \distancesetname{} of $[a,b]$.
	Then, the interval transformation of the interval $[a,b]$ is computed by
	\begin{alignat}{1}
		\intervaltransformationupperupper{+}{a}{b} 
		&
		=
		\left\{ \left(\intervalduration, \heavisidedurationupperupper{\Sh}{\intervalduration}{+}\right) \big| \intervalduration \in \distanceset{\setofimpulsetimepoint}{a}{b} \right\}
		\label{eq:graph1}
		\\
		\intervaltransformationupperlower{+}{a}{b} 
		&
		=
		\left\{ \left(\intervalduration, \heavisidedurationupperlower{\Sh}{\intervalduration}{+}\right) \big| \intervalduration \in \distanceset{\setofimpulsetimepoint}{a}{b} \right\}
		\label{eq:graph2}
		\\
		\intervaltransformationlowerupper{+}{a}{b} 
		&
		=
		\left\{ \left(\intervalduration, \heavisidedurationlowerupper{\Sh}{\intervalduration}{+}\right) \big| \intervalduration \in \distanceset{\setofimpulsetimepoint}{a}{b} \right\}
		\label{eq:graph3}
		\\
		\intervaltransformationlowerlower{-}{a}{b} 
		&
		=
		\left\{ \left(\intervalduration, \heavisidedurationlowerlower{\Sh}{\intervalduration}{-}\right) \big| \intervalduration \in \distanceset{\setofimpulsetimepoint}{a}{b} \right\}
		\label{eq:graph4}
		\\
		\intervaltransformationlowerupper{-}{a}{b} 
		&
		=
		\left\{ \left(\intervalduration, \heavisidedurationlowerupper{\Sh}{\intervalduration}{-}\right) \big| \intervalduration \in \distanceset{\setofimpulsetimepoint}{a}{b} \right\}
		\label{eq:graph5}
		\\
		\intervaltransformationupperlower{-}{a}{b} 
		&
		=
		\left\{ \left(\intervalduration, \heavisidedurationupperlower{\Sh}{\intervalduration}{-}\right) \big| \intervalduration \in \distanceset{\setofimpulsetimepoint}{a}{b} \right\}
		\label{eq:graph6}
	\end{alignat}
\end{theorem}
\begin{proof}
	By Lemma \ref{lemma:piecewiseconstant}, the \heavisidedurationname{} is constant and only increases at shifts of $\impulseinterference{\impulseshift}{\impulsedegree}{\impulsevariable}$. 
	This means, if 		$X(a,b) =	\setofimpulsetimepoint \cup \{a,b\} \cap [a,b] $ is the set of shifts with $a$ and $b$ in $[a,b]$, then all function values of the \heavisidedurationname{} in the interval $[a,b]$ are covered by  the set 
	\begin{alignat}{1}
		\heavisideduration{\Sh}{[a,b]}{}
		&
		=
		\{ \heavisideduration{\Sh}{x_{1},x_{2}}{} \, \big| \, x_{1},x_{2} \in X(a,b) \}
		\\
		&
		=
		\{ \heavisideduration{\Sh}{x_{1},x_{1} + x_{2} - x_{1}}{} \, \big| \, x_{1},x_{2} \in X(a,b) \}
		\\
		&
		=
		\{ \heavisideduration{\Sh}{x_{1},x_{1} + \intervalduration}{} \, \big| \, x_{1} \in X(a,b), \intervalduration \in \distanceset{\setofimpulsetimepoint}{a}{b}, x_{1} + \intervalduration \leq b \}
		\\
		&
		=
		\bigcup_{\intervalduration \in \distanceset{\setofimpulsetimepoint}{a}{b}} \{ \heavisideduration{\Sh}{x_{1},x_{1} + \intervalduration}{} \, \big| \, x_{1} \in X(a,b), x_{1} + \intervalduration \leq b \}
	\end{alignat}
	which means we can describe the function values of the \heavisidedurationname{} as a function of $\intervalduration$. For a given $\intervalduration$, we can now compute a local extremum by 
	\begin{alignat}{1}
		\bigcup_{\intervalduration \in \distanceset{\setofimpulsetimepoint}{a}{b}} \max\{ \heavisideduration{\Sh}{x_{1},x_{1} + \intervalduration}{} \, \big| \, x_{1} \in X(a,b), x_{1} + \intervalduration \leq b \}
		\\
		\bigcup_{\intervalduration \in \distanceset{\setofimpulsetimepoint}{a}{b}} \min\{ \heavisideduration{\Sh}{x_{1},x_{1} + \intervalduration}{} \, \big| \, x_{1} \in X(a,b), x_{1} + \intervalduration \leq b \}
	\end{alignat}
	But this is exactly computed by $\heavisidedurationupperupper{\Sh}{\intervalduration}{+}$, $\heavisidedurationupperlower{\Sh}{\intervalduration}{+}$, $\heavisidedurationlowerlower{\Sh}{\intervalduration}{-}$ and $\heavisidedurationlowerupper{\Sh}{\intervalduration}{-}$, by Theorem \ref{theorem:analysisofmaximumheavisideduration} and \ref{theorem:analysisofminimumheavisideduration} .
	Similarly, we have 
	\begin{alignat}{1}
		\heavisideduration{\Sh}{[a,b]}{}
		&
		=
		\{ \heavisideduration{\Sh}{x_{1},x_{2}}{} \, \big| \, x_{1},x_{2} \in X(a,b) \}
		\\
		&
		=
		\{ \heavisideduration{\Sh}{x_{2} + x_{1} - x_{2}, x_{2} }{} \, \big| \, x_{1},x_{2} \in X(a,b) \}
		\\
		&
		=
		\{ \heavisideduration{\Sh}{x_{2} - \intervalduration,x_{2} }{} \, \big| \, x_{2} \in X(a,b), \intervalduration \in \distanceset{\setofimpulsetimepoint}{a}{b}, x_{2} - \intervalduration \geq a \}
		\\
		&
		=
		\bigcup_{\intervalduration \in \distanceset{\setofimpulsetimepoint}{a}{b}} \{ \heavisideduration{\Sh}{x_{2} - \intervalduration,x_{2}}{} \, \big| \, x_{2} \in X(a,b), x_{2} - \intervalduration \geq a \}
	\end{alignat}
	in case of $\heavisidedurationlowerupper{\Sh}{\intervalduration}{+}$ and $\heavisidedurationupperlower{\Sh}{\intervalduration}{-}$.
	As $\heavisideduration{\Sh}{[a,b]}{}$ includes all function values of the \heavisidedurationname{} in $[a,b]$, it follows that Equation \eqref{eq:graph1} to \eqref{eq:graph6} compute the whole graph.
\end{proof}

If the \heavisidedurationname{} is a periodic function, i.e. if $\heavisideduration{\Sh}{\impulsevariable, \impulsevariable + \intervalduration}{} = \heavisideduration{\Sh}{\impulsevariable + np , \impulsevariable + \intervalduration + np}{} $ where $p \in \positivrealnumbers$ is the period of the function and $n \in \naturalnumbers$, then we can find the global maximum and minimum of the \heavisidedurationname{} in the interval $[\localextremevalueintervalleft,\localextremevalueintervalright]$ where $\localextremevalueintervalright - \localextremevalueintervalleft = p$, because
\begin{equation}
	\forall \impulsevariable \in [\localextremevalueintervalleft, \localextremevalueintervalleft + p]\colon \heavisideduration{\Sh}{\impulsevariable , \impulsevariable + \intervalduration }{} = \heavisideduration{\Sh}{\impulsevariable + np , \impulsevariable + \intervalduration + np}{}
\end{equation}
so that all the different function values of the \heavisidedurationname{} already occur in one period.

%wenn wir das interval um eine periode erhöhen, verdoppelt sich der funktionswert. wir können also eine impulsdichte aufstellen, die sich periodisch wiederholt und durch das integral oder die diskretisierung wieder die eriegnisse periodisch zählt. jetzt muss man noch herleiten, wie man die heaviside duration zu einer dichte umwandeln kann.

This means we can limit the number of interval durations $\intervalduration$ required to compute the global maximum and minimum of a \heavisidedurationname{}. 
Generalizing, if the \heavisidedurationname{} is periodic or if the number of impulses of the \impulseinterferencenameshort{} is finite, then the interval transformation finds a global extremum based on a finite set of interval durations.
We note that if the number of impulses is infinite and if the \heavisidedurationname{} is not periodic, then we cannot find the global extrema.
\newpage
\appendix
\section{Notation}\label{appendix:notation}
The linear space of impulses requires a lot of composed summations and convolutions. Therefore, we introduce some short-form notations. First, we apply the well-known Einstein notation \cite{einstein1916foundation}
\begin{equation}
	\sum_{\impulseindex = 0}^{\impulsedegree - 1} c_{\impulseindex}x^{\impulseindex} = c_{0}x^{0} + c_{1}x^{1} + c_{2}x^{2} + c_{3}x^{3} + \ldots + c_{\impulsedegree - 1}x^{\impulsedegree - 1} \eqqcolon c_{\impulseindex}x^{\impulseindex}
\end{equation}
for a short-form notation of summations. We have 
\begin{alignat}{1}
	\impulsespectraldensityeinsteinmono{\amplitude}{\impulseindex}{\impulseshift}{\impulsevariable} 
	=
	\sum_{\impulseindex = 0}^{\impulsedegree - 1} \amplitudevector{\impulseindex} \diracdelta(\impulsevariable - \impulseshift_{\impulseindex})
	=
	\sum_{\impulseindex = 0}^{\impulsedegree - 1} 
	\amplitudevector{\impulseindex} \diracdelta(\impulsevariable - \impulseindex \cdot \impulseshift)
\end{alignat}
If multiple summations of the form $\impulsespectraldensityeinsteinmono{\amplitude}{\impulseindex}{\impulseshift}{\impulsevariable} $ are convolved, we represent their convolution by the enclosing angles $\langle \rangle$ as follows
\begin{alignat}{1}
	\impulsespectraldensityeinsteinmulti{\amplitude}{\impulseindex}{\impulseshift}{\impulsevariable} = \sum_{\impulseindex_{1} = 0}^{\degreevectorcomponent{1} - 1} 
	\amplitudevector{\impulseindex_{1}} \diracdelta(\impulsevariable - \impulseindex_{1} \phasevectorcomponent{1}) 
	\ast
	\ldots 
	\ast 
	\sum_{\impulseindex_{\impulseinterferenceindexone} = 0}^{\degreevectorcomponent{\impulseinterferenceindexone} - 1} 
	\amplitudevector{\impulseindex_{\impulseinterferenceindexone}} \diracdelta(\impulsevariable - \impulseindex_{\impulseinterferenceindexone} \phasevectorcomponent{\impulseinterferenceindexone}) 
\end{alignat}
Note that the enclosing angles $\langle \rangle$ to represent convolution are to be distinguished from the notation $\langle , \rangle$ to represent a distribution. Moreover, greek and latine letters are used to describe scalars and vectors, respectively. Further notations are listed in Table \ref{table:listofsymbols}.
\renewcommand{\arraystretch}{1.3}
\renewcommand{\cellalign}{cl}
\begin{longtable}{| p{.27\textwidth} | p{.73\textwidth}  |} 	
	\hline
	Symbol & Meaning  \\ [0.5ex]
	\hline\hline
	$\naturalnumbers$, $\naturalnumberswithzero$,$\naturalnumbersuntil{n}$
	&
	\makecell{set of natural numbers, natural numbers with zero, \\ natural numbers $\{1,2,\dots n\}$}
	\\
	\hline
	$\realnumbers, \realnumbersvector{n}$
	&
	set of real numbers, $n$-dimensional vectors
	\\
	\hline
	\impulseshift 
	&
	shift
	\\
	\hline
	$\amplitude $
	&
	amplitude
	\\
	\hline
	$\impulsedegree $
	&
	degree
	\\
	\hline
	$\shiftedimpulse{\impulseindex}{\impulsevariable}$ 
	&
	\makecell{Dirac impulse function $\diracdelta(\impulsevariable - \impulseindex s)$ of variable $\impulsevariable$ \\ shifted by $\impulseindex \cdot s$}
	\\ 
	\hline
	$\shiftedimpulseset{\impulsevariable}$ & set of all shifted Dirac impulse functions of variable $\impulsevariable$  \\
	\hline
	$\impulsegroupoperation$ &  addition operator on two impulse functions \\
	\hline
	$\impulsegroup{\impulsevariable}  = (\shiftedimpulseset{\impulsevariable}, \impulsegroupoperation)$ &  abelian group of impulse functions $\shiftedimpulseset{\impulsevariable}$ and group operation $\impulsegroupoperation$ \\
	\hline
	$\impulsespectralspacemult$ &  multiplication operator on a real number and an impulse function \\
	\hline
	$\impulsespectralspace{\impulsevariable} =  (\shiftedimpulseset{\impulsevariable}, \impulsegroupoperation, \impulsespectralspacemult) $& \makecell{\impulsespectralspacename{} with vector addition $\impulsegroupoperation$ \\ and scalar multiplication $\impulsespectralspacemult$ } \\
	\hline
	$\impulsespectraltrain{\impulseshift}{\impulsedegree}{\impulsevariable}$
	&
	\impulsespectraltrainname{} (\impulsespectraltrainshort): series of impulse functions where each is impulse is multiplied an amplitude $\amplitude$
	\\
	\hline
	$\impulsespectraldensity{\impulseshift}{\impulsedegree}{\impulsevariable}$ & \impulsespectraldensityname{} (\impulsespectraldensitynameshort): convolution of $\impulseinterferencedegreetwo \in \naturalnumbers$ \impulsespectraltrainshort s
	\\
	\hline
	$\impulseinterference{\impulseshift}{\impulsedegree}{\impulsevariable}$
	&
	\impulseinterferencename{} (\impulseinterferencenameshort): series of $\impulseinterferencedegreeone \in \naturalnumbers$ \impulsespectraldensitynameshort s
	\\
	\hline
	$\amplitudematrix$, 
	$\amplitudematrixentry{\impulseindex}{\impulseinterferenceindexone}{\impulseinterferenceindextwo}$
	&
	\amplitudematrixname,
	\amplitudevectorname{} at matrix entry $(\impulseinterferenceindexone, \impulseinterferenceindextwo)$
	\\
	\hline 
	$\shiftmatrix{\impulseindex}{\impulsevariable}$,
	$\shiftmatrixentry{\impulseindex}{\impulsevariable}{\impulseinterferenceindexone}{\impulseinterferenceindextwo} $
	&
	\shiftmatrixname{}, 
	\shiftvectorname{} at matrix entry $(\impulseinterferenceindexone, \impulseinterferenceindextwo)$
	\\
	\hline
	$\amplitudematrixentry{\impulseindex}{\impulseinterferenceindexone}{\impulseinterferenceindextwo} \matrixdotproduct 
	\shiftmatrixentry{\impulseindex}{\impulsevariable}{\impulseinterferenceindexone}{\impulseinterferenceindextwo} $
	&
	\dotproductname{} of an \amplitudevectorname{} and an \shiftvectorname{}
	\\
	\hline
	$\amplitudematrix \matrixdotproduct  \shiftmatrix{\impulseindex}{\impulsevariable}$
	&
	\matrixdotproductname{} 
	\\
	\hline
	$\impulsespectraldensityeinsteinmono{\amplitude}{\impulseindex}{\impulseshift}{\impulsevariable}$
	&
	\impulsespectraltrainshort{} in Einstein notation
	\\
	\hline
	$\impulsespectraldensityeinsteinmulti{\amplitude}{\impulseindex}{\impulseshift}{\impulsevariable}$
	&
	\impulsespectraldensitynameshort{} in Einstein notation
	\\
	\hline
	$\innerconvolution{}{}$
	&
	\innerconvolutionname{} that convolves the  columns of the two input matrices
	\\
	\hline
	$\innerconvolution{\amplitudematrix}{\shiftmatrix{\degreevector}{\impulsevariable} }$
	&
	\impulseconvolutionvectorname{} resulting from the \innerconvolutionname{} of the \amplitudematrixname{} and the \shiftmatrixname{}
	\\
	\hline 
	$\impulsespectralinterferenceeinstein{\amplitudematrix \matrixdotproduct \shiftmatrix{\impulsedegree}{\impulsevariable}}$
	&
	\impulseinterferencenameshort{} as a result of the dot product $\innerconvolution{\amplitudematrix}{\shiftmatrix{\degreevector}{\impulsevariable} } \cdot \boldsymbol{1}$
	\\
	\hline
	$	\amplitudevector{\impulseindex_{\impulseinterferenceindexone,1} \dots \impulseindex_{\impulseinterferenceindexone,\multiperiodicimpulsedimension}} \overline{| \langle \phasevectorcomponent{\impulseindex_{\impulseinterferenceindexone,1}} \dots \phasevectorcomponent{\impulseindex_{\impulseinterferenceindexone,\multiperiodicimpulsedimension}} \rangle  |}_{a,b}$
	&
	\heavisidedurationname{}: sum of amplitudes of shifted impulse functions in the interval from a to b
	\\
	\hline
	$\heavisideduration{\Sh}{a,b}{} $
	&
	short-form of \heavisidedurationname
	\\
	\hline
	$\heavisidedurationupperupper{\Sh}{a, b}{}$, $\heavisidedurationupperlower{\Sh}{a, b}{}$, $\heavisidedurationlowerupper{\Sh}{a, b}{}$
	&
	\heavisidedurationname{} of the interval $[a,b]$,$[a,b)$,$(a,b]$
	\\
	\hline
	$\heavisideduration{\Sh}{\intervalduration}{+} $, $\heavisideduration{\Sh}{\intervalduration}{-} $
	&
	\maximumheavisidedurationname{} and \minimumheavisidedurationname{}
	\\ 
	\hline
	$\heavisidedurationupperupper{\Sh}{\intervalduration}{+} $, 				$\heavisidedurationupperlower{\Sh}{\intervalduration}{+} $, 	$\heavisidedurationlowerupper{\Sh}{\intervalduration}{+} $
	&
	\maximumheavisidedurationname{} of all intervals $[\impulsevariable, \impulsevariable +\intervalduration]$, $[\impulsevariable, \impulsevariable +\intervalduration)$, $(\impulsevariable, \impulsevariable +\intervalduration]$, $\impulsevariable \in \realnumbers$
	\\
	\hline
	$\heavisidedurationupperupper{\Sh}{\intervalduration}{-} $, $\heavisidedurationupperlower{\Sh}{\intervalduration}{-} $,  $\heavisidedurationlowerupper{\Sh}{\intervalduration}{-} $
	&
	\minimumheavisidedurationname{} of all intervals $[\impulsevariable, \impulsevariable +\intervalduration]$, $[\impulsevariable, \impulsevariable +\intervalduration)$, $(\impulsevariable, \impulsevariable +\intervalduration]$ $\impulsevariable \in \realnumbers$    \label{table:listofsymbols}
	\\ \hline \caption{List of symbols}
\end{longtable}
	\newpage
	\bibliographystyle{plain}
	\bibliography{bib.bib}

\begin{thebibliography}{10}

\bibitem{albers2006hierarchical}
Karsten Albers, Frank Bodmann, and Frank Slomka.
\newblock Hierarchical event streams and event dependency graphs: A new
  computational model for embedded real-time systems.
\newblock In {\em 18th Euromicro Conference on Real-Time Systems (ECRTS'06)},
  pages 10--pp. IEEE, 2006.

\bibitem{bracewell2000fourier}
Ronald~N. Bracewell.
\newblock {\em The Fourier Transform and Its Applications}.
\newblock Electrical engineering series. McGraw Hill, 2000.

\bibitem{buttazzo2011hard}
Giorgio~C Buttazzo.
\newblock {\em Hard real-time computing systems: predictable scheduling
  algorithms and applications}, volume~24.
\newblock Springer Science \& Business Media, 2011.

\bibitem{clarke1997model}
Edmund~M Clarke.
\newblock Model checking.
\newblock In {\em Foundations of Software Technology and Theoretical Computer
  Science: 17th Conference Kharagpur, India, December 18--20, 1997 Proceedings
  17}, pages 54--56. Springer, 1997.

\bibitem{dirac1981principles}
Paul A.~M. Dirac.
\newblock {\em The principles of quantum mechanics}.
\newblock Oxford university press, 1948.
\newblock 3rd edition.

\bibitem{einstein1916foundation}
Albert Einstein.
\newblock {\em The Foundation of the General Theory of Relativity}, volume~49.
\newblock 1916.

\bibitem{fattohigeneral}
Iwan~Feras Fattohi, Christian Prehofer, and Frank Slomka.
\newblock A general job model for real-time analysis.

\bibitem{grubb2008distributions}
Gerd Grubb.
\newblock {\em Distributions and operators}, volume 252.
\newblock Springer Science \& Business Media, 2008.

\bibitem{hefferon2016linear}
Jim Hefferon.
\newblock {\em Linear algebra}.
\newblock https://hefferon.net/linearalgebra/, 2016.

\bibitem{josephs1950heaviside}
Henry~John Josephs.
\newblock Heaviside's electric circuit theory.
\newblock {\em (No Title)}, 1950.

\bibitem{keisler2013elementary}
H~Jerome Keisler.
\newblock {\em Elementary calculus: An infinitesimal approach}.
\newblock Courier Corporation, 2013.

\bibitem{le2001network}
Jean-Yves Le~Boudec and Patrick Thiran.
\newblock {\em Network calculus: a theory of deterministic queuing systems for
  the internet}.
\newblock Springer, 2001.

\bibitem{richter2004compositional}
Kai Richter.
\newblock {\em Compositional scheduling analysis using standard event models}.
\newblock PhD thesis, 2004.

\bibitem{shortle2018fundamentals}
John~F Shortle, James~M Thompson, Donald Gross, and Carl~M Harris.
\newblock {\em Fundamentals of queueing theory}, volume 399.
\newblock John Wiley \& Sons, 2018.

\bibitem{slomka2021beyond}
Frank Slomka and Mohammadreza Sadeghi.
\newblock Beyond the limitations of real-time scheduling theory: a unified
  scheduling theory for the analysis of real-time systems.
\newblock {\em SICS Software-Intensive Cyber-Physical Systems}, 35(3):201--236,
  2021.

\bibitem{stigge2011digraph}
Martin Stigge, Pontus Ekberg, Nan Guan, and Wang Yi.
\newblock The digraph real-time task model.
\newblock In {\em 2011 17th IEEE Real-Time and Embedded Technology and
  Applications Symposium}, pages 71--80. IEEE, 2011.

\bibitem{strichartz2003guide}
Robert~S. Strichartz.
\newblock {\em A guide to distribution theory and Fourier transforms}.
\newblock World Scientific Publishing Company, 2003.

\bibitem{tao2009analysis}
Terence Tao.
\newblock {\em Analysis i}, volume 185.
\newblock Springer, 2009.

\bibitem{thiele2000real}
Lothar Thiele, Samarjit Chakraborty, and Martin Naedele.
\newblock Real-time calculus for scheduling hard real-time systems.
\newblock In {\em 2000 IEEE International Symposium on Circuits and Systems
  (ISCAS)}, volume~4, pages 101--104. IEEE, 2000.

\end{thebibliography}
\end{document}